\documentclass{revtex4}
\usepackage{epsfig}
\usepackage{amssymb}
\usepackage{amsmath}
\usepackage{amsfonts}
\usepackage{graphicx}
\usepackage{mathrsfs}
\usepackage[dvips]{color}
\usepackage{multirow}

\newcommand{\fa}{\mathfrak{a}}

\newcommand{\fu}{\mathfrak{u}}

\newcommand{\fn}{{\mathfrak{n}}}

\newcommand{\fz}{\mathfrak{z}}

\newcommand{\bM}{\mathbf{M}}

\newcommand{\cH}{\mathcal{H}}
\newcommand{\cE}{\mathcal{E}}

\newcommand{\cP}{\mathcal{P}}

\newcommand{\cT}{\mathcal{T}}

\newcommand{\be}{\begin{equation}}
\newcommand{\ee}{\end{equation}}
\newcommand{\bea}{\begin{eqnarray}}
\newcommand{\eea}{\end{eqnarray}}
\newcommand{\nn}{\nonumber}

\newcommand{\ed}{\end{document}}

\newcommand{\bi}{\begin{itemize}}
\newcommand{\ei}{\end{itemize}}

\newcommand{\bce}{\begin{center}}
\newcommand{\ece}{\end{center}}

\newcommand{\sE}{\mathscr{E}}
\newcommand{\sF}{\mathscr{F}}

\newcommand{\sH}{\mathscr{H}}

\newcommand{\sT}{\mathscr{T}}

\begin{document}


\title{Unidirectional Reflectionlessness and Invisibility in the TE and TM modes of a $\cP\cT$-Symmetric Slab System}

\author{Mustafa~Sar{\i}saman}
\address{Departments of Mathematics and Physics, Ko\c{c}
University,\\ Sar{\i}yer 34450, Istanbul, Turkey\\
msarisaman@ku.edu.tr}

\begin{abstract}
Unidirectional invisibility of a $\cP\cT$-symmetric optical system is of great interest, but challenging as well since it is infeasible to fulfill it through wide optical frequency ranges in all angular directions. Accordingly we study reflectionless and invisible patterns in the TE and TM modes of an optical slab system consisting of adjacent or separated pair of balanced gain and loss layers with a gap. We provide a comprehensive study of one of the simplest experimentally accessible examples of a unidirectionally reflectionless and invisible $\cP\cT$-symmetric optical slab system. We obtain the physically optimal conditions for the realization of these phenomena. We derive analytic expressions, and show that only certain gain amounts restricted to take values between certain minimum and maximum values give rise to uni/bi-directionally invisible configurations. The size of gap decides the measure of reflectionlessness and invisibility parameters, especially on gain value and incident angle.
\medskip

\noindent {Pacs numbers: 03.65.Nk, 42.25.Bs, 42.60.Da, 24.30.Gd}
\end{abstract}

\maketitle

\section{Introduction}

Since the first debut of $\cP\cT$-symmetric quantum mechanics~\cite{bender}, there has been tremendous efforts towards understanding the non-Hermition operators which give rise to real energies. In this respect,  psedo-Hermiticity~\cite{ijgmmp-2010} has revealed numerous mysteries which found many substantial applications especially in quantum field theories~\cite{bender2}, Lie algebras~\cite{bagchi}, optical and condensed matter systems~\cite{PT1, PT2, PT3, PT4, PT5, PT6, PT7, PT8}. It is a prominent feature of a $\cP\cT$-symmetric Hamiltonian that the potential associated with it obeys $V(x) = V^{\star} (-x)$~\cite{bender,PT1, PT4, PT5}. In this respect, realizing complex $\cP\cT$-symmetric potentials in the realm of optics is achieved by the formal equivalence between the quantum mechanical Schr\"{o}dinger equation and the optical wave equation derived from Maxwell's equations. Optics is the field that can provide a fertile ground where $\cP\cT$-related notions including the nonreciprocal responses, the power oscillations, optical transparency, optical solitons and unidirectional invisibility can be implemented and experimentally investigated~\cite{PT1, PT2, PT3}. By exploiting optical modulation of the refractive index in the complex dielectric permittivity plane and engineering both optical absorption and amplification, $\cP\cT$-symmetric optical systems can lead to a series of intriguing optical phenomena and devices, such as dynamic power oscillations of light propagation, coherent perfect absorber lasers~\cite{CPA, pra-2015d, lastpaper} and unidirectional invisibility~\cite{PT1, PT6, PT7}.

It is revealed that the evolution of parity-time symmetry becomes measurable through the quantum-optical analogue. An interesting phenomenon named unidirectional invisibility was theoretically proposed at exceptional point where amplitudes of the real and imaginary parts of the modulated refractive index are identical, and $\cP\cT$-symmetry is spontaneously broken, in parity-time metamaterials. Inspired from this idea we investigate the feasibility of realizing unidirectional reflectionlessness and invisibility properties of a $\cP\cT$-symmetric optical structure by means of a optically active real material using the impressive power of transfer matrix in the framework of quantum scattering formalism. In this context, it is presented that spectral singularities and unidirectional invisibility are leading issues encountered in scattering states of electromagnetic fields~\cite{naimark, ss-math, p123, pra-2012a, longhi4, longhi3}. The role of $\cP\cT$-symmetry in the context of unidirectional indivisibility is similar to its role in the study of spectral singularities~\cite{lin1}

In a scattering problem, all the scattering data is contained in transfer matrix~\cite{prl-2009}. It is highly advantageous and thus more preferable rather than scattering matrix due to its composition property. It is a magnificent feature of transfer matrix that exploits spectral singularities and invisibility of electromagnetic fields interacted with an optically active medium~\cite{CPA, pra-2015d, lastpaper, p123}. Spectral singularities correspond to zero with resonance states giving the real and positive energies~\cite{naimark, ss-math}. They produce purely outgoing waves and have connection with the lasing threshold conditions. However, invisibility is the point that arose most curiosity about the transfer matrix, and requires a lot of work to do.

Studies about invisibility problem in literature is two-fold. On one side, one exploits the beauty of transformation optics and stunning competency of metamaterials~\cite{pendry1, leonhardt1}. This approach employs the truth that object being invisible is to be concealed behind an artificially manufactured material~\cite{metamaterials}. The complication in the process of fabrication on account of geometrical requirements and challenges in the applicability process highlights the second approach benefited from interferometric methods heading the transfer matrix, which has found a growing interest in recent years~\cite{pra-2012a, longhi4, longhi3, pra-2015b, pra-2015a, longhi1, pra-2014a, soto, midya, jpa-2014a, pra-2013a, longhi2, lin1}.

\begin{figure}
    \begin{center}
    \includegraphics[scale=.60]{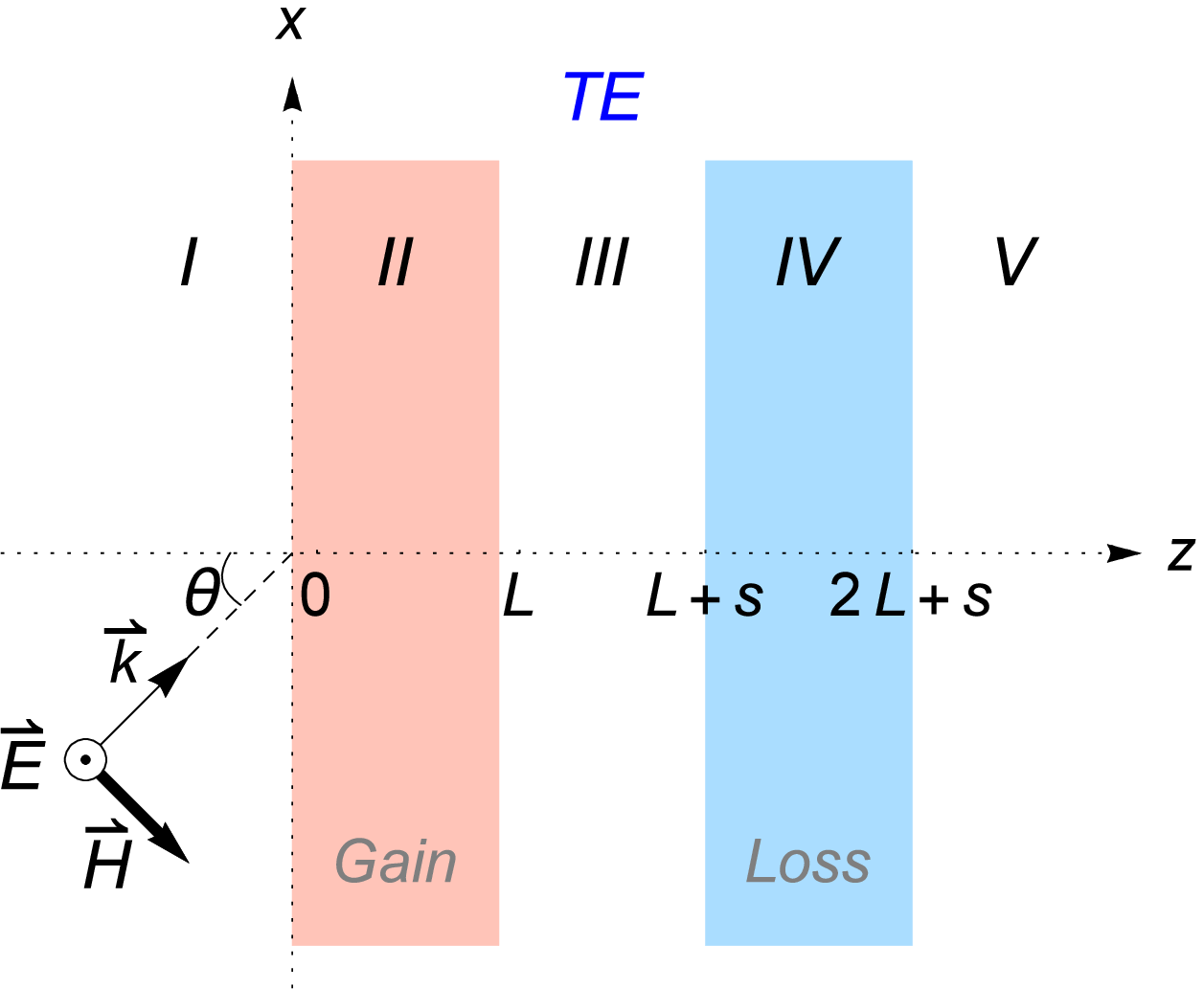}~~~~~
    \includegraphics[scale=.60]{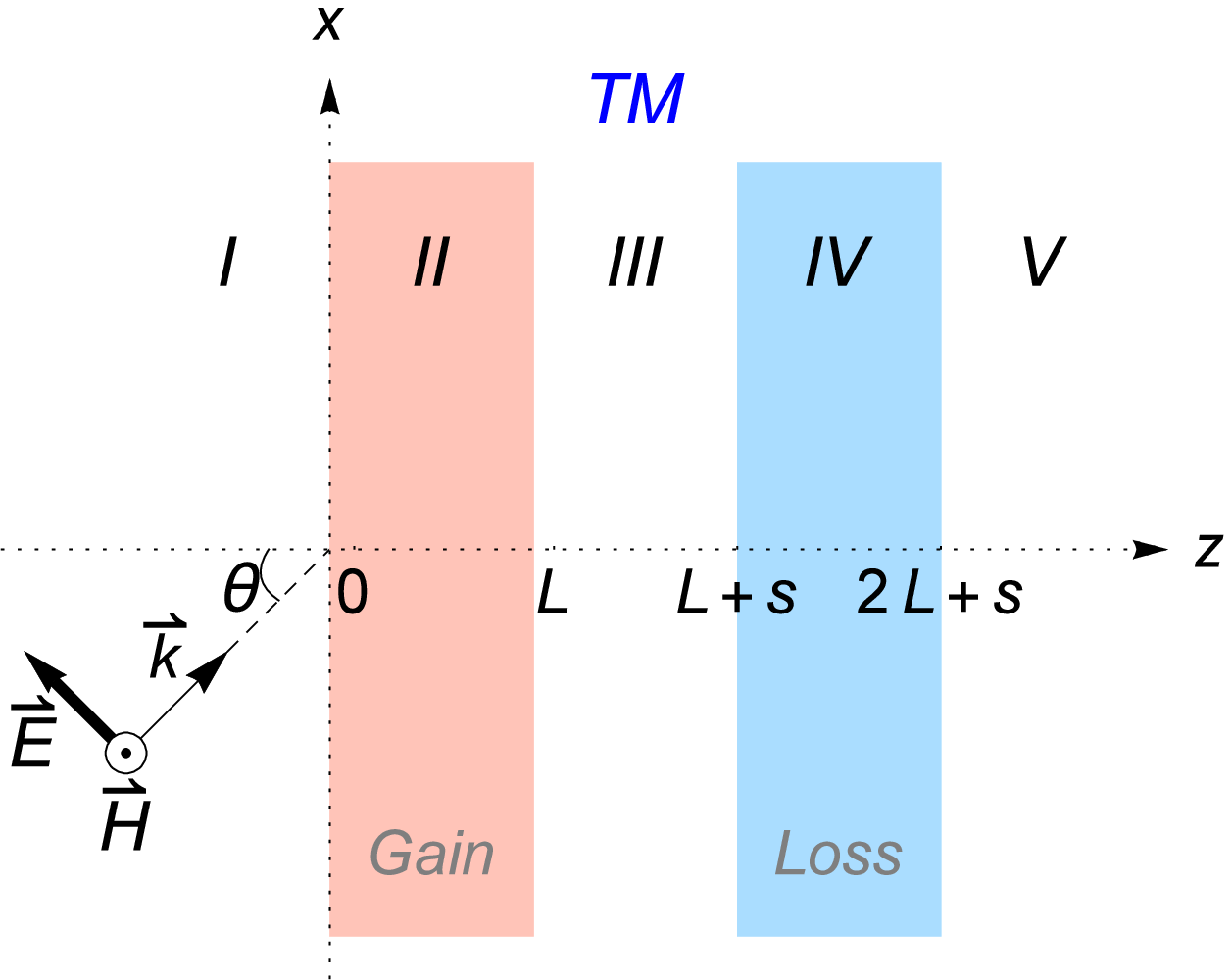}
    \caption{(Color online)  TE (on the left) and TM (on the right) modes of a slab system consisting of a pair of loss and gain layers of thickness $L$ that are placed a distance $s$ apart in vacuum. The symbols $I$, $I\!I$, $I\! I\! I$, $I\!V$, and $V$ respectively label the regions of the space corresponding to $z<0$, $0<z<L$, $L<z<L+s$, $L+s<z<2L+s$, and $z>2L+s$. $I\!I$ and $I\!V$ respectively correspond to the gain and loss layers while  $I$, $I \!I\! I$, and $V$ represent the vacuum.}
    \label{fig1}
    \end{center}
    \end{figure}

In~\cite{jones1}, the effect of oblique incidence directed upon a exponential potential region on the property of invisibility is investigated. In~\cite{lastpaper}, we throughly studied the spectral singularities and coherent perfect absorption (CPA) features in the oblique TE and TM modes of a $\cP\cT$-symmetric planar slab system that consists of a pair of balanced gain and loss layers of thickness $L$ separated by a distance $s\geq 0$. We showed the optimal conditions of realizing a CPA laser due to the power of transfer matrix formalism.  In the present article we conduct a comprehensive study of unidirectional reflectionlessness and invisibility in the oblique TE and TM modes of the same system to unveil the intriguing traits of transfer matrix as the complementary to~\cite{lastpaper}. Our system is depicted in Fig.~\ref{fig1}.

Our analysis of a $\cP\cT$-symmetric slab system in its TE and TM modes reveals all possible configurations of this system that support unidirectional reflectionlessness and invisibility. In order to determine the practically most desirable choices among these, we calculate complete solutions and schematically demonstrate their behaviors using various parameter choices. This provides valuable information about unidirectional reflectionlessness and invisibility for a possible experimental realization of a $\cP\cT$-symmetric slab system.

In particular, we obtain analytic expressions for reflectionless and invisible configurations, examine the behavior of  right and left reflection amplitudes for the TE and TM waves. We explicitly show that optimal control of parameters such as gain coefficient, incident angle, slab thickness and gap width give rise to a desired outcome of achieving wide wavelength range of unidirectional reflectionlessness and invisibility. In fact, we provide a concrete sounding grounds that reveal the range of gain coefficient to be restricted between minimum and maximum values.

\section{TE and TM Modes of a Parallel Pair of Slabs}
\label{S2}

Consider two layer gain-loss slab system with a gap between gain and loss regions as sketched in Fig.~\ref{fig1}. Assume that our problem is one dimensional and complex refractive indices identifying gain and loss regions of space respectively by $\fn_2$ and $\fn_4$ are $z$-dependent. This optically active system obeys the Maxwell's equations in time-harmonic forms\footnote{In time harmonic forms $\vec\cE (\vec{r}, t)$ and $\vec\cH (\vec{r}, t)$ fields are respectively given by $\vec\cE (\vec{r}, t) = e^{-i\omega t} \vec{E}(\vec{r})$ and $\vec\cH (\vec{r}, t)= e^{-i\omega t} \vec{H}(\vec{r})$.} and therefore leads to a couple of TE and TM mode solutions of the Helmholtz equations respectively in the form
    \begin{align}
    &\left[\nabla^{2} +k^2\fz_j(z)\right] \vec{E}^j(\vec{r}) = 0, &&
    \vec{H}^j(\vec{r}) = -\frac{i}{k Z_{0}} \vec{\nabla} \times \vec{E}^j(\vec{r}),
    \label{equation4}\\
    &\left[\nabla^{2} +k^2\fz_j(z)\right] \vec{H}^j(\vec{r}) = 0,&&
    \vec{E}^j(\vec{r}) = \frac{i Z_{0}}{k \fz_j(z)} \vec{\nabla} \times \vec{H}^j(\vec{r}),
    \label{equation5}
    \end{align}
where $\vec r:=(x,y,z)$ symbolizes the cartesian coordinate system, $k:=\omega/c$ is the wavenumber, $Z_{0}:=\sqrt{\mu_{0}/\varepsilon_{0}}$ is the impedance of the vacuum, $c:=1/\sqrt{\mu_{0}\varepsilon_{0}}$ is the the speed of light in vacuum,  and $\varepsilon_0$ and $\mu_0$ are respectively the permittivity and permeability of the vacuum. Here the subindex $j = 1, 2, 3, 4$ and $5$ represents the regions of space depicted in Fig.~\ref{fig1}. The function $\fz_j(z)$ is given by
    \be
    \fz_j(z):= \fn_j^2 ~~~~~~~ {\rm for}~  z\in z_j
    \label{e1}
    \ee
We observe that $\fn_1 = \fn_3= \fn_5 = 1$ and all denotes the vacuum. We remark that the indice $j$ in the field vectors denotes the field components in the specified $j^{\textrm{th}}$-region of space.

It is the demonstrative feature of transverse electric (TE) and transverse magnetic (TM) waves that they correspond to the solutions of (\ref{equation4}) and (\ref{equation5}) for which $\vec E(\vec{r})$ and $\vec H(\vec{r})$ are respectively parallel to the surface of the slab, which is aligned along the $y$-axis. Suppose that plane wave is incident on the gain-loss slab system with an angle $\theta\in[-90^\circ,90^\circ]$ from the left-hand side (See Fig.~\ref{fig1}). Thus wave vector $\vec k$  has the components in the $x$-$z$ plane denoted by $k_x$ and $k_x$ respectively as follows
    \begin{align}
 &k_x:=k\sin\theta, &&k_z:=k\cos\theta,
    \end{align}
In this geometrical set-up, the electric field for the TE waves and the magnetic field for the TM waves are respectively given by
    \begin{align}
    &\vec E^j(\vec{r})=\sE^j(z)e^{ik_{x}x}\hat e_y,
    && \vec H^j(\vec{r})=\sH^j (z)e^{ik_{x}x}\hat e_y,
    \label{ez1}
    \end{align}
where $\hat e_x,\hat e_y,$ and $\hat e_z$ are respectively the unit vectors along the $x$-, $y$- and $z$-axes, and $\sE^j$ and $\sH^j$ are solutions of the Schr\"odinger equation
    \be
    -\psi^{j''}(z)+v_j(z)\psi^j(z)=k^2\psi^j(z),~~~~~~~~~~z\notin\{ 0,L,L+s,2L+s\},
    \label{sch-eq}
    \ee
for the potential $v_j(z):=k^2[1+\sin^2\theta-\fz_j(z)]$. Because $v_j(z)$ is a constant potential in the $j^{th}$-division of relevant space, we can easily solve (\ref{sch-eq}) to obtain
    \be
    \psi^j(z):=
    a_j\,e^{i\tilde{k}_j z} + b_j\,e^{-i\tilde{k}_j z} ~~~~ {\rm for} ~ z\in z_j
    \label{E-theta}
    \ee
where $a_j$ and $b_j$, with $i=1,2,3,4,5$, are complex valued amplitudes, possibly $k$-dependent, and
    \begin{align}
    &{\tilde k}_{j}:=k\sqrt{\fn_{j}^2-\sin^2\theta}=k_{z}\tilde\fn_{j},
    &&\tilde\fn_{j}:=\frac{\sqrt{\fn_{j}^2 -\sin^2\theta}}{\cos\theta}.
    \label{tilde-parm}
    \end{align}

Substituting (\ref{ez1}) in the second equation in (\ref{equation4}) and (\ref{equation5}), we can find the magnetic field for the TE waves and the electric field for the TM waves inside and outside of the slabs. We then impose the appropriate boundary conditions for the problem to relate the coefficients $a_j$ and $b_j$. These amount to the requirement that the tangential components of $\vec E$ and $\vec H$ must be continuous functions of $z$ at $z=0$, $z=L$, $z=L+s$ and $z=2L+s$. Table~\ref{table01} gives explicit expressions for the components of the the electric and magnetic fields, and relation (\ref{boundaryconds}) exhibits the corresponding set of boundary conditions in compact form.%
    \begin{table}[!htbp]
    \begin{center}
	{
    \begin{tabular}{|c|c|}
    \hline
    TE-Fields & TM-Fields \\
    \hline & \\[-6pt]
    $\begin{aligned}
    & E_{x}^j=E_{z}^j=H_{y}^j=0\\[3pt]
    & E_{y}^j=\sE^j(z)\,e^{ik_{x}x}\\[3pt]
    &H_{x}^j =-\frac{\sF^j(z)}{Z_0}\,\sT^j(x,z)\\[3pt]
    &H_{z}^j =\frac{\sin\theta\, e^{ik_{x}x}\sE^j(z)}{Z_0}\\[3pt]
    \end{aligned}$ &
    $\begin{aligned}
    & E_{y}^j=H_{x}^j=H_{z}^j=0\\[2pt]
    & E_{x}^j =\frac{Z_{0}\,\sF^j(z)}{\fz_j(z)}\,\sT^j(x,z)\\[3pt]
    & E_{z}^j =- \frac{Z_{0}\sin\theta\, e^{ik_{x}x}\sH^j(z)}{\fz_j(z)}\\
    & H_{y}^j =\sH^j(z)\,e^{ik_{x}x}\\[-8pt]
    &
    \end{aligned}$\\
    \hline
    \end{tabular}}
    \vspace{6pt}
    \caption{Components of the TE and TM fields in cartesian coordinates. Here $\sE^j(z)$ and $\sH^j(z)$ are given by the right-hand side of (\ref{E-theta}), and $\sF^j(z)$ and $\sT^j(x,z)$ are respectively defined by (\ref{F-theta}) and (\ref{T1-2}).}
    \label{table01}
    \end{center}
    \end{table}%

   \be
   a_j e^{i \tilde{k}_j \xi} \pm b_j e^{-i \tilde{k}_j \xi} = [\fu_{j+1}]^{(1 \mp 1)\frac{c}{2}} \left\{a_{j+1} e^{i \tilde{k}_{j+1} \xi} \pm b_{j+1} e^{-i \tilde{k}_{j+1} \xi}\right\} \label{boundaryconds}
   \ee
where $\xi$ takes values at boundaries $z= 0, L, L+s$ and $2L+s$, and
    \be
    c :=\left\{\begin{array}{ccc}
    +1 & {\rm for~odd~layers}\\
    -1 & {\rm for~even~layers}.
    \end{array}\right.
    \label{c-value}
     \ee

They involve the following quantities.
    \bea
    \sF^j(z)&:=&
     a_j\,e^{i\tilde{k}_j z} - b_j\,e^{-i\tilde{k}_j z} ~~~~ {\rm for} ~ z\in z_j,
    \label{F-theta}\\[6pt]
    \sT^j(x,z)&:=& \tilde{\fn}_j e^{i k_x x}\cos\theta ,
    \label{T1-2}\\[6pt]
    \fu_{j}&:=&\frac{\tilde{\fn}_j}{\fn_j^{\ell}}.
    \label{u=}
    \eea
\be
 \ell :=\left\{\begin{array}{cc}
    0 & \mbox{for TE waves},\\[6pt]
    2 & \mbox{for TM waves}.
    \end{array}\right.
    \label{ell=}
    \ee

\section{Transfer Matrix Formalism}
\label{S3}

Scattering properties of any multi-component system can be best understood by means of transfer matrix formalism. Advantage of using transfer matrix in a multi-layer system lies in the fact that its composition property lets the resultant transfer matrix to be figured out in terms of individual matrices that comprise the multi-layer system. For our two-layer system~\cite{lastpaper}, transfer matrix can be described as

    \begin{align}
    &\left[\begin{array}{c}
    a_3\\ b_3\end{array}\right]=\bM_1 \left[\begin{array}{c}
    a_1\\ b_1\end{array}\right],
    &&\left[\begin{array}{c}
    a_5\\ b_5\end{array}\right]=\bM_2 \left[\begin{array}{c}
    a_3\\ b_3\end{array}\right],
    &&\left[\begin{array}{c}
    a_5\\ b_5\end{array}\right]=\bM \left[\begin{array}{c}
    a_1\\ b_1\end{array}\right].
    \nn
    \end{align}
where $\bM_1$, $\bM_2$ are $2\times 2$ matrices corresponding to the slabs placed in regions $I\!I$ and $I\!V$, and $\bM=[M_{ij}]$ is the transfer matrix of the composite system. They all satisfy the composition property $\bM=\bM_2\bM_1$. Transfer matrix can also be expressed by means of (right and left) reflection and transmission coefficients~\cite{prl-2009} via
\begin{align}
\bM=\left(
  \begin{array}{cc}
    T-\frac{R^{l} R^{r}}{T} & \frac{R^{r}}{T} \\
    -\frac{R^{l}}{T} & \frac{1}{T} \\
  \end{array}
\right)\label{transfermatrix}
\end{align}
Ref.~\cite{pra-2015a} gives an explicit expression for $\bM_1$. We can easily compute $\bM_2$ using this expression and the transformation property of the transfer matrices under translations $z\stackrel{T_a}{\longrightarrow}z-a$ which has the form \cite{pra-2014b}:
    \begin{align}
    &M_{11}\stackrel{T_a}{\longrightarrow}M_{11},
    &&M_{22}\stackrel{T_a}{\longrightarrow}M_{22},
    &&M_{12}\stackrel{T_a}{\longrightarrow}e^{-2iak_z}M_{12},
    &&M_{21}\stackrel{T_a}{\longrightarrow}e^{2iak_z}M_{21}.\nn
    \end{align}
With $\bM_1$ and $\bM_2$ computed we can determine $\bM$ using $\bM=\bM_2\bM_1$. Components of this matrix satisfy the symmetry relations in~\cite{pra-2014c}
 \begin{align}
    &M_{11}\stackrel{\mathcal{PT}}{\longleftrightarrow}M^{*}_{22},
    &&M_{12}\stackrel{\mathcal{PT}}{\longleftrightarrow}-M^{*}_{12},
    &&M_{21}\stackrel{\mathcal{PT}}{\longleftrightarrow}-M^{*}_{21}.\nn
    \end{align}
and are described as follows
    \bea
    &M_{11} &=\cos\fa_2\cos\fa_4\Big[1 + i\,\mathfrak{u}^{+}_2 \tan\fa_2+ i\,\mathfrak{u}^{+}_4  \tan\fa_4+
    (\fu_2^-\fu_4^-e^{-2ik_z s}-\fu_2^+\fu_4^+)\tan\fa_2 \tan\fa_4 \Big] e^{-2ia_0},\nn\\
    &M_{12} &=\cos\fa_2\cos\fa_4\Big[i\,\mathfrak{u}^{-}_2 \tan\fa_2+ i\,\mathfrak{u}^{-}_4  \tan\fa_4e^{-2ik_z s}+
    (\fu_2^+\fu_4^-e^{-2ik_z s} -\fu_2^-\fu_4^+)\tan\fa_2 \tan\fa_4 \Big] e^{-2ia_0},\nn\\
     &M_{21} &=-\cos\fa_2\cos\fa_4\Big[i\,\mathfrak{u}^{-}_2 \tan\fa_2+ i\,\mathfrak{u}^{-}_4  \tan\fa_4e^{2ik_z s}-
    (\fu_2^+\fu_4^-e^{2ik_z s} -\fu_2^-\fu_4^+)\tan\fa_2 \tan\fa_4 \Big] e^{2ia_0},\nn\\
    &M_{22} &=\cos\fa_2\cos\fa_4\Big[1 - i\,\mathfrak{u}^{+}_2 \tan\fa_2- i\,\mathfrak{u}^{+}_4  \tan\fa_4+
    (\fu_2^-\fu_4^-e^{2ik_z s}-\fu_2^+\fu_4^+)\tan\fa_2 \tan\fa_4 \Big] e^{2ia_0},
    \label{M22=x}
    \eea
where for $j=1,2,3,4,5$ we have introduced
    \begin{align}
    &\fa_j:=k_zL\tilde\fn_j,
    &&\fu_j^\pm:=\frac{1}{2}\left(\fu_{j} \pm \fu^{-1}_{j}\right).\nn
    \end{align}
and singled out the identification $\fa_1 = \fa_3 = \fa_5 := a_0$. Transfer matrix (\ref{transfermatrix}) gives rise to many intriguing phenomena. We already studied spectral singularities in \cite{lastpaper}. In this paper we focus on its another fascinating feature, unidirectional reflectionlessness and in turn invisibility, which can be seen directly from the form of transfer matrix. We observe distinct cases from transfer matrix (\ref{transfermatrix}) knowledge as follows~\cite{pra-2012a}:
\begin{enumerate}
  \item If $R^{l} = 0$ and $R^{r}\neq 0$, then the prescribed potential is called ``reflectionless from left". This in turn implies that $M_{12}\neq 0$ together with $M_{21}= 0$.
  \item If $R^{r} = 0$ and $R^{l}\neq 0$, then the potential given is called ``reflectionless from right". This condition yields that $M_{21}\neq 0$ together with $M_{12}= 0$.
  \item If the potential which is reflectionless from left comes along with the situation $T = 1$, potential is named ``invisible from left". This condition implies $M_{11} = M_{22} = 1$ in addition to the results in case (1).
  \item Likewise if the potential which is reflectionless from right, then it is named ``invisible from right". Accordingly this condition implies $M_{11} = M_{22} = 1$ in addition to the results in case (2).
\end{enumerate}
We analyze all these cases alternately in the following sections.

\section{Unidirectionally Reflectionless Potentials}
\label{S4}
We realize that the conditions $R^{l} = 0$ and $R^{r} = 0$ give rise to the following pair of equations
\begin{align}
i\mathfrak{u}^{-}_2 \tan\fa_2 + i\mathfrak{u}^{-}_4 \tan\fa_4 e^{\pm 2ik_zs} \mp (\mathfrak{u}^{+}_2 \mathfrak{u}^{-}_4 e^{\pm 2ik_zs} - \mathfrak{u}^{-}_2 \mathfrak{u}^{+}_4)\tan\fa_2 \tan\fa_4 = 0 \label{unireflectionlesswiths}
 \end{align}
where equation with upper sign belongs to $R^{l} = 0$ and the one with lower sign is relevent to $R^{r} = 0$. Notice that these two relations are the negations of each other if one desires to emanate the unidirectional reflectionlessness.

\subsection{Unidirectional Reflectionlessness: A Perturbative Analysis Approach}
\begin{figure}
	\begin{center}
    \includegraphics[scale=0.7]{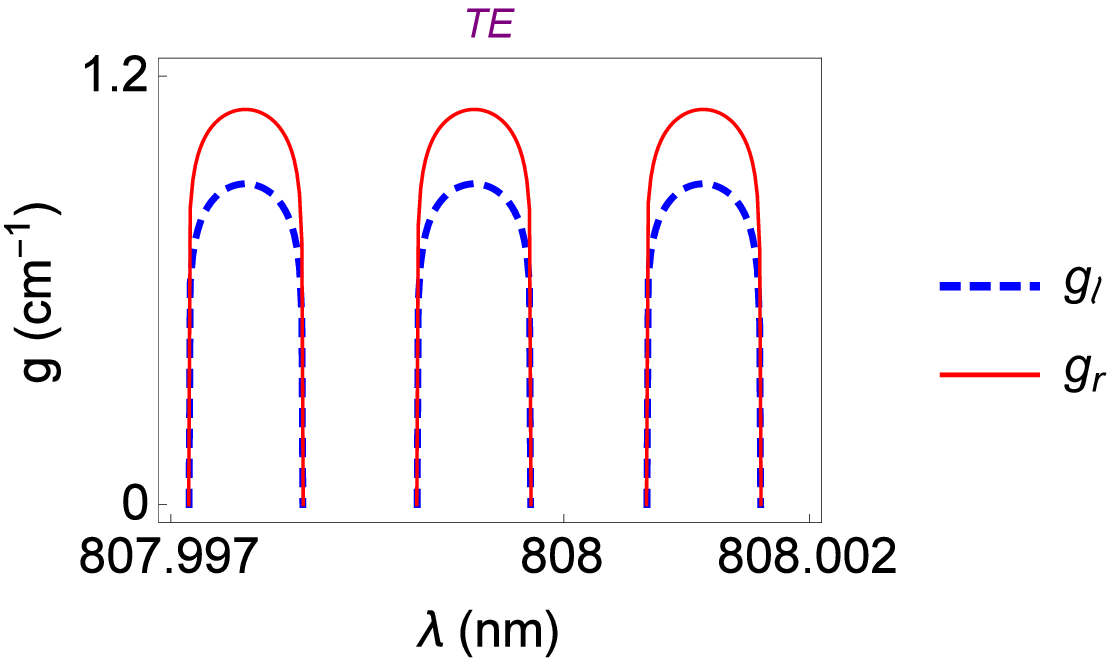}~~~
    \includegraphics[scale=0.7]{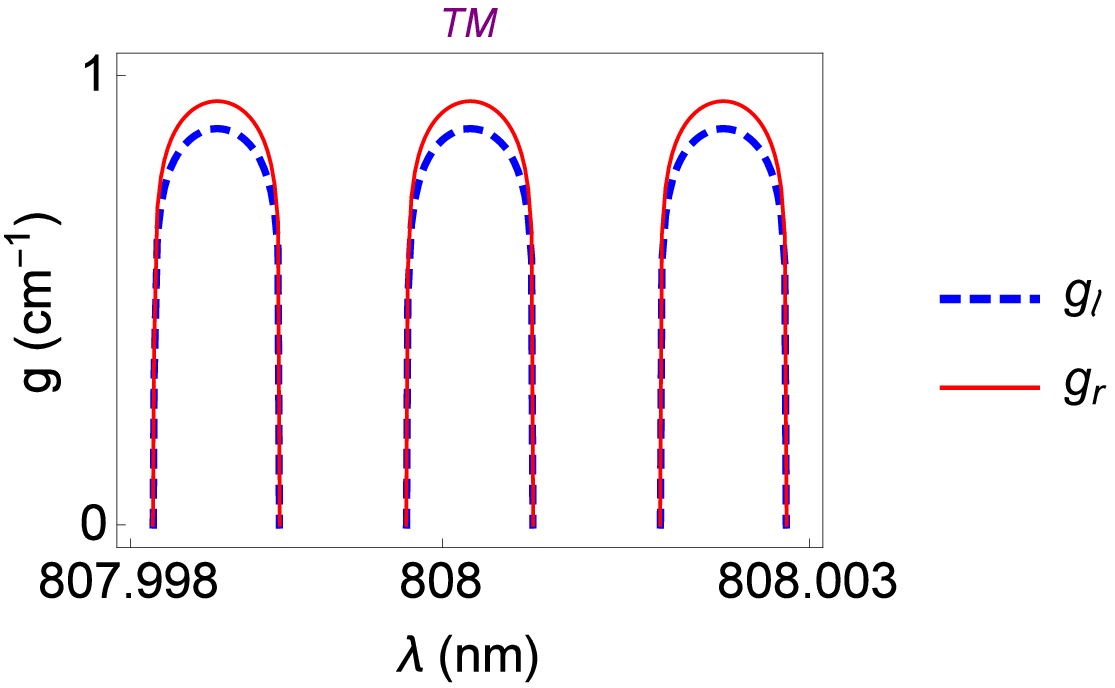}
	\caption{(Color online) Plots of gain coefficient $g$ as a function of wavelength $\lambda$ corresponding to TE and TM wave solutions of a $\cP\cT$-symmetric bilayer in~(\ref{reflectionlessgain}), which gives the uni/bi-directionally reflectionless potential configurations at incident angle of $30^\circ$. Thick dashed blue curves represent gain values that yields the left reflectionless case, while thin solid red curves the right reflectionless situation which are excluded from left-reflectionless situation for the case of only left reflectionless case. }
    \label{g01TE}
    \end{center}
    \end{figure}

\begin{figure}
	\begin{center}
    \includegraphics[scale=0.7]{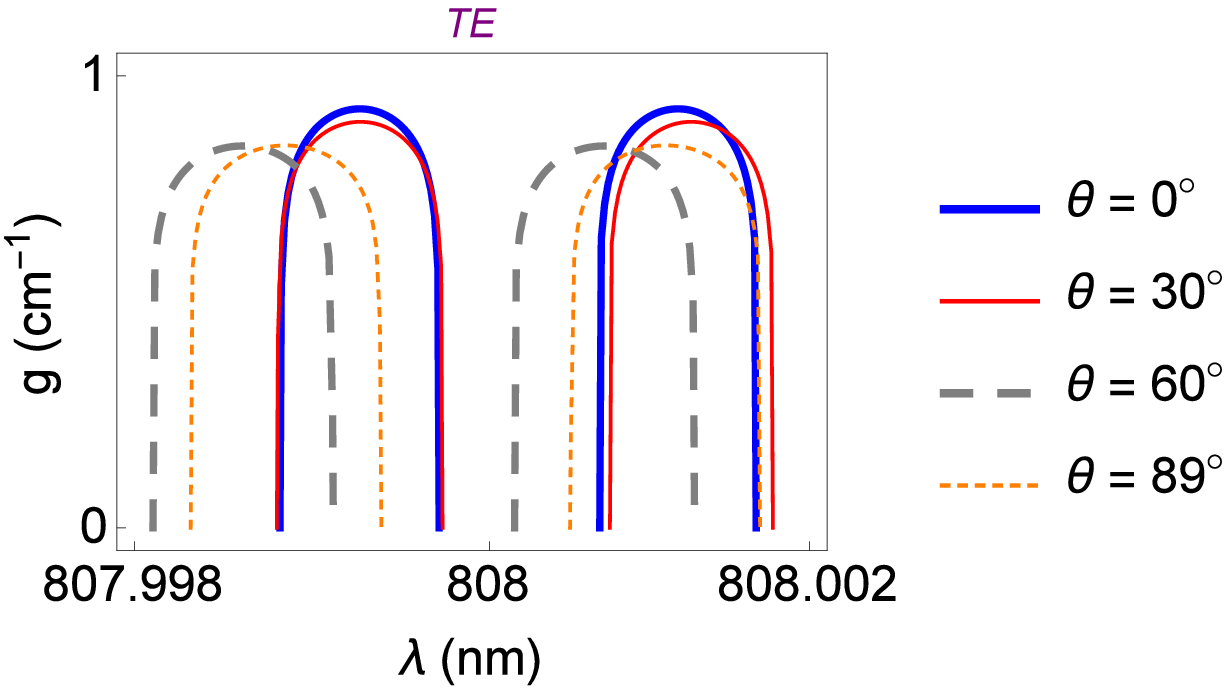}~~~
    \includegraphics[scale=0.7]{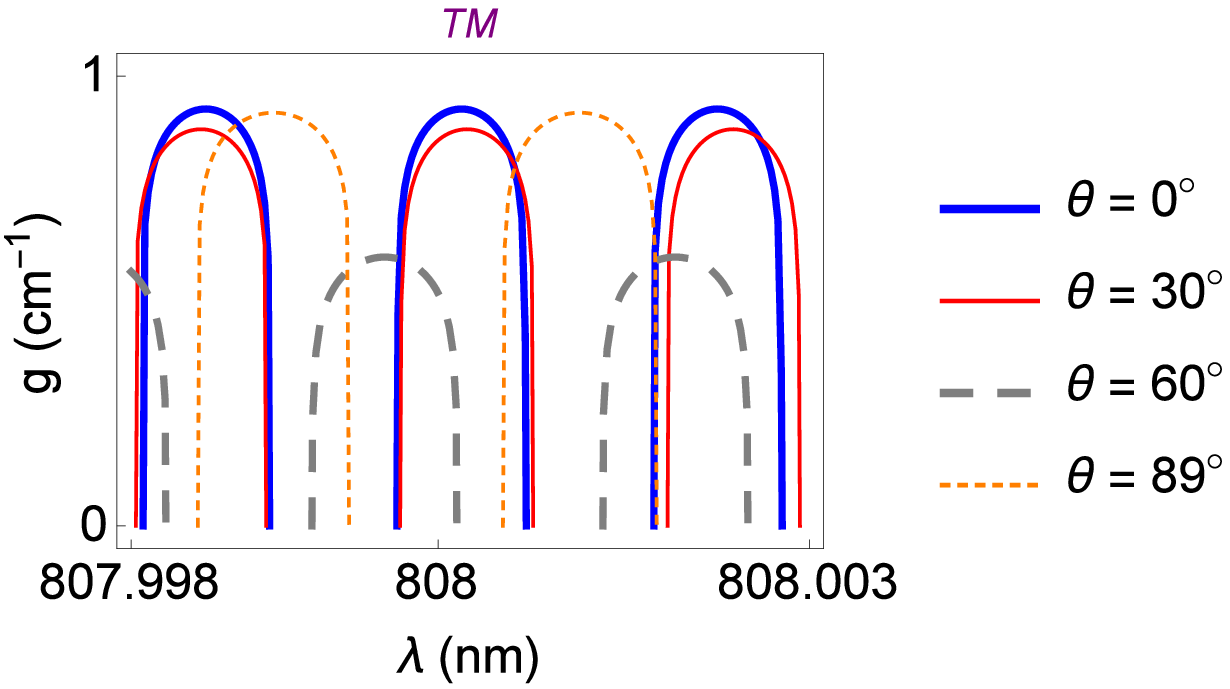}
	\caption{(Color online) Plots of gain coefficient $g$ as a function of wavelength $\lambda$ at various incidence angles, which show left side reflectionlessness for TE and TM wave solutions respectively of $\cP\cT$-symmetric Nd:YAG bilayer in~(\ref{reflectionlessgain}). }
    \label{g02TE}
    \end{center}
    \end{figure}

\begin{figure}
	\begin{center}
    \includegraphics[scale=0.6]{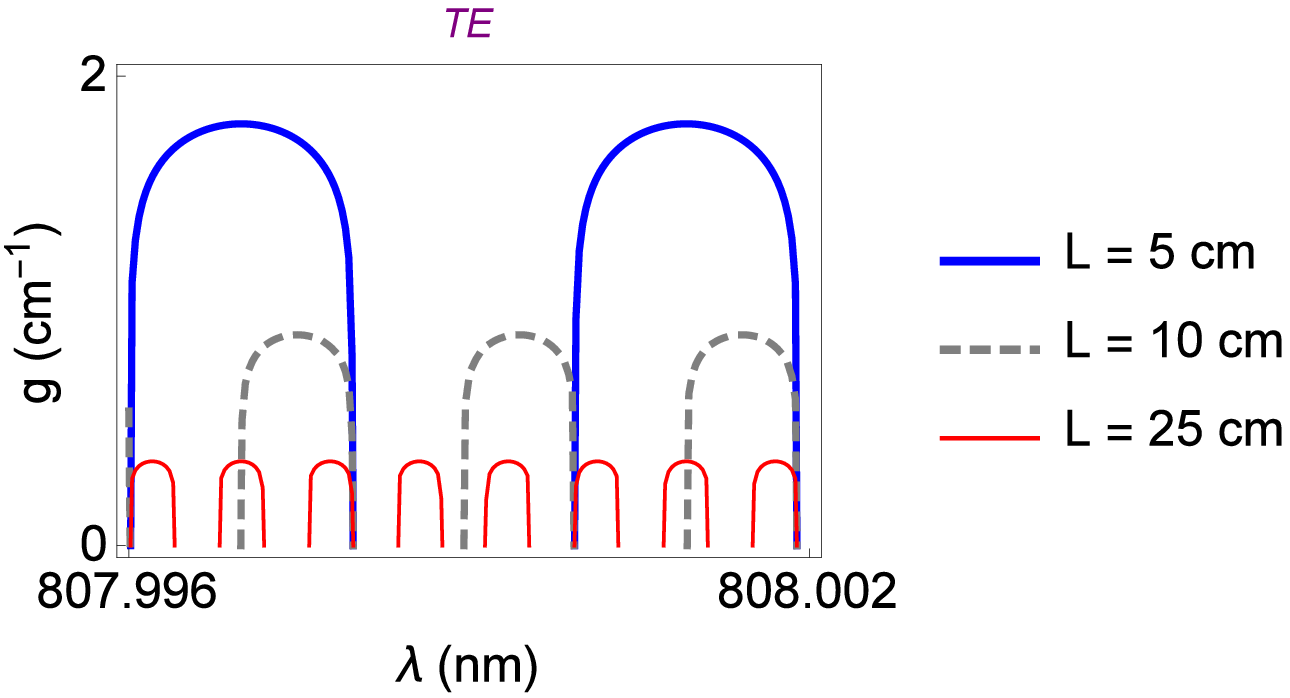}~~~
    \includegraphics[scale=0.6]{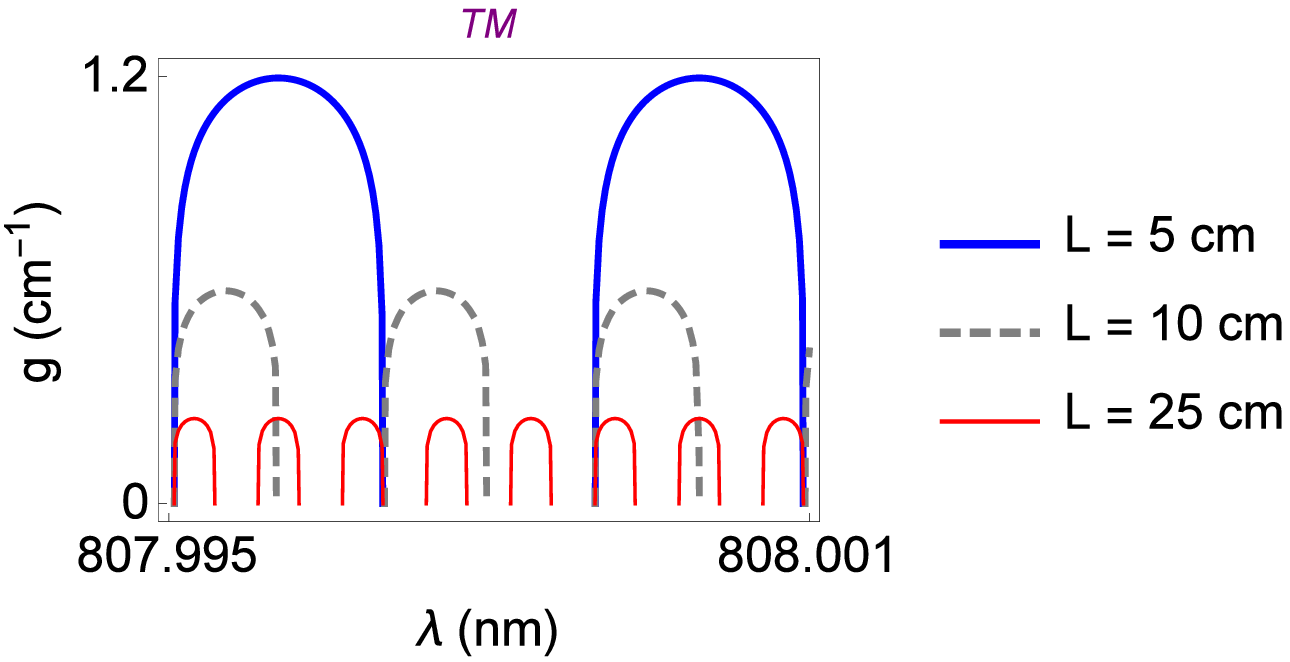}
	\caption{(Color online) Plots of gain coefficient $g$ as a function of wavelength $\lambda$ at various slab thicknesses at $\theta = 30^\circ$, which show left-reflectionlessness for TE and TM wave solutions respectively of $\cP\cT$-symmetric Nd:YAG bilayer crystals in~(\ref{reflectionlessgain}). }
    \label{g03TE}
    \end{center}
    \end{figure}

\begin{figure}
	\begin{center}
    \includegraphics[scale=0.4]{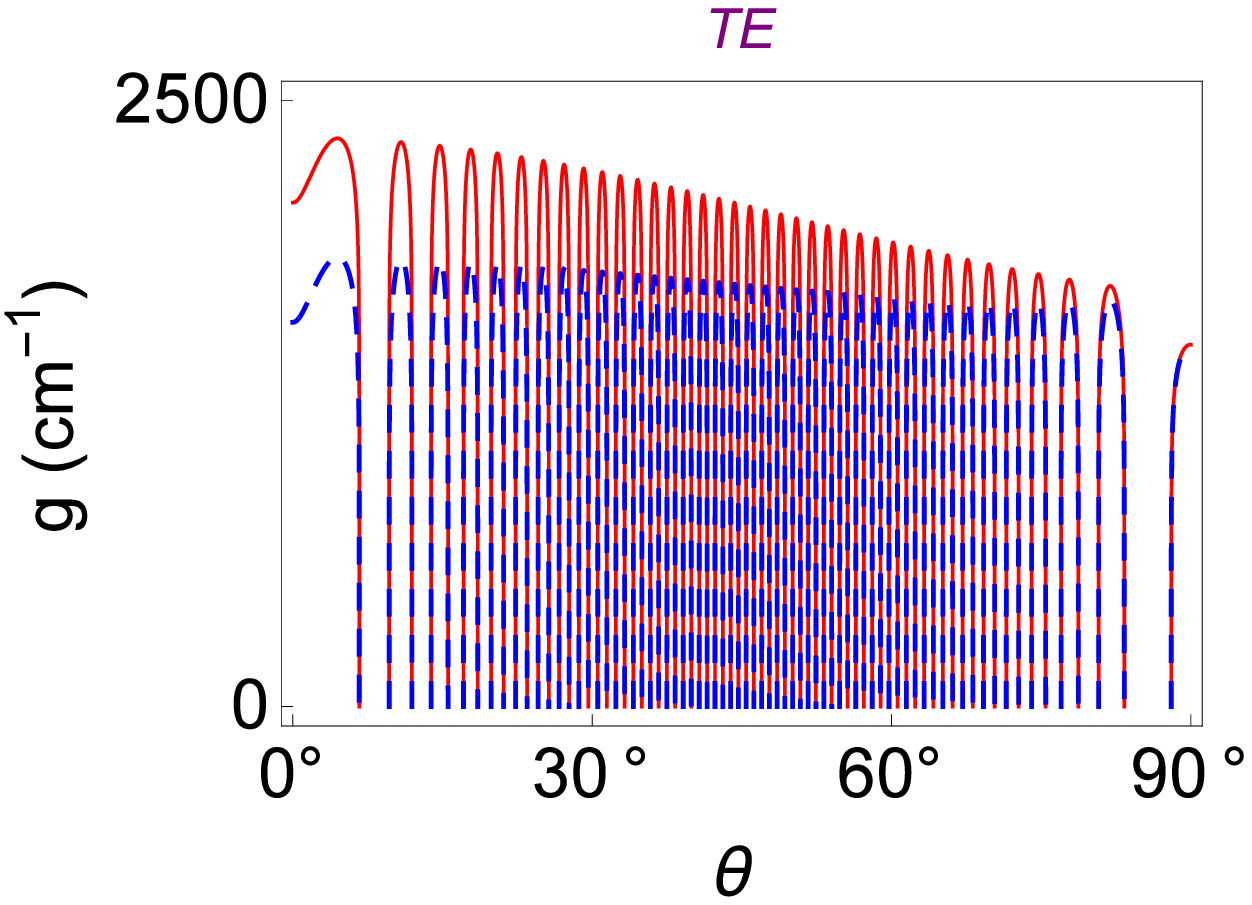}~~~
    \includegraphics[scale=0.4]{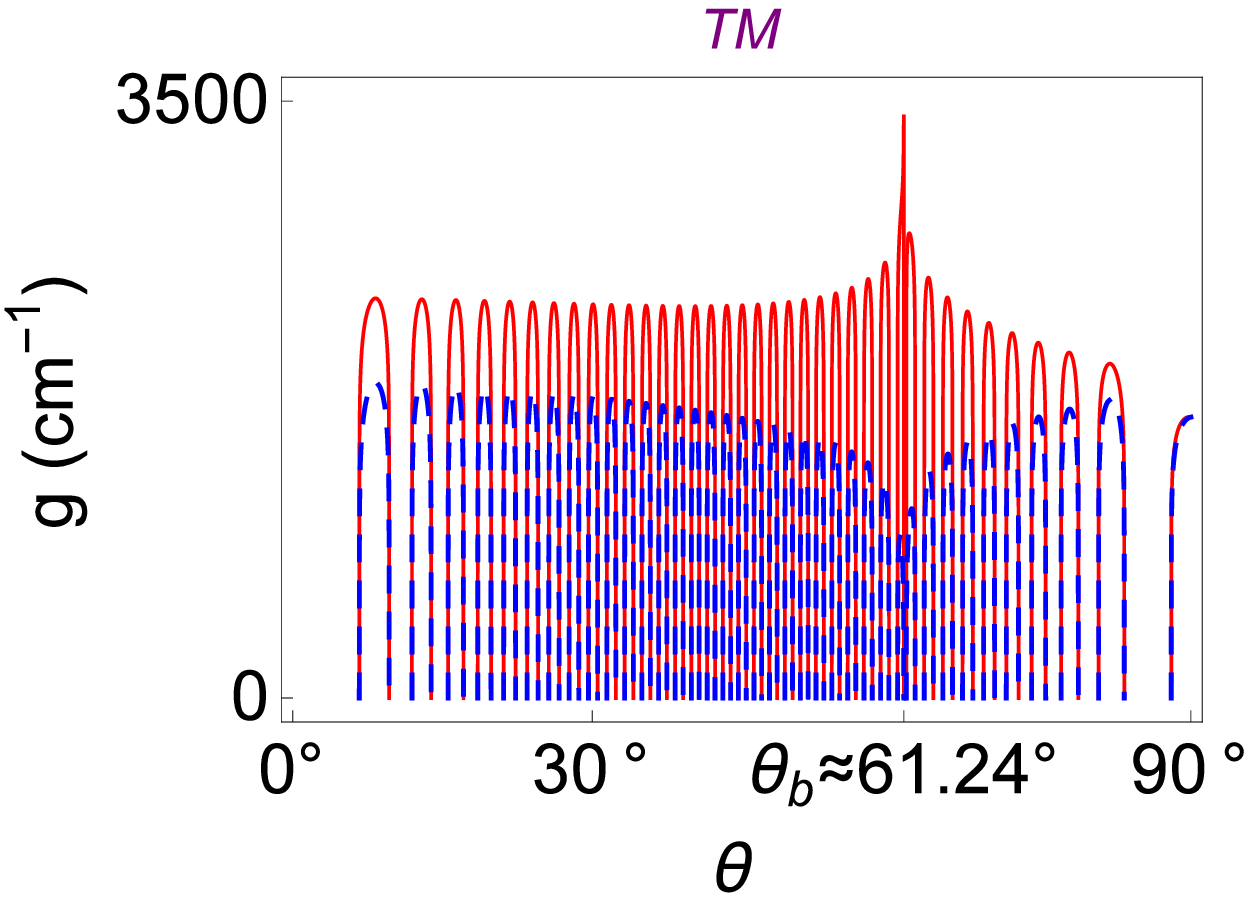}~~~
    \includegraphics[scale=0.4]{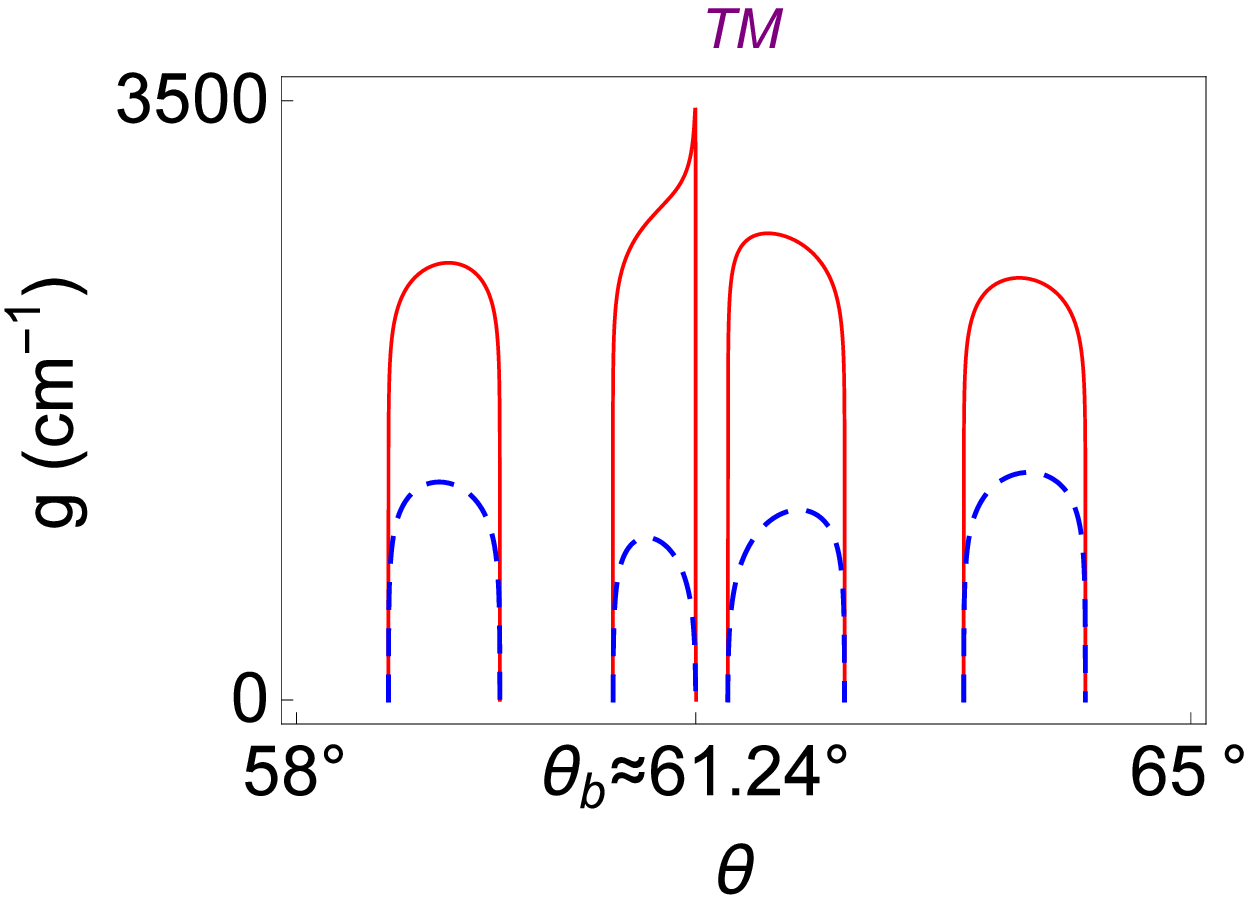}
	\caption{(Color online) Plots of gain coefficient $g$ as a function of incident angle $\theta$ for TE and TM wave solutions respectively. Blue dashed curves represent left zero-reflection amplitude while red solid curves represent right one. Brewster angle $\theta_b$ is clearly seen in the last two Figs. of TM case. }
    \label{g04TE}
    \end{center}
    \end{figure}

We realize that (\ref{unireflectionlesswiths}) is sufficient to provide the necessary conditions for unidirectional reflectionlessness. It is therefore needed to further reduce it to get a better understanding of the physical parameters that generate the desired unidirectionally reflectionless potentials. Consider the left (right) reflectionless situation. This case leads to a constraint on transfer matrix components given by $M_{21} = 0$ ($M_{12} = 0$) and $M_{12} \neq 0$ ($M_{21} \neq 0$). Thus, one obtains the equation
\begin{align}
i[\mathfrak{u}^{-}_2 \tan\fa_2 + \mathfrak{u}^{-}_4 \tan\fa_4 \cos(2k_zs) - \mathfrak{u}^{+}_2 \mathfrak{u}^{-}_4 \tan\fa_2 \tan\fa_4\sin(2k_zs)]=\notag\\
\pm[\mathfrak{u}^{-}_4 \tan\fa_4 \sin(2k_zs)+(\mathfrak{u}^{+}_2 \mathfrak{u}^{-}_4 \cos(2k_zs) - \mathfrak{u}^{-}_2 \mathfrak{u}^{+}_4)\tan\fa_2 \tan\fa_4] \label{leftreflectionlesswiths}
 \end{align}
where again upper (lower) sign denotes the left (right) reflectionless situation while each of which requires the invalidity of the other equation. For simplicity, we first consider the special case of $\cP\cT$-symmetric bilayer system satisfying $s = 0$. Therefore, (\ref{leftreflectionlesswiths}) reduces to a simpler form
\begin{align}
i[\mathfrak{u}^{-}_2 \tan\fa_2 + \mathfrak{u}^{-}_4 \tan\fa_4]=
\pm(\mathfrak{u}^{+}_2 \mathfrak{u}^{-}_4 - \mathfrak{u}^{-}_2 \mathfrak{u}^{+}_4)\tan\fa_2 \tan\fa_4 \label{leftreflectionless}
 \end{align}
Equality can be stated more clearly in an expanded form as follows
\be
\frac{(\fu_4\mp 1)}{(\fu_4\pm1)}e^{2i\fa_4} = \frac{(\fu_2 \pm 1)(\fu_4-\fu_2)e^{2i\fa_2} + (\fu_2 \mp 1)(\fu_2+\fu_4)}{(\fu_2 \pm 1)(\fu_2+\fu_4)e^{2i\fa_2} + (\fu_2 \mp 1)(\fu_4-\fu_2)}\label{reflectionlesspotential1}
\ee
of which the actual solution for the left (right) reflectionlessness is obtained by means of distracting the counterpart of (\ref{reflectionlesspotential1}). It may be of interest to point out the analogy of this expression with the spectral singularity relation of the same optical system as described in \cite{lastpaper} except for the factor $(\fu_2\pm 1)$ in spectral singularity case now turns to $(\fu_2\mp 1)$ and vice versa. For our $\cP\cT$-symmetric bilayer system whose reflectionlessness condition is given in (\ref{reflectionlesspotential1}), we identify the following
    \begin{align}
    &\tilde\fn:=\tilde\fn_4^*=\tilde\fn_2,
    &&\fu:=\fu_4^*=\fu_2,
    &&\fa:=\fa_4^*=\fa_2.
    \label{eq251}
    \end{align}
Therefore, (\ref{reflectionlesspotential1}) come down to
    \be
    e^{2i\fa^\ast} = \frac{[\tilde{\fn}^{\ast} \pm (\fn^{\ast})^{\ell}]}{[\tilde{\fn}^{\ast} \mp (\fn^{\ast})^{\ell}]} \frac{\left\{e^{2i\fa}[\tilde{\fn}^\ast \fn^{\ell}-\tilde{\fn}(\fn^\ast)^{\ell}]
    (\tilde{\fn} \pm \fn^{\ell})+[\tilde{\fn}(\fn^\ast)^{\ell}+\tilde{\fn}^\ast \fn^{\ell}](\fn^{\ell} \mp \tilde{\fn})\right\}}{\left\{e^{2i\fa}[\tilde{\fn}^\ast \fn^{\ell}+\tilde{\fn}(\fn^\ast)^{\ell}](\tilde{\fn} \pm \fn^{\ell})+[\tilde{\fn}^\ast \fn^{\ell}-\tilde{\fn}(\fn^\ast)^{\ell}](\fn^{\ell} \mp \tilde{\fn})\right\}}
    \label{reflectionless2}
    \ee
To get a concrete understanding of (\ref{reflectionless2}), we describe the real and imaginary parts of $\fn$ as $\eta$ and $\kappa$ respectively such that $\fn = \eta + i\kappa$. For the most materials of concern, one does have
    \be
    |\kappa|\ll \eta-1<\eta,
    \label{condi-pert}
    \ee
Thus, in this limit of refractive index components one can safely write down the approximations
    \begin{align}
    &\tilde{\eta}\approx\sec\theta\sqrt{\eta^2 -\sin^2\theta}, && \tilde{\kappa}\approx\frac{\sec\theta\,\eta\,\kappa}{\sqrt{\eta^2-\sin^2\theta}}.
    \label{approx-001}
    \end{align}
in the leading order of $\kappa$. Furthermore, we employ the definition of gain coefficient g, which is a physically applicable parameter, as given by
\be
g:= -\frac{4\pi\kappa}{\lambda} \label{gaindefinition}
\ee
in the expression of $\fa$ (\ref{eq251}). Thus, one can restate $\fa$ as $\fa = a_0\tilde{\eta}-i\frac{\tilde{g}L}{2}$ with  $\tilde{g} := \frac{\eta g}{\sqrt{\eta^2-\sin^2\theta}}$. Therefore, the ultimate physical consequence of (\ref{reflectionless2}) can be deduced by splitting real and imaginary parts. Luckily, the real part of the expression cancels out and the remaining imaginary part yields
\be
\gamma_{\ell}\{\pm \alpha_{\ell}\cos(2a_0\tilde{\eta})-\sin(2a_0\tilde{\eta})\} = \alpha_{\ell}\{\sinh(\tilde{g}L) \pm \gamma_{\ell}\cosh(\tilde{g}L)\} \label{reflectionless3}
\ee
where the parameters $\alpha_{\ell}$ and $\gamma_{\ell}$ are described as follows
 \begin{align}
    &\alpha_{\ell} :=\frac{2\tilde{\kappa}\sigma_{\ell}}{\tilde{\eta}^2 -\eta^{2\ell}},
    &&\gamma_{\ell} := \frac{2\tilde{\eta}\eta^{\ell}}{\tilde{\eta}^2 +\eta^{2\ell}}.\nn
   \end{align}
with $\sigma_\ell$ providing great convenience in notations
    \[\sigma_\ell:=
    \left\{\begin{array}{ccc}
    1&{\rm for}&\ell=0,\\
    2\sin^2\theta-\eta^2&{\rm for}&\ell=2.
    \end{array}\right.\]
One can figure out gain coefficient up to the leading order of $\kappa$, which leads to a unidirectionally reflectionless configuration, from (\ref{reflectionless3}) as follows
\be
g \approx \frac{\sqrt{\eta^2 -\sin^2\theta}}{\eta L}\ln\left[\frac{\mathcal{A}_{\mp}-\sqrt{\mathcal{A}_{\mp}^2+(\tilde{\eta}^2-\eta^{2\ell})^2}}{(\tilde{\eta} \mp \eta^{\ell})^2}\right]\label{reflectionlessgain}
\ee
where
\be
\mathcal{A}_{\mp} := \frac{\tilde{\eta}\eta^{\ell}(\tilde{\eta}^2-\eta^{2\ell})\sin(2a_0\tilde{\eta})}{\tilde{\kappa}\sigma_{\ell}} \mp 2\tilde{\eta}\eta^{\ell}\cos(2a_0\tilde{\eta})\notag
\ee
But notice that not all values of $g$ satisfying (\ref{reflectionlessgain}) gives rise to a left (right) reflectionless configuration. One must account for the excluded points arising from the invalidity of the counter-direction reflectionlessness in (\ref{leftreflectionless}). To get a physical meaning of the expressions in (\ref{reflectionlessgain}) depending on various parameters, we refer to Figs.~\ref{g01TE}, \ref{g02TE}, \ref{g03TE}, \ref{g04TE}. In Fig.~\ref{g01TE}, TE and TM plots of gain coefficient $g$ as a function of wavelength $\lambda$ obeying left (right) reflectionlessness are displayed for $\cP\cT$-symmetric bilayer which contains the Nd:YAG crystals\footnote{We ignore to realize equal amounts of gain and loss in the relevant components of Nd:YAG crystals bilayer systems in experimental setup. We use Nd:YAG crystals just to demonstrate the validity of our model.} with specifications
 \begin{align}
    &\eta = 1.8217,
    &&L= 10~\textrm{cm},
    &&\theta = 30^{\circ} \label{ndyag1}
    \end{align}
In these graphs thin solid red curves establish the points that constitute the right zero-reflection amplitude situations whereas thick dashed blue curves form up the left zero-reflection amplitudes. Obviously, bidirectional reflectionlessness occurs provided that both curves overlap. In these graphs one can safely declare that nonoverlapping single graphs denote unidirectional reflectionlessness since we can explicitly see all alternate conditions. Therefore, we will use this convention for the rest of the paper. We observe that only certain periodically determined range of wavelengths allow uni/bi-directional reflectionlessness. In fact, curves that belong to the left and right zero-reflection amplitude curves never cross each other except for zero gain value such that the actual curves of gain coefficients for each curve determine the unidirectional reflectionlessness above the positive $g$-axes. However, one can not precisely determine the actual wavelength range for extremely small values of gains separately, therefore, in a moderate wavelength range, curves for the left and right zero-reflection amplitude situations seem to coincide after some gain values while we reduce its amount, i.e they get closer to each other in such a way that they appear to be overlapping each other. As we increase the precision of measurement, bilateral reflectionlessness turns into unidirectional one. This realization is reflected in the decimal parts of wavelength. For example, If we are sensitive to thousandths after decimal point, only gain values $g \gtrapprox 0.3~\textrm{cm}^{-1}$ for TE mode and $g \gtrapprox 0.35~\textrm{cm}^{-1}$ for TM are allowed for the same parameters of our choice in (\ref{ndyag1}). If the sensitivity of measurement decreases so that only tenths or hundreds after decimal point of wavelengths are measured, then required gain values for unidirectional reflectionlessness gets larger than about $0.8~\textrm{cm}^{-1}$.

In Fig.~\ref{g02TE}, TE and TM graphs of gain coefficients as a function of wavelength for various incidence angles are shown for the left reflectionless case. In these Figs. we do not explicitly show the excluded curves arising from right reflectionlessness for the clarification purpose, but depending on the precision of measurement they can be shown around the sides of each curve. Again periodic structures leading to reflectionlessness are observed clearly for each angle. Also, wavelength range that leads to reflectionlessness increases when the angle of incidence increases. We observe that once the incident wave angle is increased, the allowed wavelength range shifts to the left (for wavelengths less than resonance wavelength) and right (for those greater than resonance wavelength) for each periodic entity, and the peak of the curve and thus the required gain value is slightly lowered for TE case. In TM case, Brewster's angle plays a special role at which peak takes the minimum value and then increases with the increase of angle above Brewster's angle. This shows that incidence angles nearby the Brewster's angle is favorable for unidirectional reflectionlessness in TM mode.

In Fig.~\ref{g03TE}, the role of thickness of the bilayer slab on the gain coefficient and wavelength graph is displayed for the left TE and TM reflectionless case. We again assume that excluded points (due to right reflectionlessness) are intended although they are not explicity shown on graphs. Again periodic structures leading to unidirectional reflectionlessness are observed clearly for each slab thickness. We see that required gain value decreases with the increase of thickness, but the range of wavelength allowing reflectionlessness reduces.

In Fig.~\ref{g04TE}, the dependence of gain coefficient on the incident angle is shown for the left and right zero-reflection situations corresponding to both TE and TM cases. In these graphs we use slabs with $L = 10~\mu\textrm{m}$ at wavelength $\lambda = 808~\textrm{nm}$ for a better view. For TE case, we see an almost steady periodic behaviour with the rise of angles, however gain coefficient gets minimum value at Brewster's angle in the TM case. It is noted that the range of gain coefficients are most obtained around incident angles $\theta = 0^\circ $.

We next focus on the situation of $\cP\cT$-symmetric two-layer slab where there is a gap between balanced gain and loss. Equation  (\ref{leftreflectionlesswiths}) leads to

\be
\frac{(\fu_4 \mp 1)}{(\fu_4 \pm 1)}e^{2i\fa_4} = \frac{(e^{2i\fa_2}-1)[(\fu_2^2-1)(\fu_4 \pm 1)-(\fu_2^2+1)(\fu_4 \mp 1)e^{2ik_z s}] \mp 2\fu_2(\fu_4 \mp 1)e^{2ik_z s}(e^{2i\fa_2}+1)}{(e^{2i\fa_2}-1)[(\fu_2^2-1)(\fu_4 \mp 1)-(\fu_2^2+1)(\fu_4 \pm 1)e^{2ik_z s}] \mp 2\fu_2(\fu_4 \pm 1)e^{2ik_z s}(e^{2i\fa_2}+1)}\label{reflectionlesspotential2}
\ee

In view of the identifications (\ref{eq251}) one gets the expression for (\ref{reflectionlesspotential2})

\footnotesize
\be
e^{2i\fa^{\ast}} = \frac{(e^{2i\fa}-1)[(\tilde{\fn}^2-\fn^{2\ell})(\tilde{\fn}^{\ast} \pm (\fn^{\ast})^{\ell})^2-(\tilde{\fn}^2+\fn^{2\ell})(\tilde{\fn}^{\ast 2}-(\fn^{\ast})^{2\ell})e^{2ik_z s}] \mp 2\tilde{\fn}\fn^{\ell}(\tilde{\fn}^{\ast 2}-(\fn^{\ast})^{2\ell})e^{2ik_z s}(e^{2i\fa}+1)}{(e^{2i\fa}-1)[(\tilde{\fn}^2-\fn^{2\ell})(\tilde{\fn}^{\ast} \mp (\fn^{\ast})^{\ell})^2-(\tilde{\fn}^2+\fn^{2\ell})(\tilde{\fn}^{\ast 2}-(\fn^{\ast})^{2\ell})e^{2ik_z s}] \mp 2\tilde{\fn}\fn^{\ell}(\tilde{\fn}^{\ast 2}-(\fn^{\ast})^{2\ell})e^{2ik_z s}(e^{2i\fa}+1)}\notag
\ee

\normalsize
Using the complex refractive index $\fn = \eta +i\kappa$ with condition (\ref{condi-pert}), the approximations (\ref{approx-001}), (\ref{gaindefinition}) and the definition of $\fa$, one obtains a complex relation whose real and imaginary parts up to the leading order of $\kappa$ are respectively given by

\begin{align}
&\left[(1-\cos2k_zs)-\alpha_{\ell}\gamma_{\ell}\sin2k_zs\right]\cosh(\tilde{g}L) \pm \left[\gamma_{\ell}(1-\cos2k_zs)-\alpha_{\ell}\sin2k_zs\right]\sinh(\tilde{g}L)\notag\\ &= \left[(1-\cos2k_zs)-\alpha_{\ell}\gamma_{\ell}\sin2k_zs\right]\cos 2a_0\tilde{\eta} \pm \left[\alpha_{\ell}(1-\cos2k_zs)+\gamma_{\ell}\sin2k_zs\right]\sin 2a_0\tilde{\eta}\label{reflectionlesswithgap1}
\end{align}

\begin{align}
&\left[\sin2k_zs - \alpha_{\ell}\gamma_{\ell}(1+\cos2k_zs)\right]\cosh(\tilde{g}L) \pm \left[\gamma_{\ell}\sin2k_zs - \alpha_{\ell}(1+\cos2k_zs)\right]\sinh(\tilde{g}L)\notag\\ &= \left[\sin2k_zs - \alpha_{\ell}\gamma_{\ell}(1+\cos2k_zs)\right]\cos 2a_0\tilde{\eta} \pm \left[\alpha_{\ell}\sin2k_zs + \gamma_{\ell}(1+\cos2k_zs)\right]\sin 2a_0\tilde{\eta}\label{reflectionlesswithgap2}
\end{align}
Notice that once $s = 0$, (\ref{reflectionlesswithgap1}) disappears and we obtain just the imaginary part which is reduced to the simplified form in (\ref{reflectionless3}). For $s \neq 0$, equations~(\ref{reflectionlesswithgap1}) and (\ref{reflectionlesswithgap2}) yield the same equation in the form
\begin{align}
[\tan k_zs - \alpha_{\ell}\gamma_{\ell}] \left(\cosh(\tilde{g}L)-\cos 2a_0\tilde{\eta}\right) =  \mp \left[(\gamma_{\ell}\tan k_zs - \alpha_{\ell})\sinh(\tilde{g}L) - (\alpha_{\ell}\tan k_zs-\gamma_{\ell})\sin 2a_0\tilde{\eta}\right]\label{reflectionlesswithsnotzero}
\end{align}
We stress out the fact that upper and lower equations point out the conditions for left and right zero-reflection coefficients and each one is the negation of the other for the corresponding unidirectional reflectionlessness. To reveal the physical meaning of these equations in (\ref{reflectionlesswithsnotzero}) for nonzero $s$, or (\ref{reflectionlesswithgap1}) and (\ref{reflectionlesswithgap2}) including the case $s =0$ , one can plot gain coefficient with respect to the parameters of wavelength, separation between gain and loss, and angle of incidence regarding various cases. But we have to take into consideration of excluded values stemming from the zero condition of counter-direction.

\begin{figure}
	\begin{center}
    \includegraphics[scale=0.5]{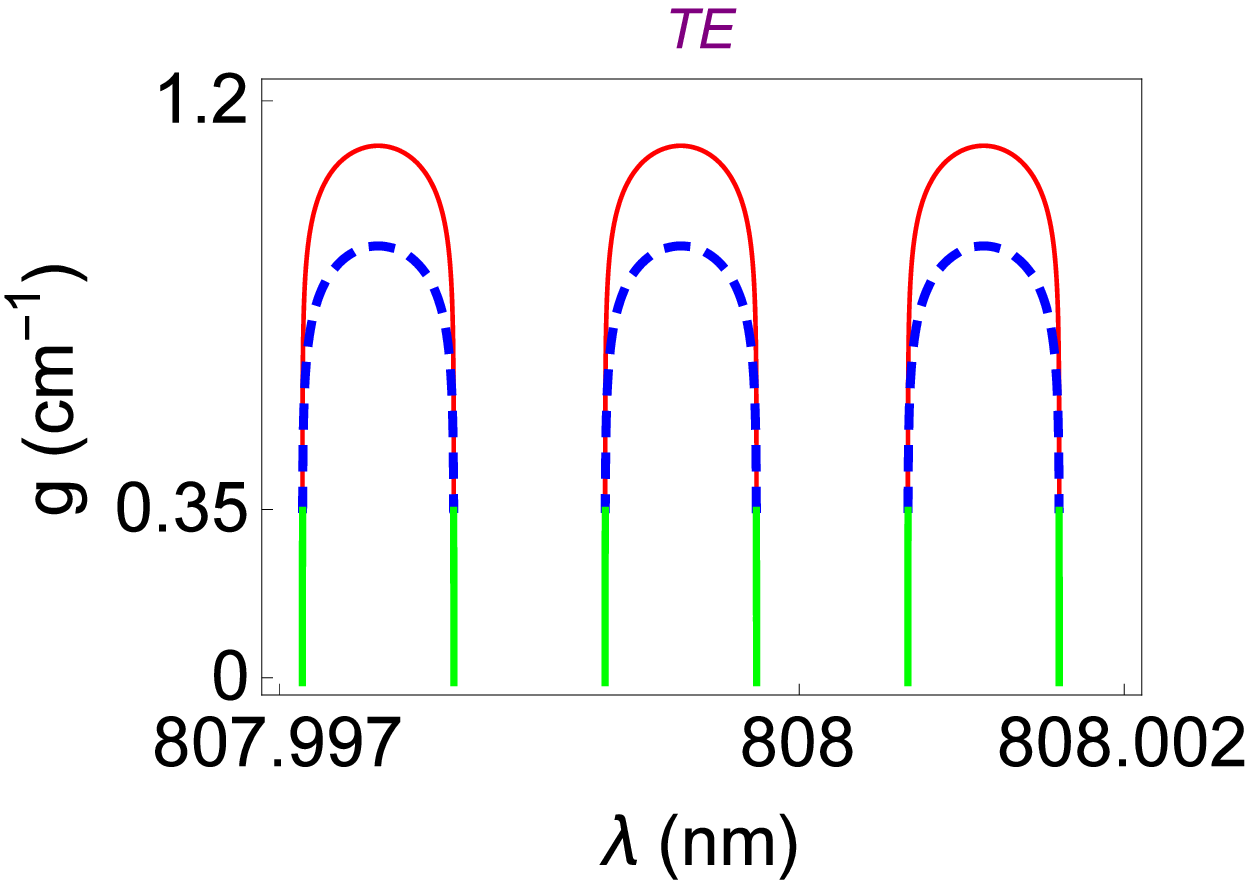}~~~
    \includegraphics[scale=0.5]{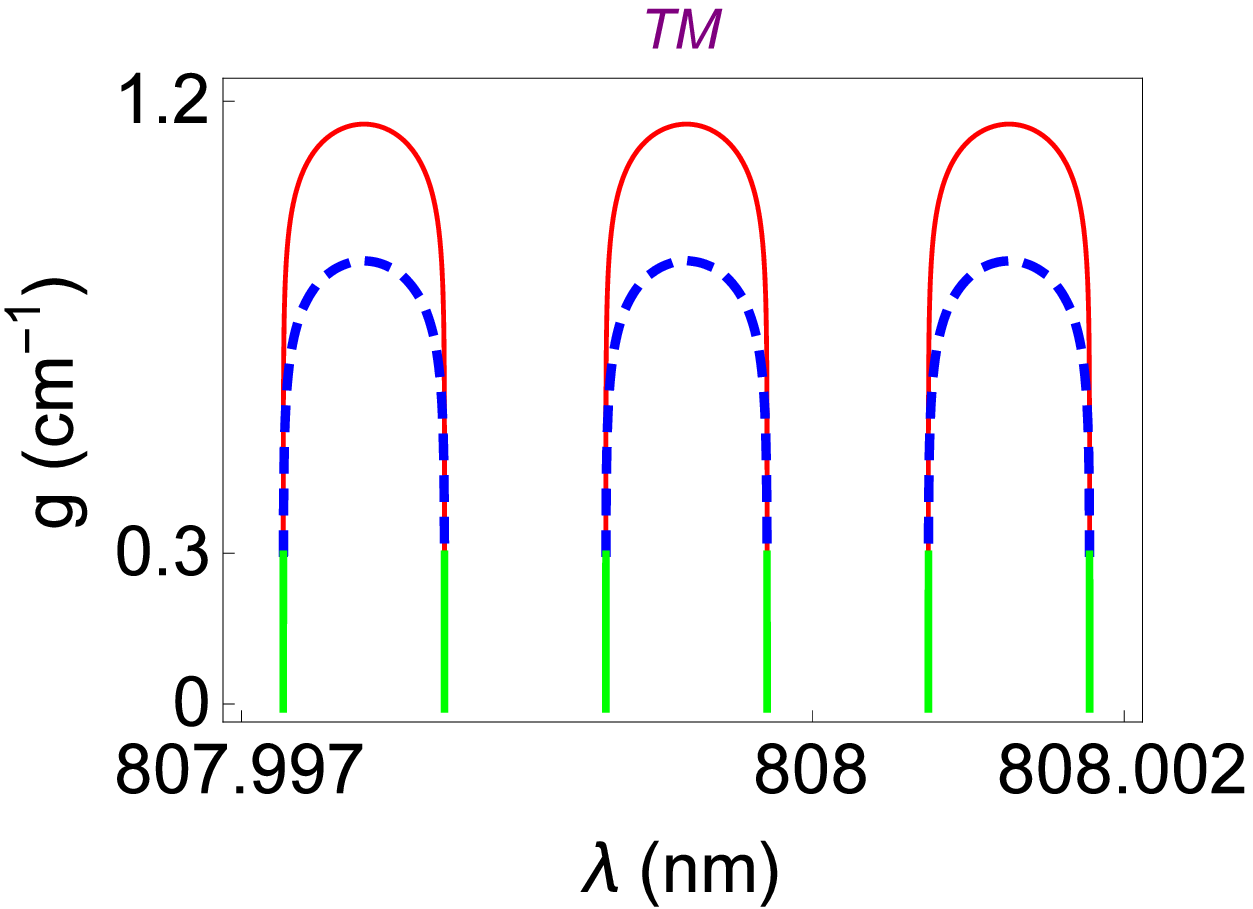}
	\caption{(Color online) Plots of gain coefficient $g$ as a function of wavelength $\lambda$ corresponding to left and right reflection-zero TE and TM wave solutions for the case of $\cP\cT$-symmetric Nd:YAG layers with a gap in~(\ref{reflectionlessgain}). Thick dashed blue curves and thin solid red curves represent gain values that yield unidirectional reflectionlessness from left and right respectively, while solid green curves indicate bidirectionally reflectionless points.}
    \label{g01TEs}
    \end{center}
    \end{figure}

\begin{figure}
	\begin{center}
    \includegraphics[scale=0.5]{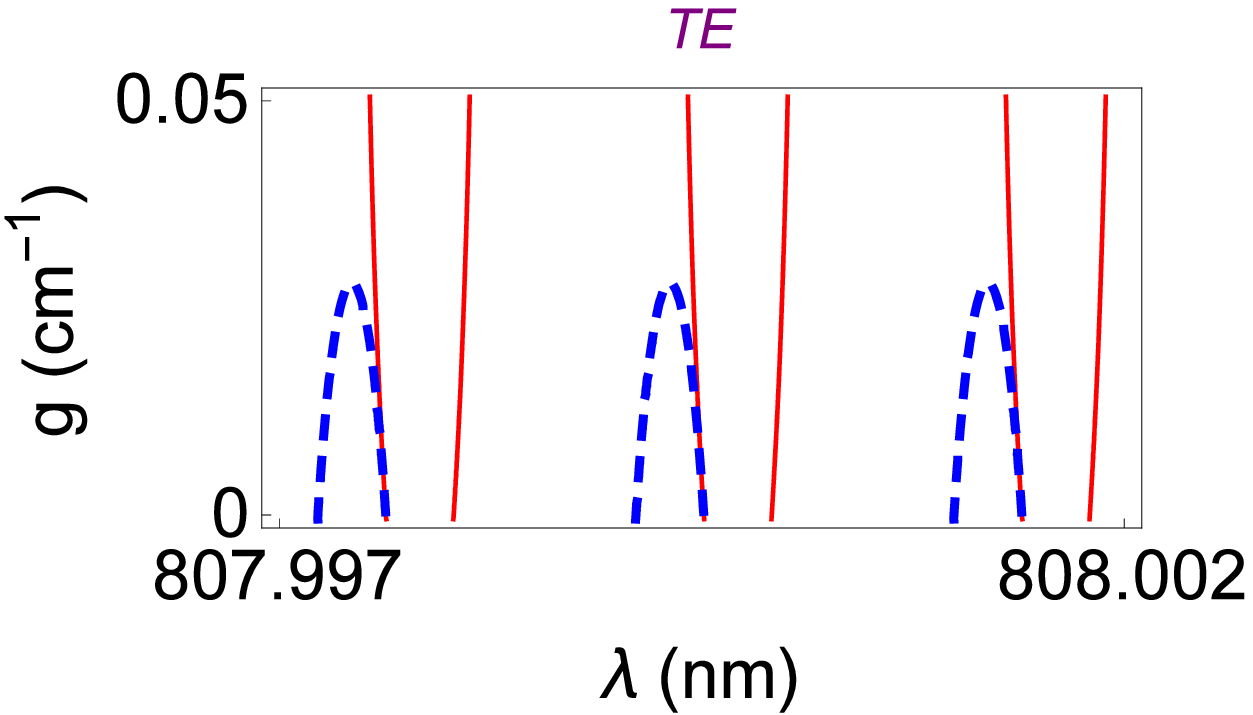}~~~
    \includegraphics[scale=0.5]{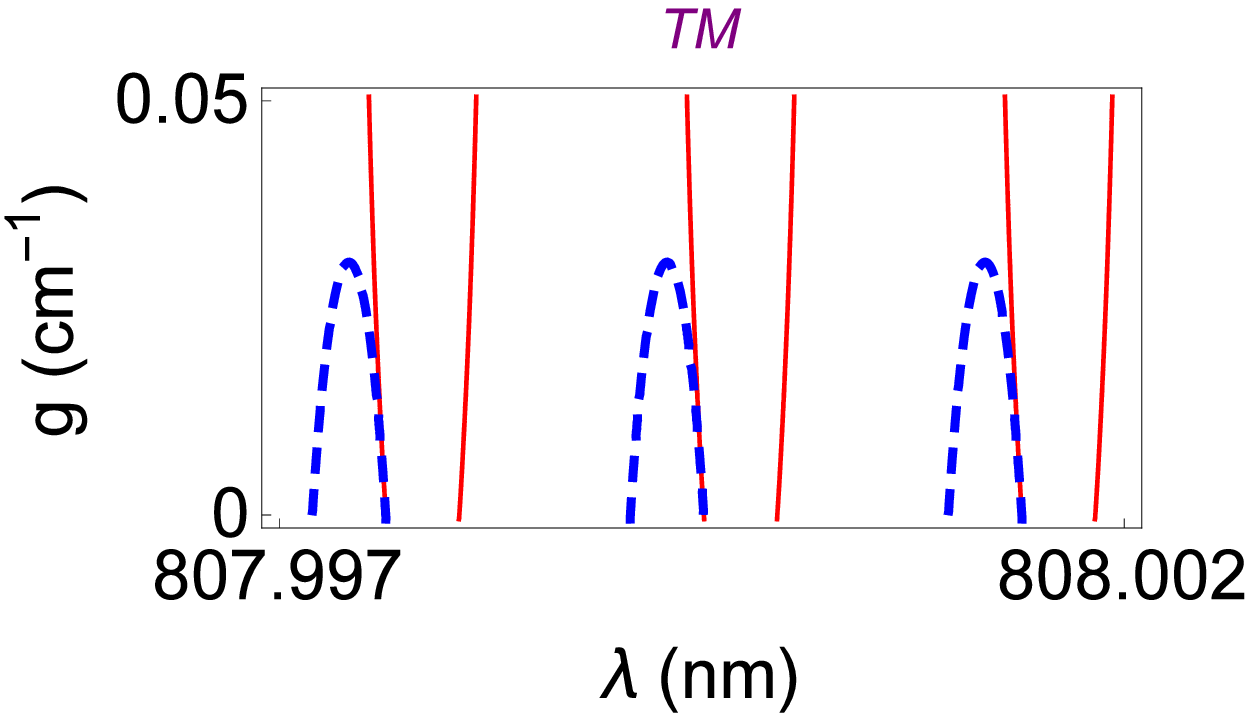}\\
    \includegraphics[scale=0.5]{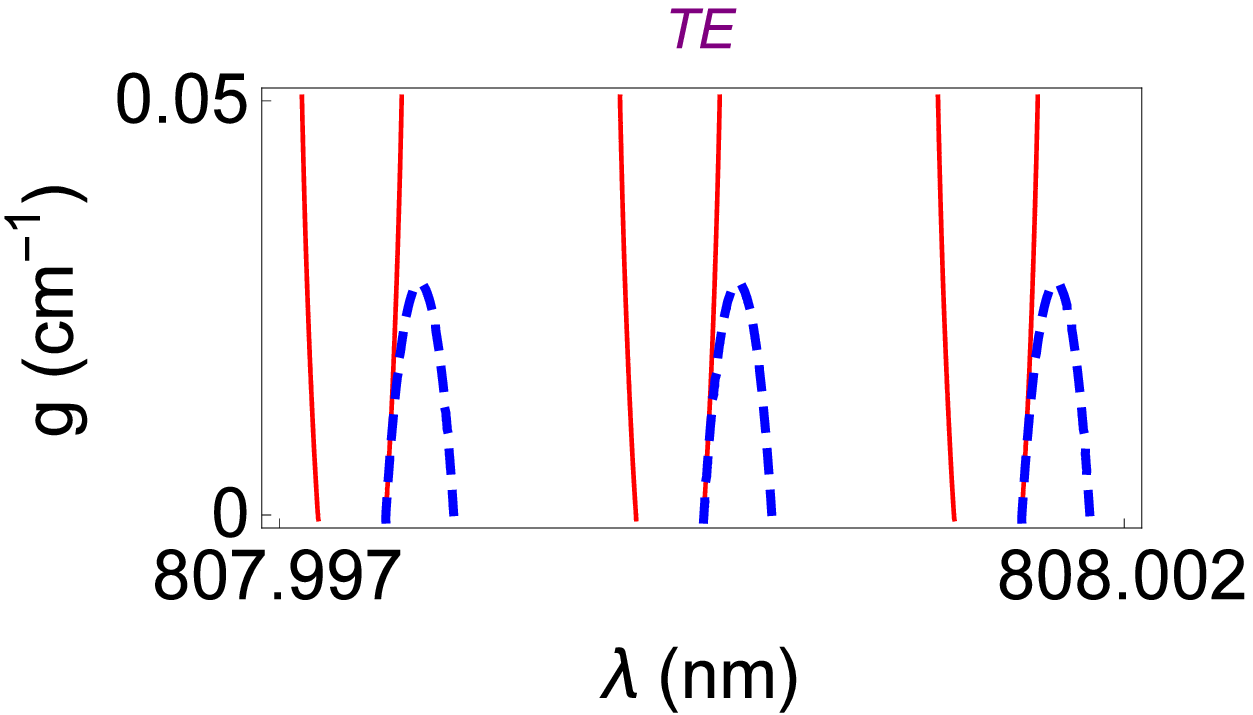}~~~
    \includegraphics[scale=0.5]{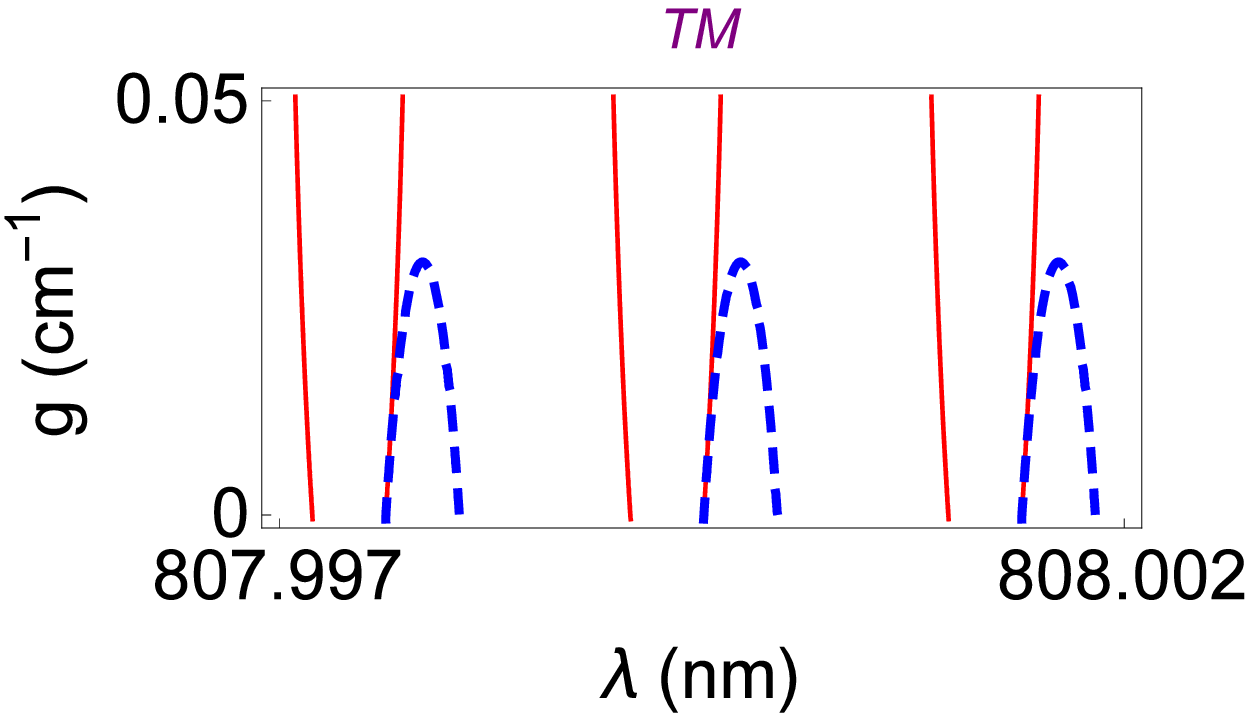}
	\caption{(Color online) Plots of gain coefficient $g$ as a function of wavelength $\lambda$ corresponding to a generic left and right reflection-zero TE and TM wave solutions for the case of $\cP\cT$-symmetric layers with a gap with parameters in~(\ref{reflectionlessgain}). Thick dashed blue curves and thin solid red curves represent left and right reflectionlessness at non-coincident points. In these plots, upper and lower figures correspond to $s/s_0 = 20.5$ and $s/s_0 = 21.5$ respectively. }
    \label{g02TEs}
    \end{center}
    \end{figure}

\begin{figure}
	\begin{center}
    \includegraphics[scale=0.5]{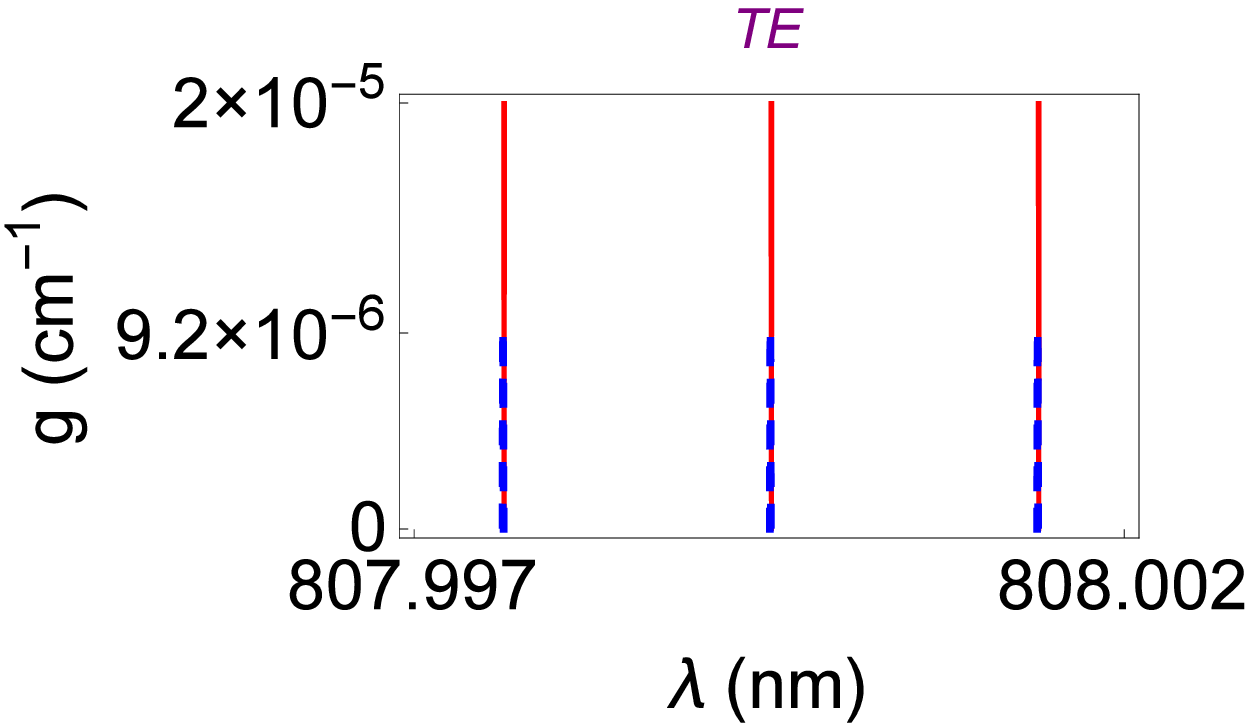}~~~
    \includegraphics[scale=0.5]{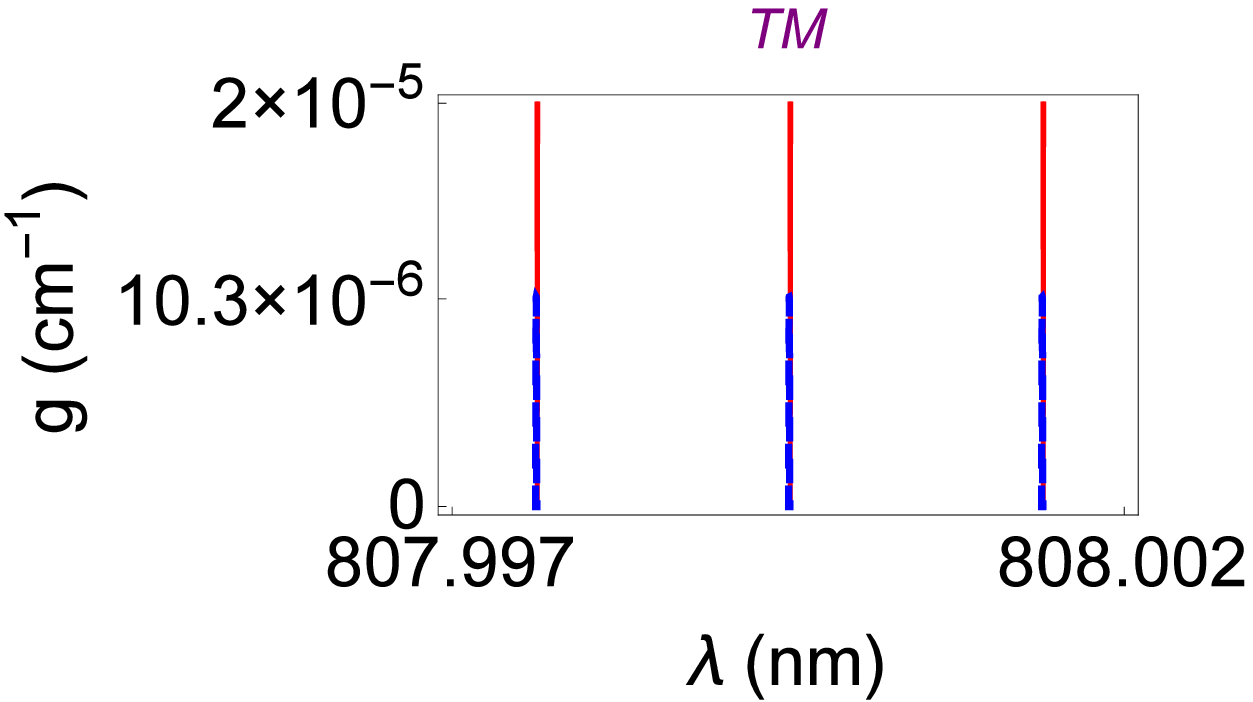}
	\caption{(Color online) Plots of gain coefficient $g$ as a function of wavelength $\lambda$ corresponding to destructive left and right reflection-zero TE and TM wave solutions for the case of $\cP\cT$-symmetric Nd:YAG crystals layers with a gap corresponding to $s/s_0 = 20.99$ with parameters in~(\ref{reflectionlessgain}). In these figures again Thick dashed blue curves and thin solid red curves represent gain values that yield zero-reflection curves from left and right respectively.}
    \label{g03TEs}
    \end{center}
    \end{figure}

One can analyze equations (\ref{reflectionlesswithgap1}) and (\ref{reflectionlesswithgap2}) for all $s$ values, or just (\ref{reflectionlesswithsnotzero}) for nonzero $s$ pictorially, corresponding to $\cP\cT$-symmetric case with a gap. In Figs.~(\ref{g01TEs}), (\ref{g02TEs}) and (\ref{g03TEs}), gain coefficient via wavelength graphs belonging to the co-called constructive, destructive and generic cases are clearly seen. If we define
\be
s_0 := \frac{\pi}{2k_z} = \frac{\lambda}{4\cos\theta} \label{s0}
\ee
so that even integer values of $s/s_0$ specify the constructive configurations, odd integer values specify the destructive configurations and values apart from these two cases correspond to the generic cases. In Fig.~(\ref{g01TEs}), $s/s_0 = 20$ for the $\cP\cT$-symmetric Nd:YAG gain-loss system with parameters in ~(\ref{reflectionlessgain}), and thus correspond to constructive cases. We immediately notice that this case yields the same situation as the gapless one in Fig.~(\ref{g01TE}). In fact, curves identifying allowed gain values for the left and right zero-reflection situations never intersect each other above the positive $g$-axis. It is just a matter of precision to discriminate the right and left reflection-zero curves. Since measurement can not be performed at the desired level, our system would appear to be reflectionless for some small positive gain values up to a certain value. For example, in Fig.~\ref{g01TEs}, thick dashed blue curve (thin red solid curve) with $g\gtrsim 0.35~\textrm{cm}^{-1}$ for TE waves and $g\gtrsim 0.30~\textrm{cm}^{-1}$ for TM waves represent the allowed gain values for left (right) reflectionlessness up to thousands of a nm-distance wavelength measurement. Thus, green curve turns out to be bidirectionally reflectionless case. As we increase the sensitivity of our measurement, the allowed gain values drop off. It is seen that TM solutions give rise to a better reflectionless situation considering the TE case.

In Fig.~\ref{g02TEs}, a generic case with $s/s_0 = 20.5$ (upper figure) and $s/s_0 = 21.5$ (lower figure) with the same parameters as the constructive case is displayed. It is manifest that reflectionless potentials occurs in prescribed ranges of wavelengths with a periodic structure. Notice that gain values requiring the reflectionlessness considerably lower and minimum gain values move along the dashed blue curves once we move from $s/s_0 = 20.5$ to $s/s_0 = 21.5$. We also note that gain values requiring left reflectionlessness is relatively much smaller than ones for right reflectionlessness. Thus, in this configuration it is easy to perform left reflectionlessness.

In Fig.~\ref{g03TEs} one attains a very small value of gain at almost very close neighborhood to the destructive case with $s/s_0 = 20.99$. Notice that left reflectionlessness is not observed while we have a perfect right reflectionlessness at very small gain values. At exact odd integer values of $s/s_0$ corresponding to destructive configuration, no positive gain value can be obtained for both left and right reflectionlessness, that is why the best way to choose a good right reflectionless system is to choose a gap width which is very close to the destructive configuration case. Finally, notice that wavelength range of left and right reflectionless situations periodically interchange around the odd integer values of $s/s_0$ in (\ref{g02TEs}).

\begin{figure}
	\begin{center}
    \includegraphics[scale=0.5]{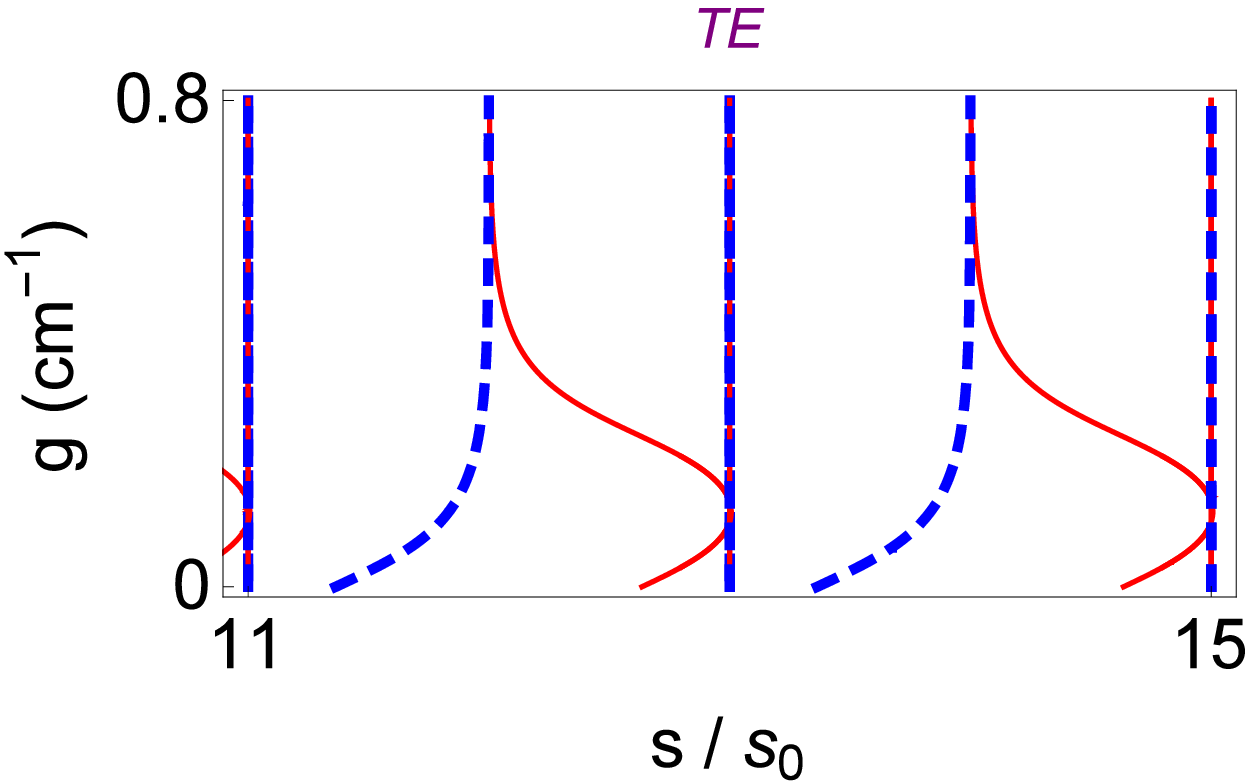}~~~
    \includegraphics[scale=0.5]{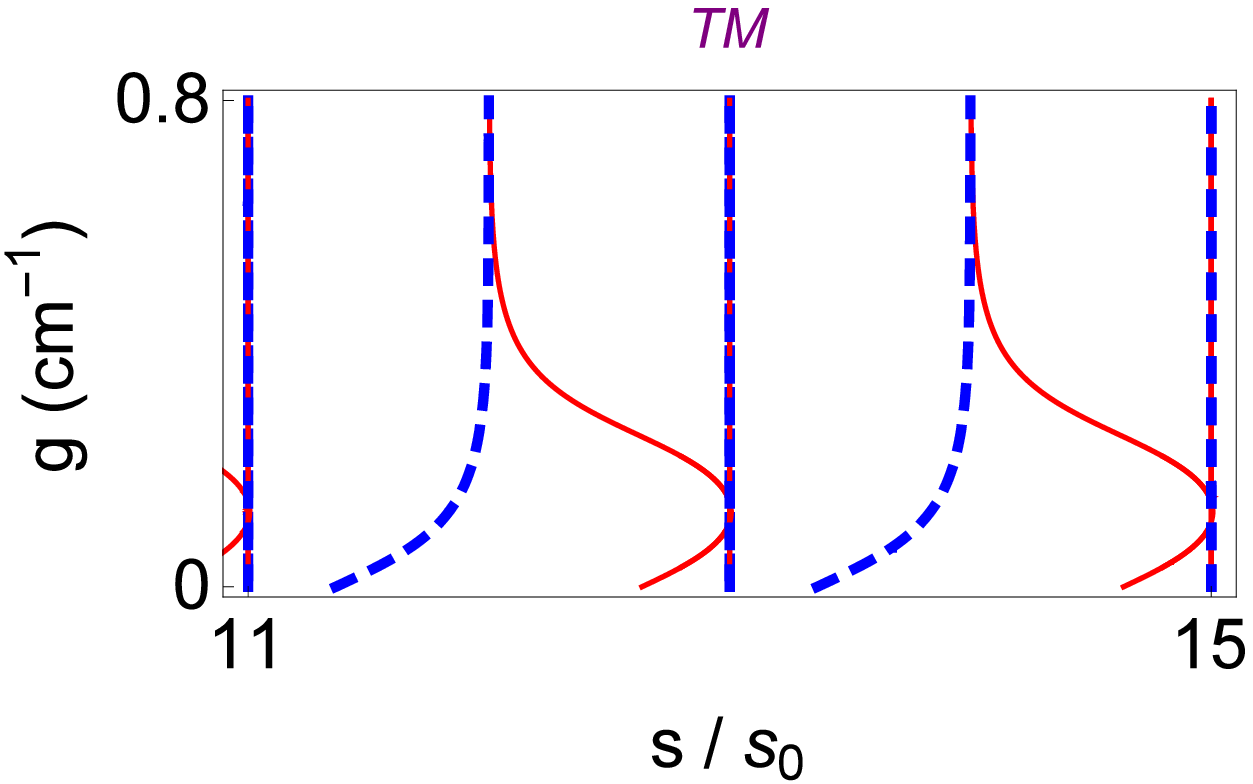}\\
    \includegraphics[scale=0.5]{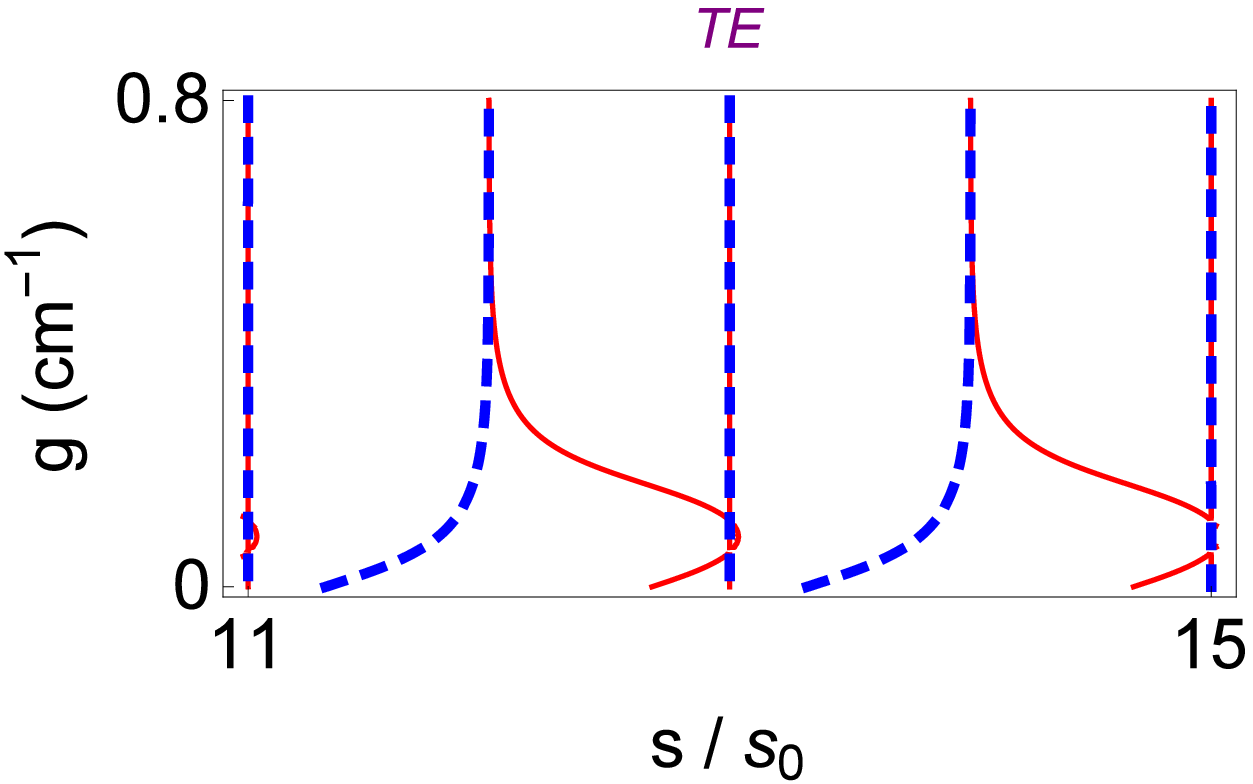}~~~
    \includegraphics[scale=0.5]{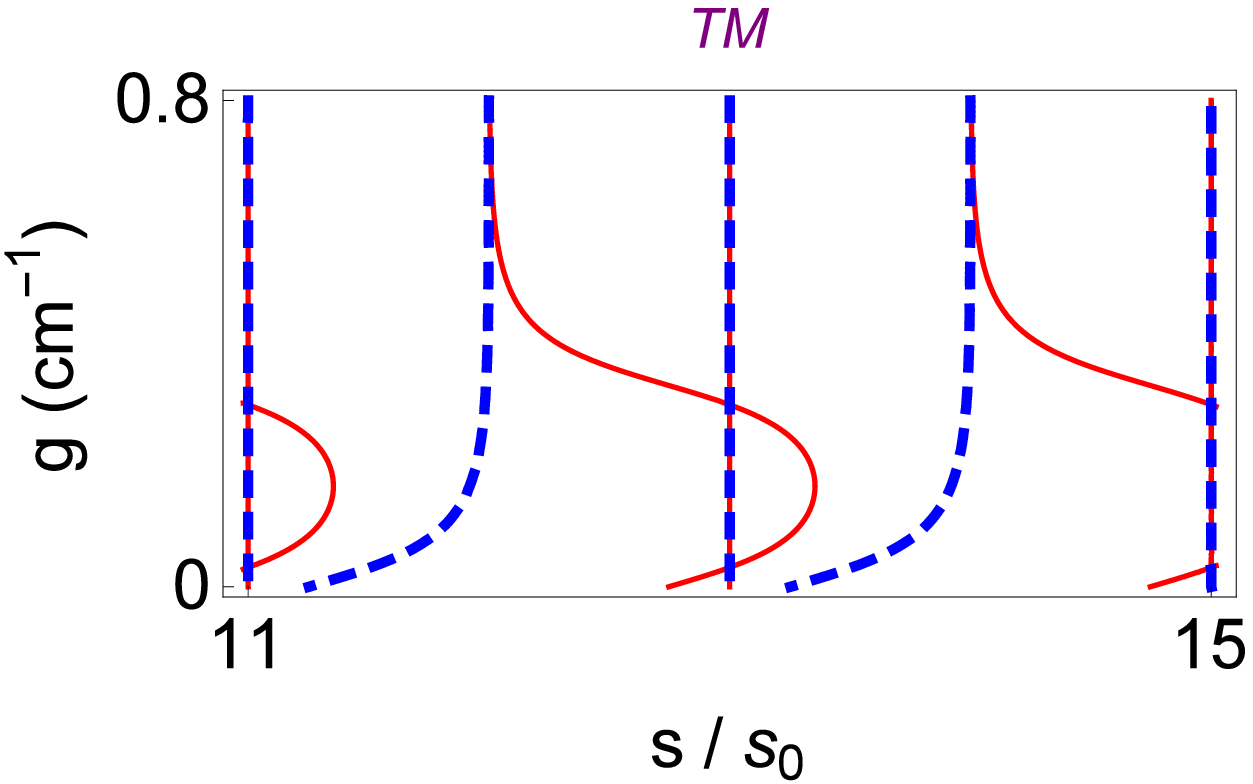}\\
    \includegraphics[scale=0.5]{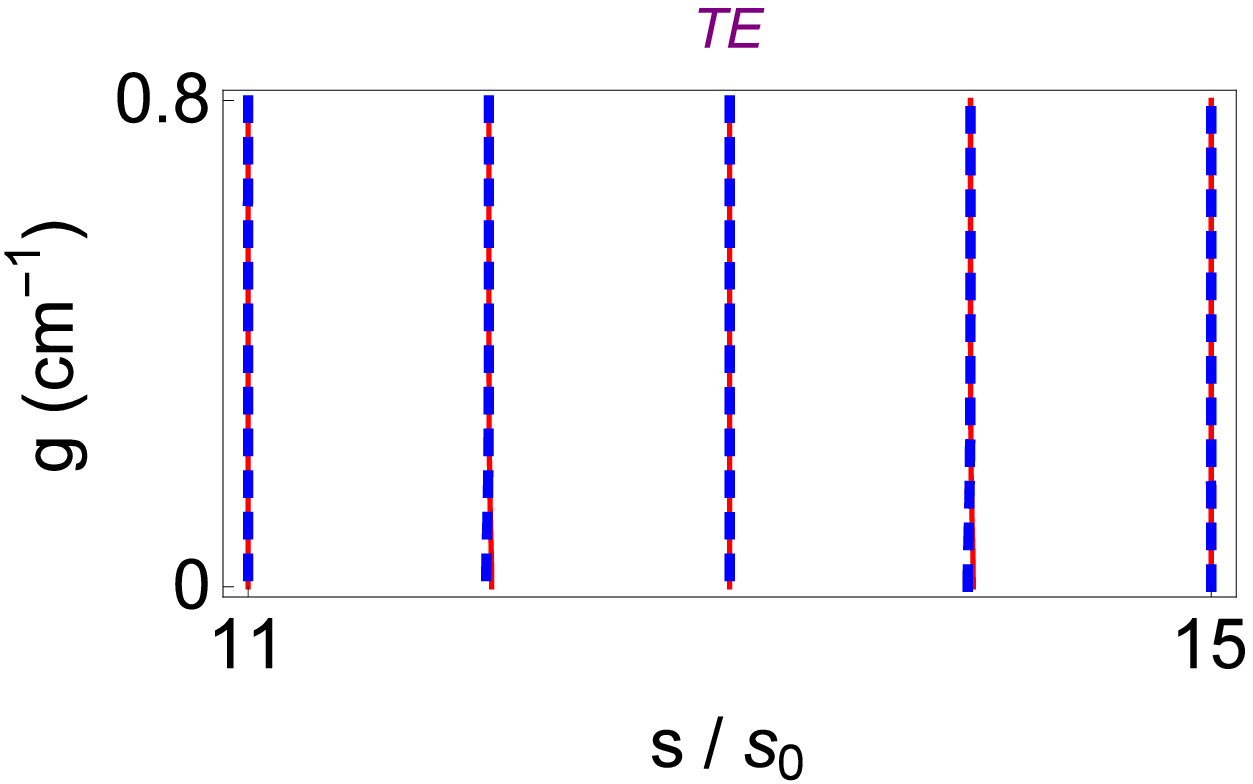}~~~
    \includegraphics[scale=0.5]{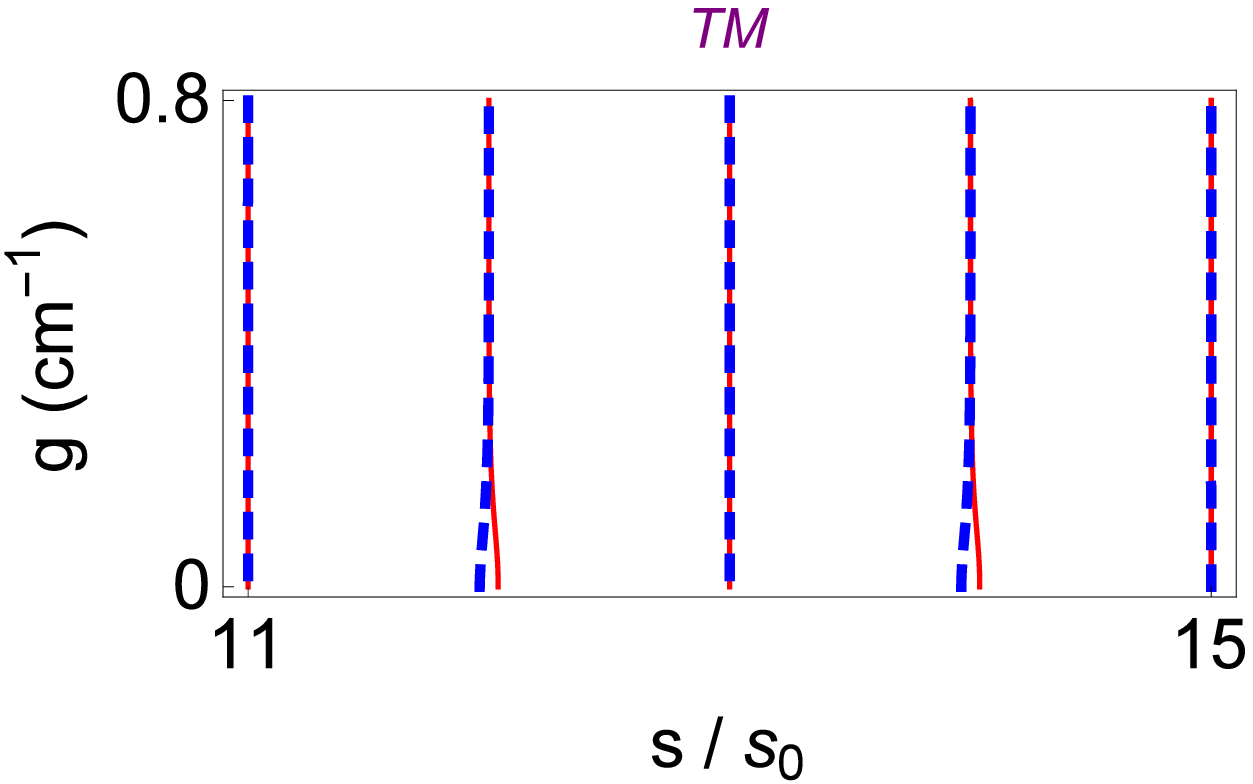}
	\caption{(Color online) Plots of gain coefficient $g$ as a function of gain-loss separation measure $s/s_0$ corresponding to TE and TM wave solutions for the case of $\cP\cT$-symmetric Nd:YAG layers with a gap at various incident angles with parameters in~(\ref{reflectionlessgain}). Thin solid red curves represent right zero-reflection amplitude, while thick dashed blue curves the left zero-reflection amplitude configurations. In these plots, upper, middle and lower rows specify incident angles of $\theta = 0^\circ$, $\theta = 45^\circ$ and $\theta = 89^\circ$ respectively. }
    \label{g04TEs}
    \end{center}
    \end{figure}

In Figs.~\ref{g04TEs}, we analyze the behaviour of gain coefficients with respect to $s/s_0$ which is a measure of gain-loss separation distance which are depicted at various incident angles. Notice that constructive configurations occur at odd integer values of $s/s_0$ while destructive configurations at even integer values of $s/s_0$. We also realize that no unidirectional reflectionlessness is observed at destructive configurations (i.e. for odd $s/s_0$ values) for all angles, verifying our previous observations. Also, at large angles very close to $\theta \approx 90^{\circ}$ again unidirectional reflectionlessness is only observed nearby constructive configurations. TE case yields a better result in this sense. Constructive and generic cases verifies our previous observation that the best choice for reflectionlessness is to pick a separation distance around odd integer values of $s/s_0$, i.e. required gain value lowers as we pass from constructive to destructive configurations. We also learn from these graphs that not all angles yields reflectionless situation at fixed wavelength and $s/s_0$.

\begin{figure}
	\begin{center}
    \includegraphics[scale=0.4]{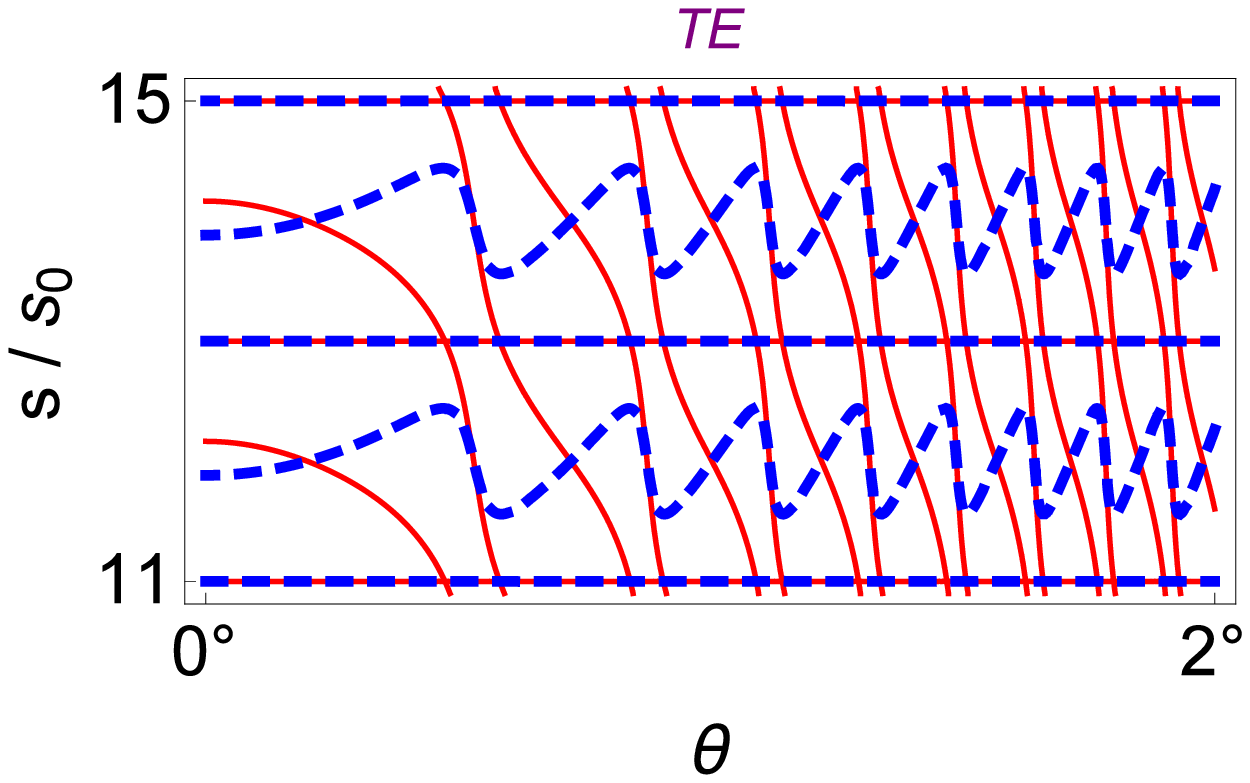}~~~
    \includegraphics[scale=0.4]{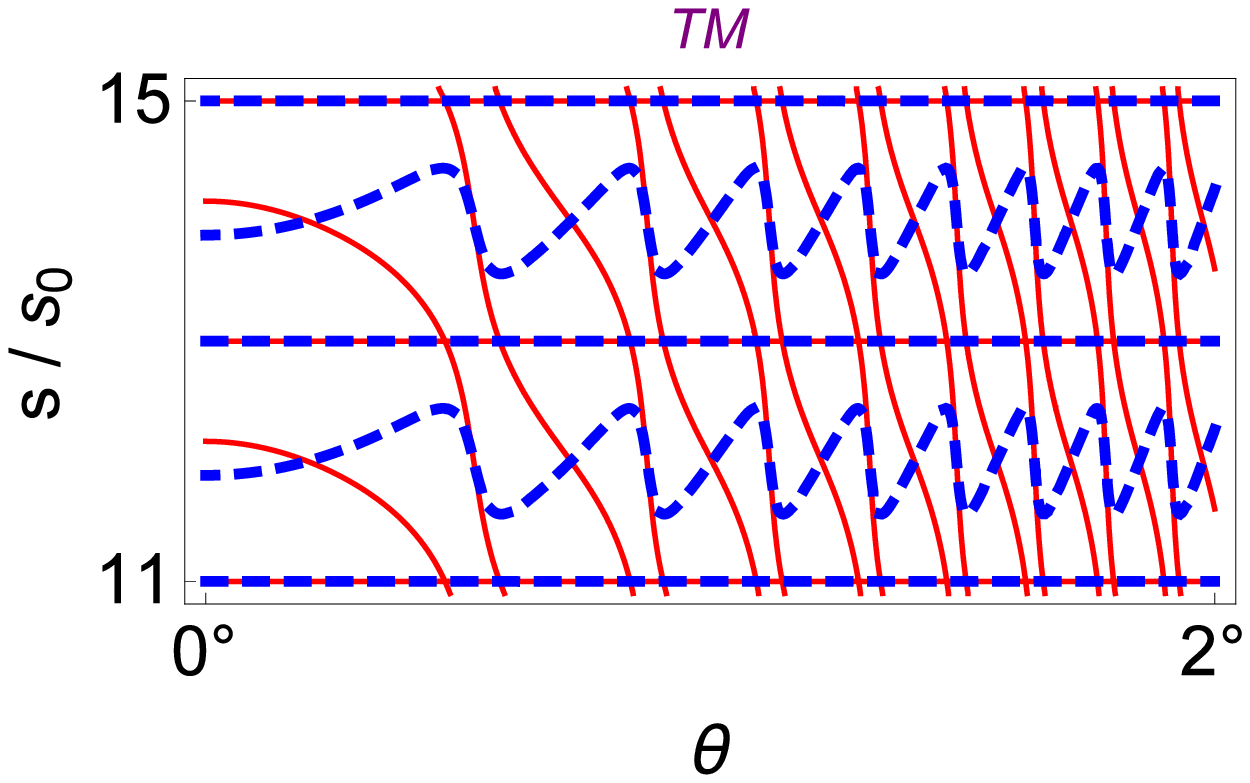}\\
    \includegraphics[scale=0.4]{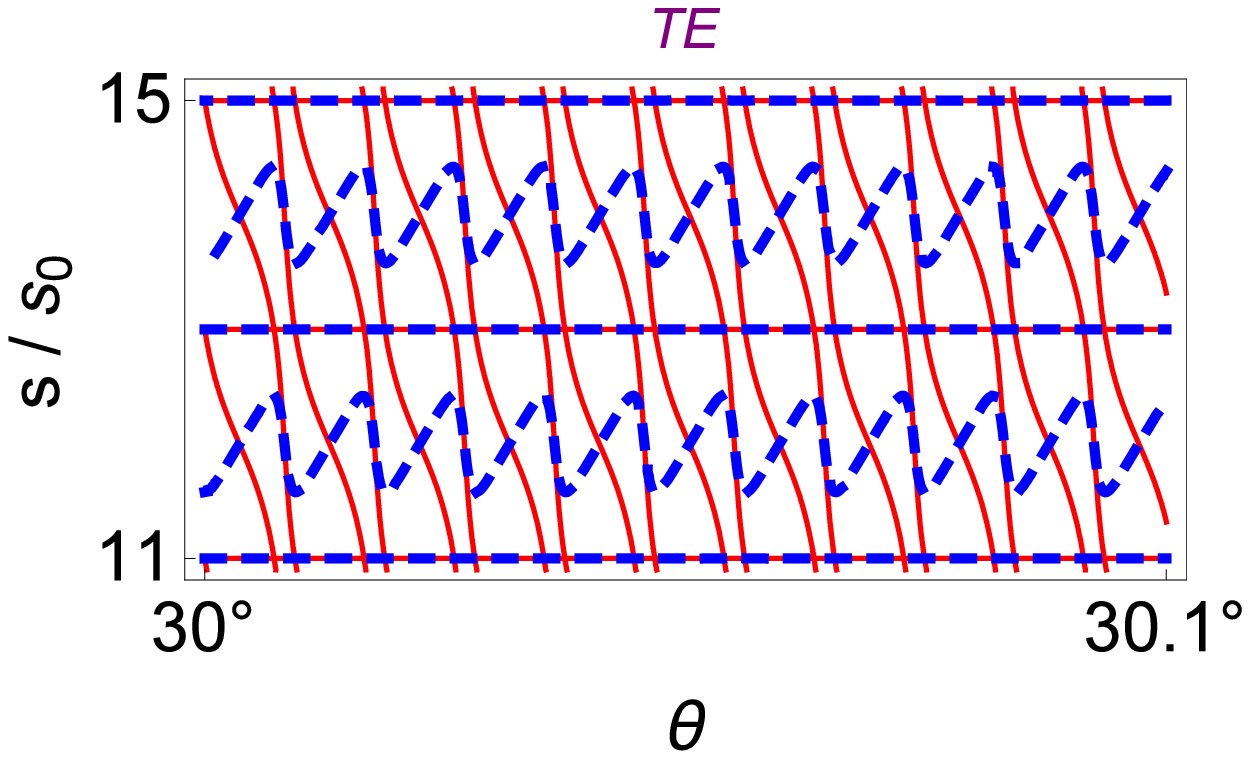}~~~
    \includegraphics[scale=0.4]{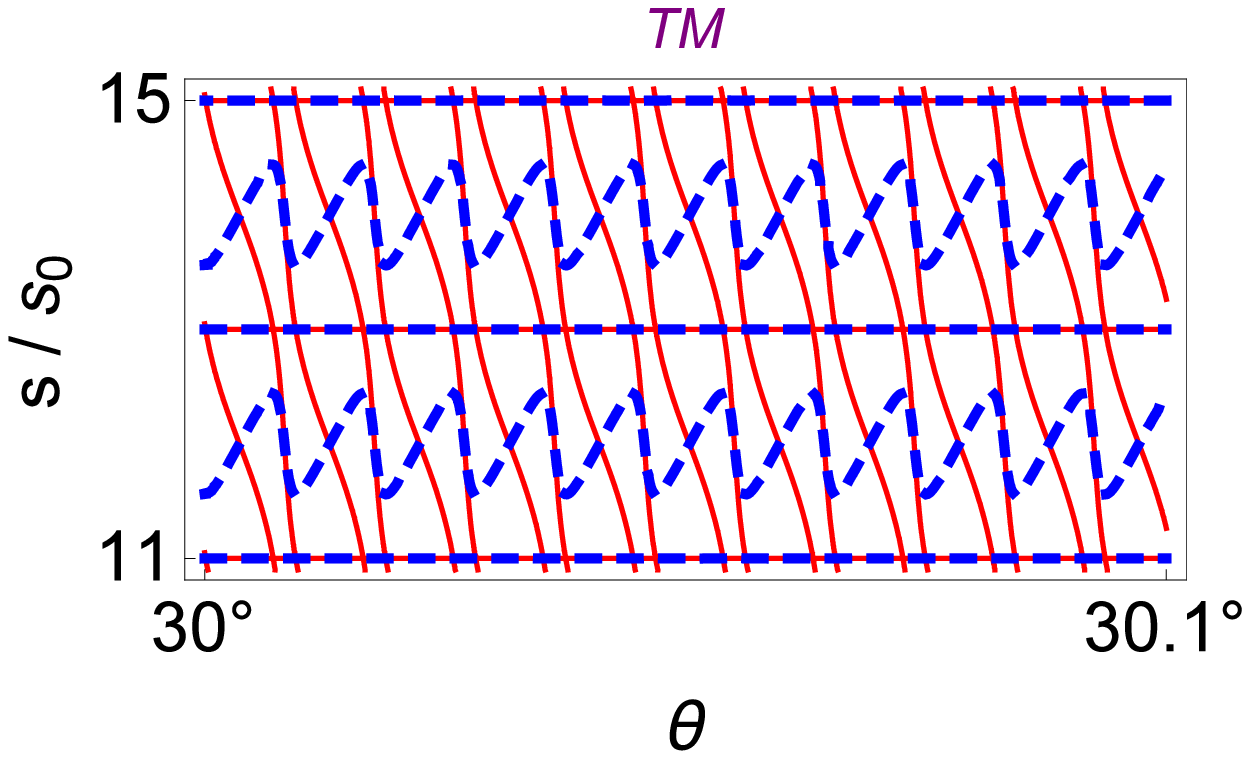}\\
    \includegraphics[scale=0.4]{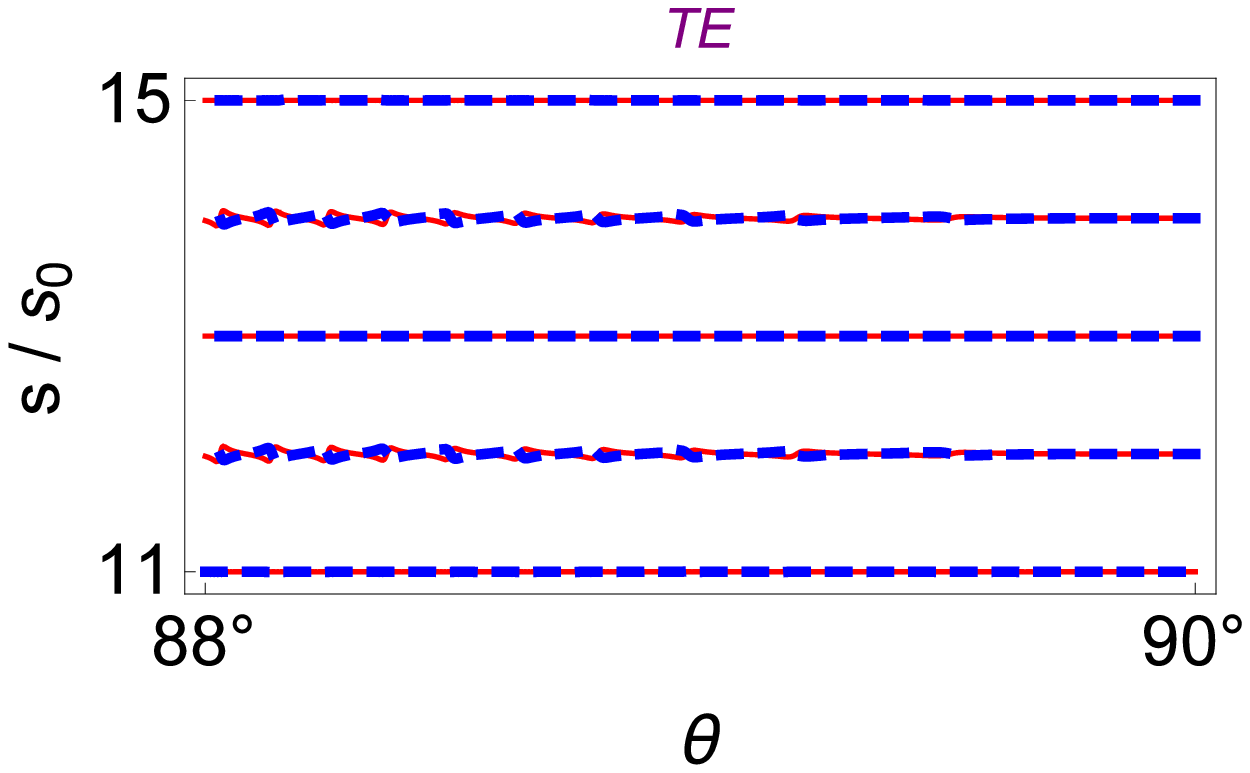}~~~
    \includegraphics[scale=0.4]{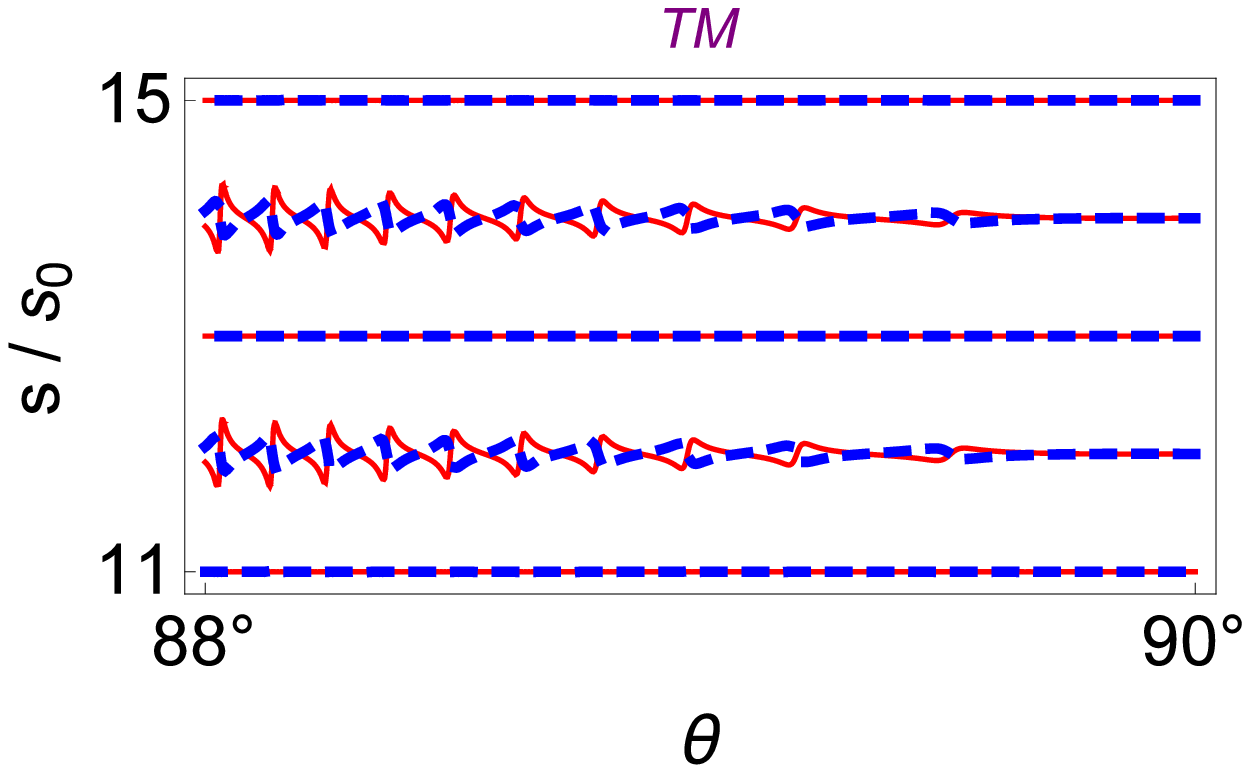}
	\caption{(Color online) Plots of gain-loss separation measure $s/s_0$ as a function of incidence angle $\theta$ corresponding to TE and TM wave solutions for the case of $\cP\cT$-symmetric layers with a gap.}
    \label{g05TEs}
    \end{center}
    \end{figure}

In Fig.~\ref{g05TEs}, the behaviour of angles on the separation distance measure of $s/s_0$ at fixed gain and wavelength is shown for various angle ranges. In these plots, $\cP\cT$-symmetric gain-loss system with a gap is made out of Nd:YAG crystals with thickness $L = 1~\textrm{cm}$, $\lambda = 808~\textrm{nm}$ and gain coefficient of $g = 0.4~\textrm{cm}^{-1}$. Notice again that destructive case for all angles never allows unidirectional reflectionlessness. At large angles, only close surroundings of constructive cases give rise to unidirectional reflectionlessness.

\begin{figure}
	\begin{center}
    \includegraphics[scale=0.35]{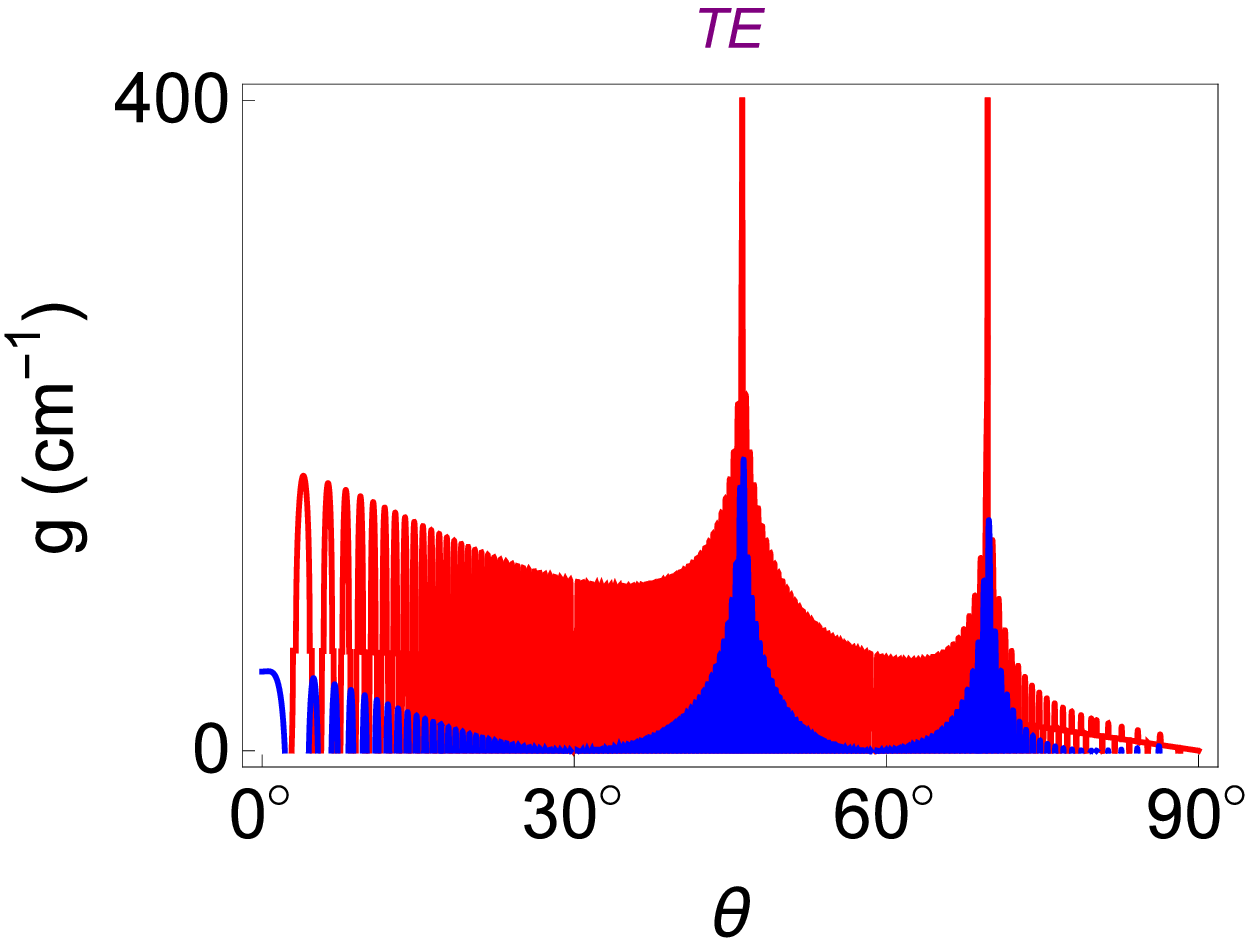}~~~
    \includegraphics[scale=0.35]{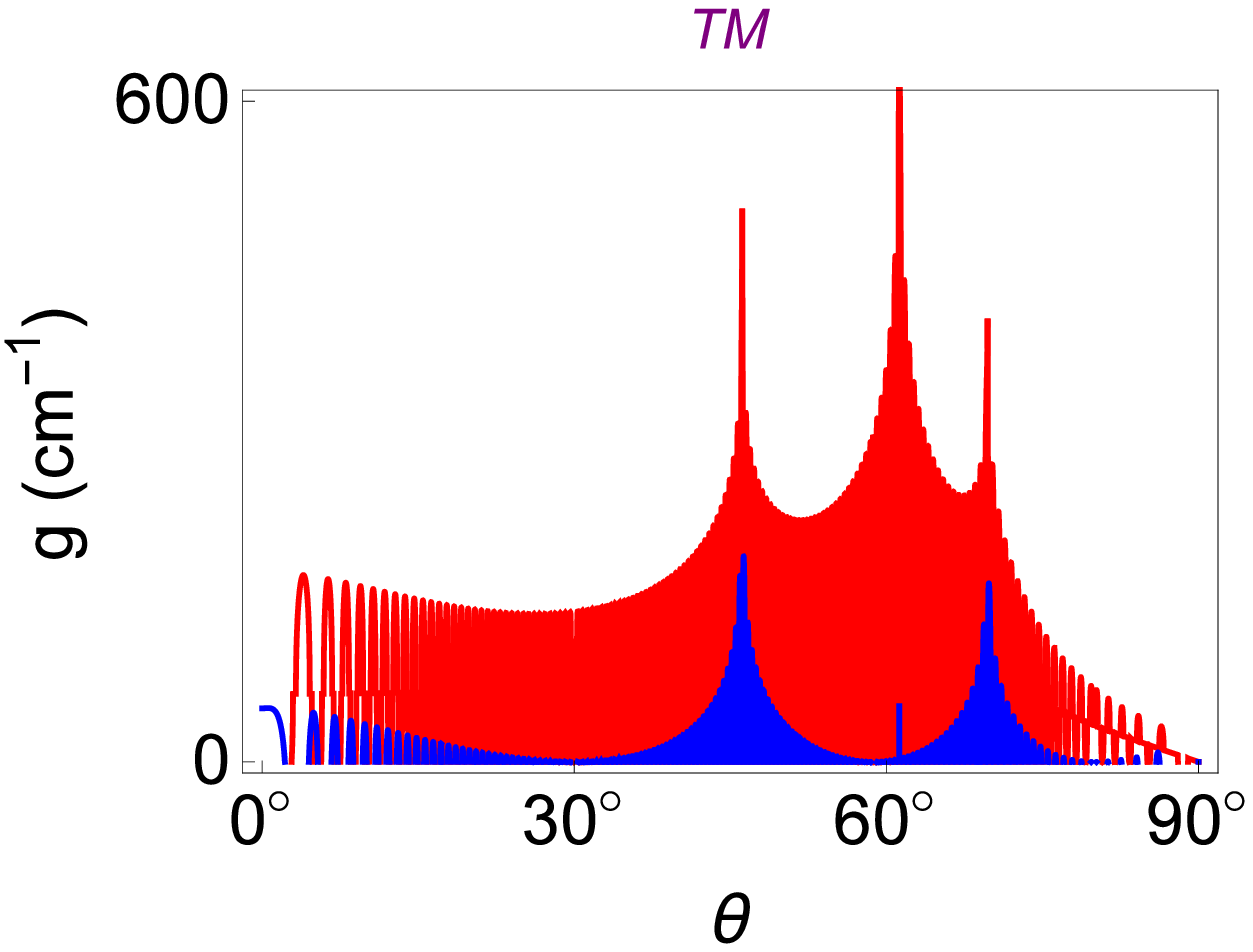}~~~
    \includegraphics[scale=0.35]{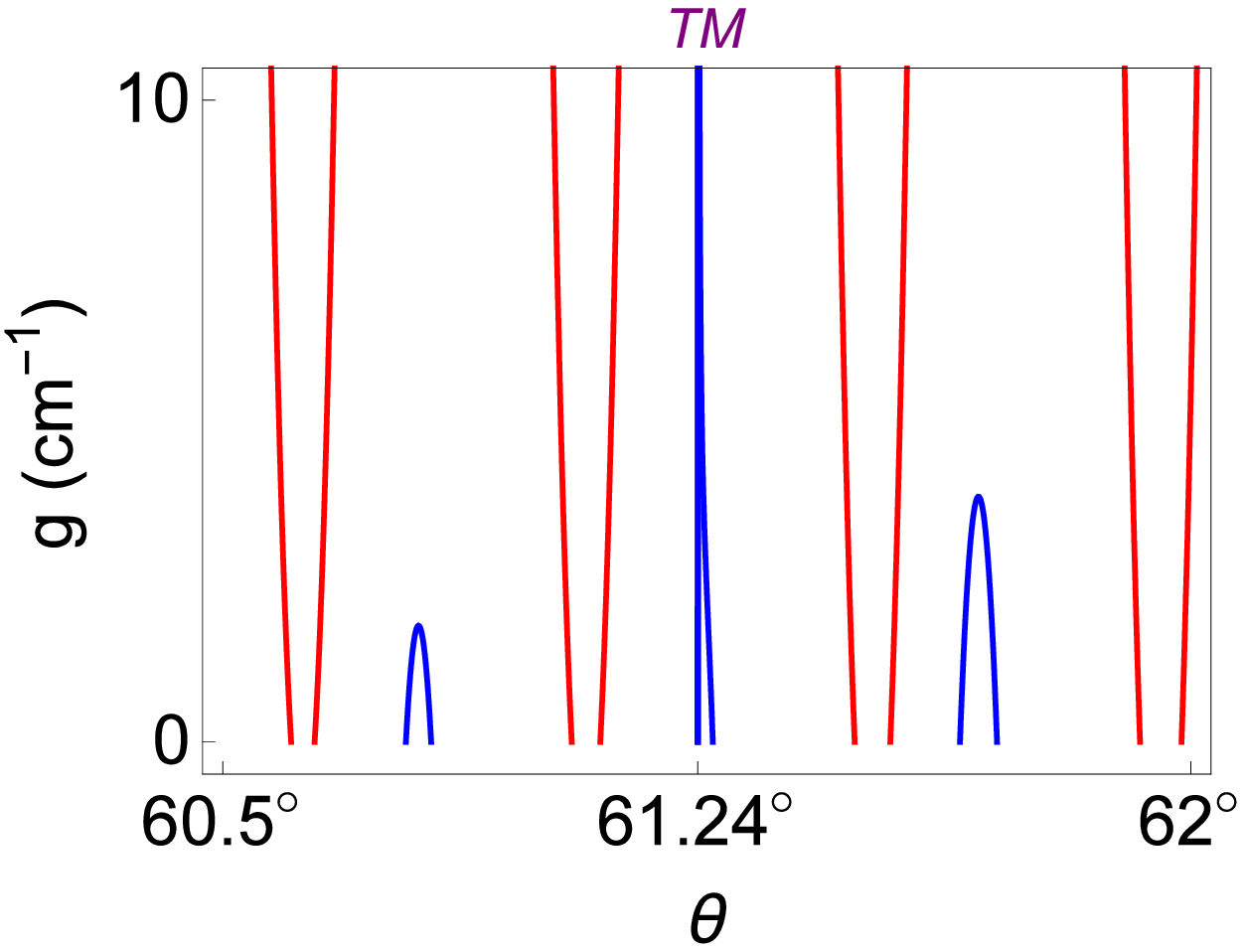}\\
    \includegraphics[scale=0.35]{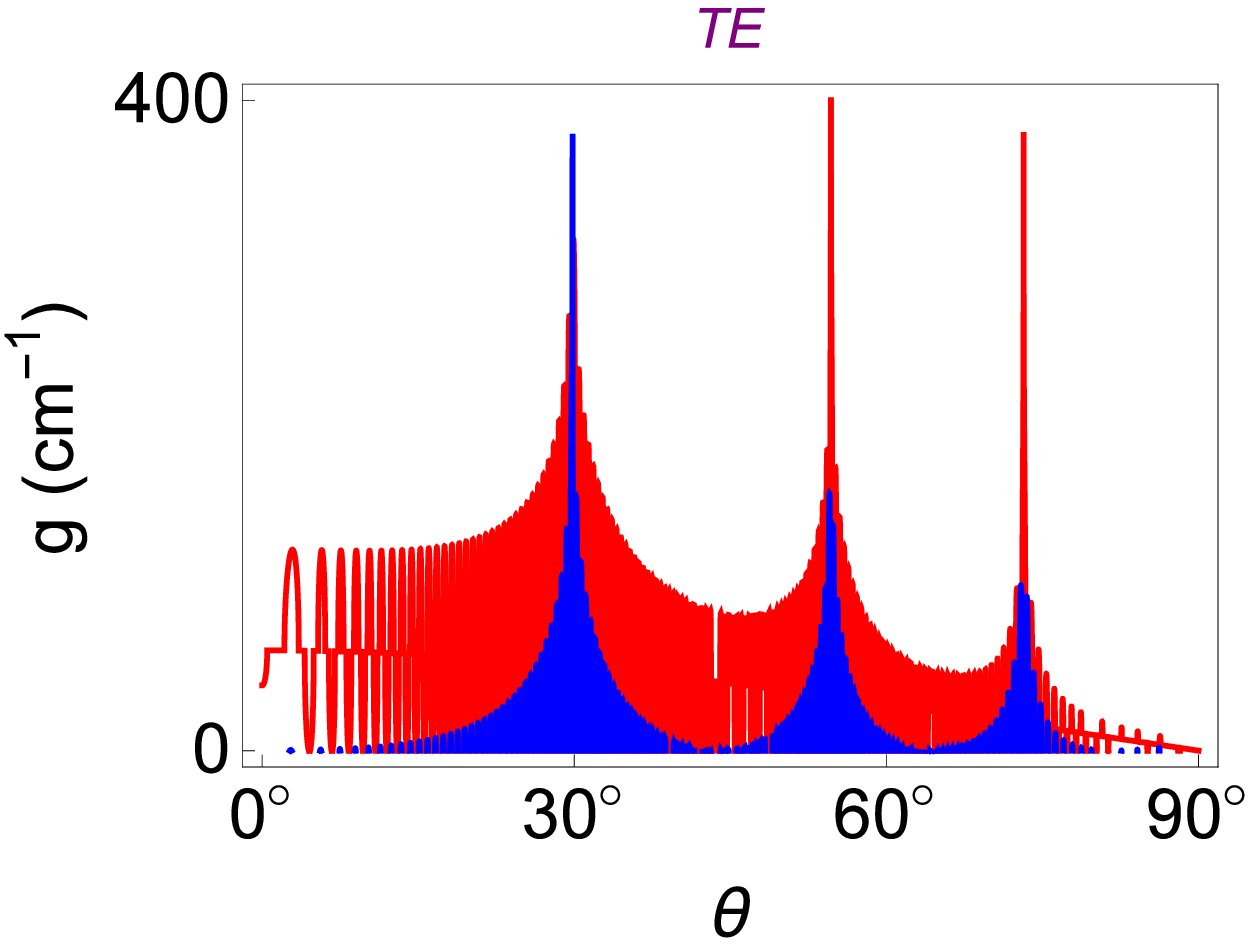}~~~
    \includegraphics[scale=0.35]{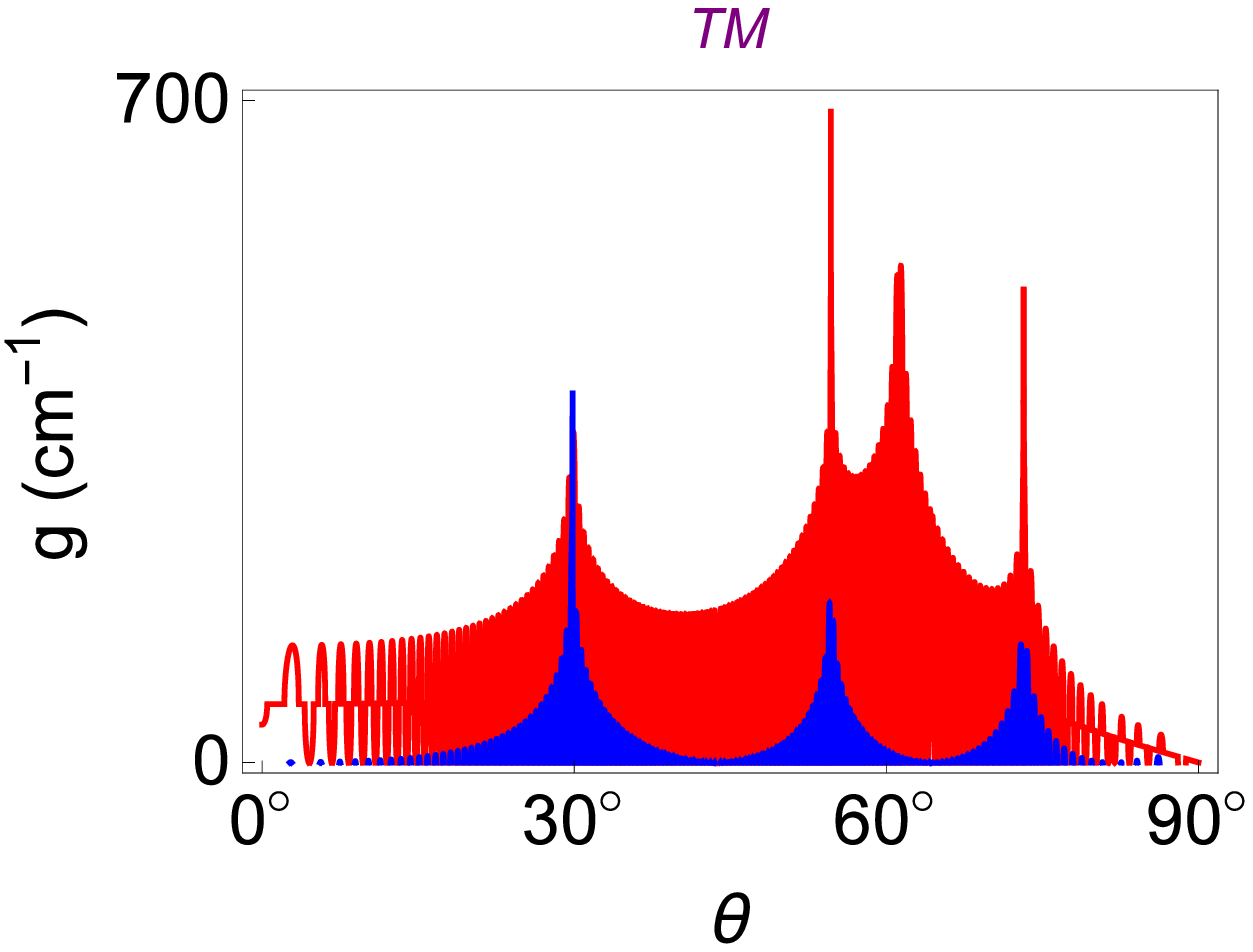}~~~
    \includegraphics[scale=0.35]{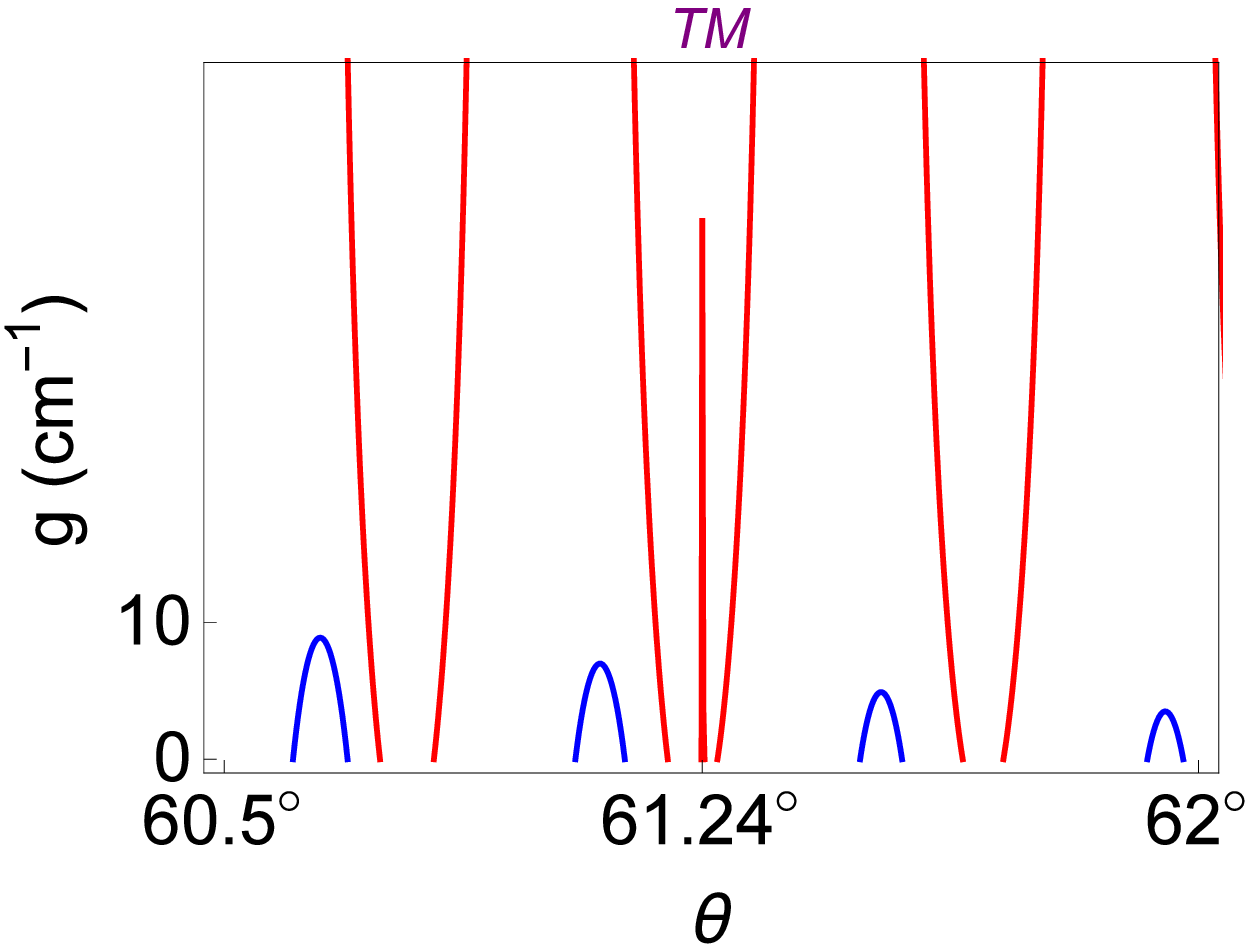}
	\caption{(Color online) Plots of gain coefficient $g$ as a function of incidence angle $\theta$ corresponding to TE and TM wave solutions for the case of $\cP\cT$-symmetric layers with a gap. In these plots, we employ slab thickness $L = 200~\mu\textrm{m}$, wavelength $\lambda = 808~\textrm{nm}$, and $s \approx 1.166~\mu \textrm{m}$ (first row) and $s \approx 1.397~\mu \textrm{m}$ (second row). }
    \label{g06TEs}
    \end{center}
    \end{figure}

In Fig.~\ref{g06TEs} behaviour of gain coefficients with respect to angle of incidence is shown for the case of $\cP\cT$-symmetric Nd:YAG layers with a gap with two different $s$ values, $s= 1.166~\mu \textrm{m}$ and $s= 1.397~\mu \textrm{m}$ corresponding to constructive and almost destructive cases respectively for $s/s_0 = 5$ and $s/s_0 = 5.99$. It is clearly seen that only certain incidence angles allows reflectionless potentials. We remark that the most convenient choice of angles for constructive case are angles around the valleys of grand patterns with two peaks. These angles correspond to around $\theta = 30^{\circ}, 60^{\circ}$ and $90^{\circ}$. Gain values at large angles around $\theta \approx 90^{\circ}$ are quite small, which is more favorable. For the almost destructive case, the favorable angles shift around $\theta = 0^{\circ}, 45^{\circ}, 65^{\circ}$ and $90^{\circ}$. In these graphs, the role of Brewster's angle in TM mode is shown that both directional reflectionlessness occur at very small gain values. Finally, these graphs clearly demonstrates that left reflectionlessness is always achieved at small gain values compared to the right one.

Finally, it is worth to mention the effect of varying thickness of gain (in turn, loss) slab(s). We verify similar observations encountered in $\cP\cT$-symmetric bilayer case that lowering thickness results in unidirectional reflectionless situation for a wide range wavelength at increased gain values.

\subsection{Unidirectional Reflectionlessness: Exact Analysis}

\begin{figure}
	\begin{center}
    \includegraphics[scale=0.4]{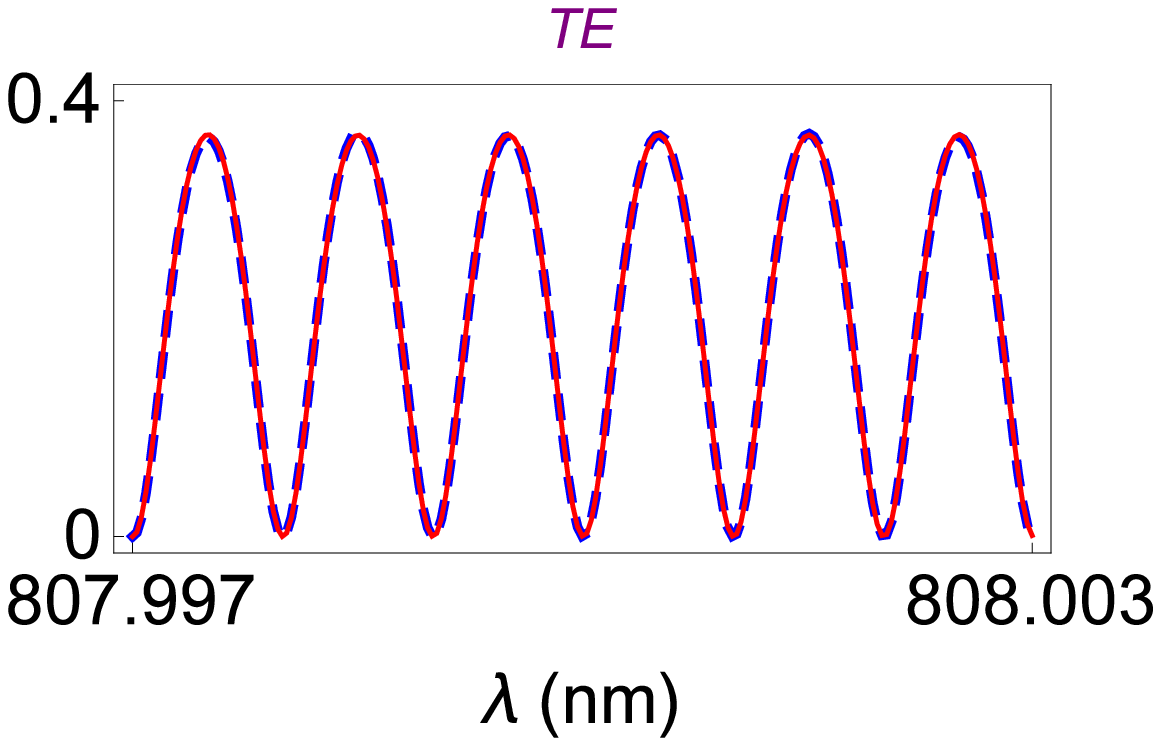}~~~
    \includegraphics[scale=0.4]{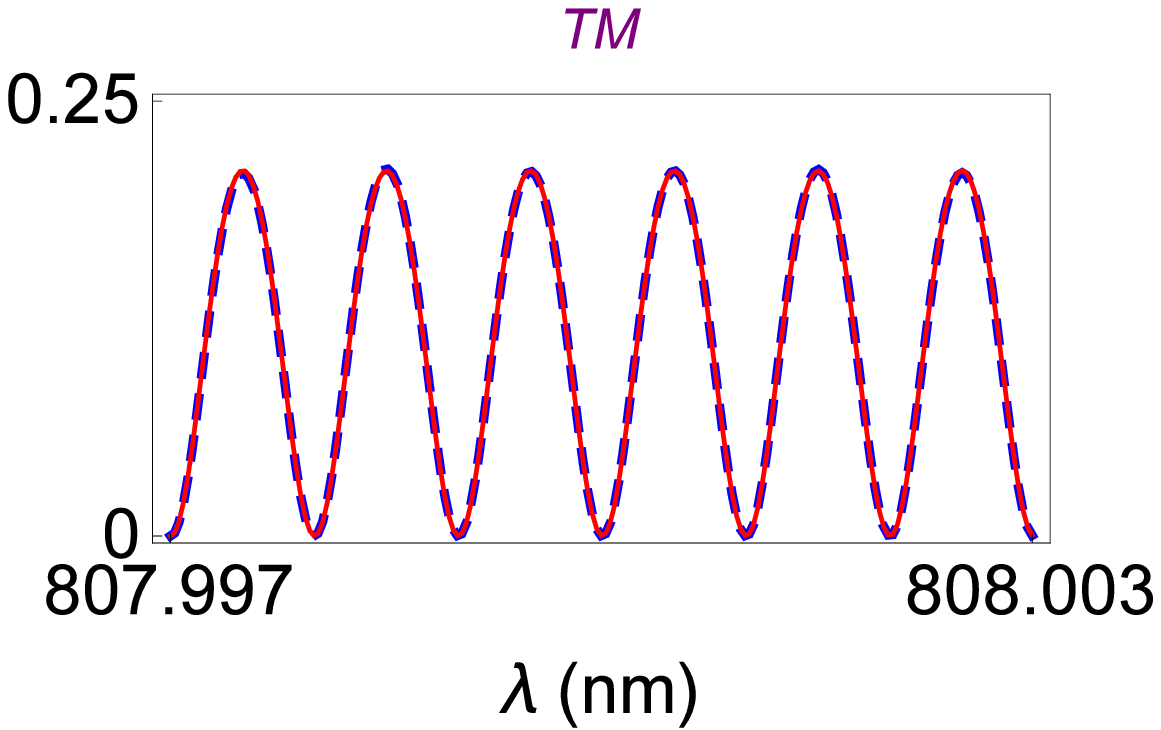}\\
    \includegraphics[scale=0.4]{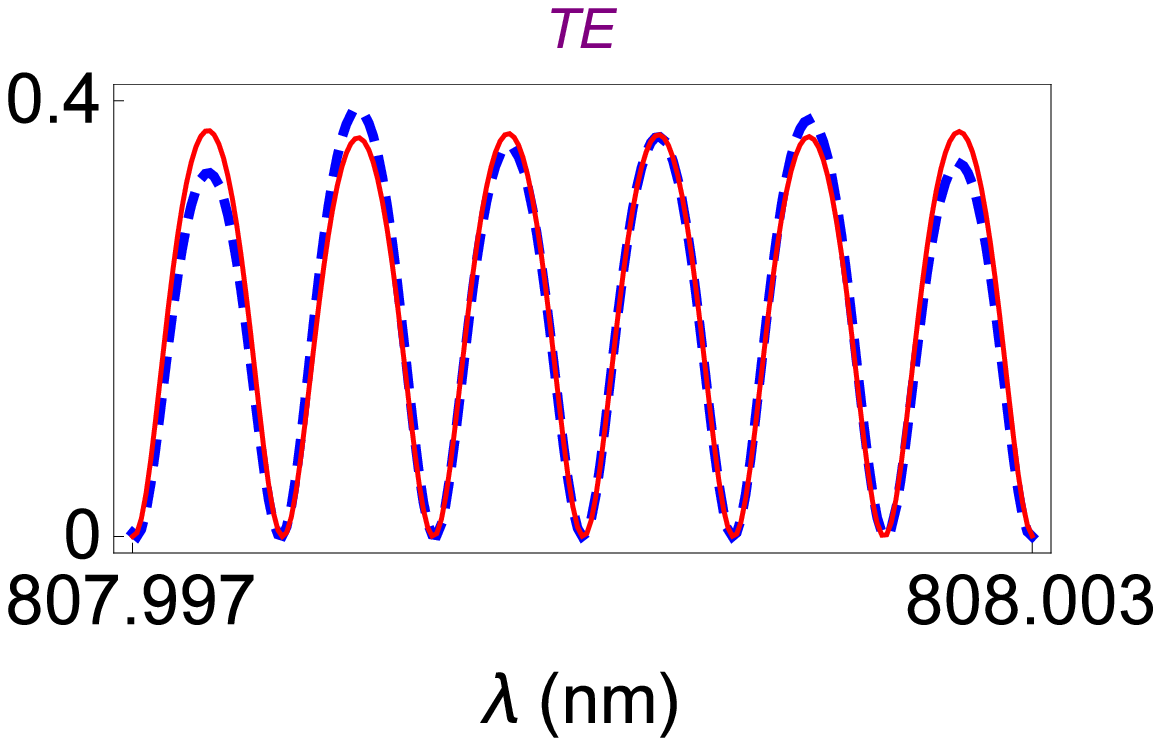}~~~
    \includegraphics[scale=0.4]{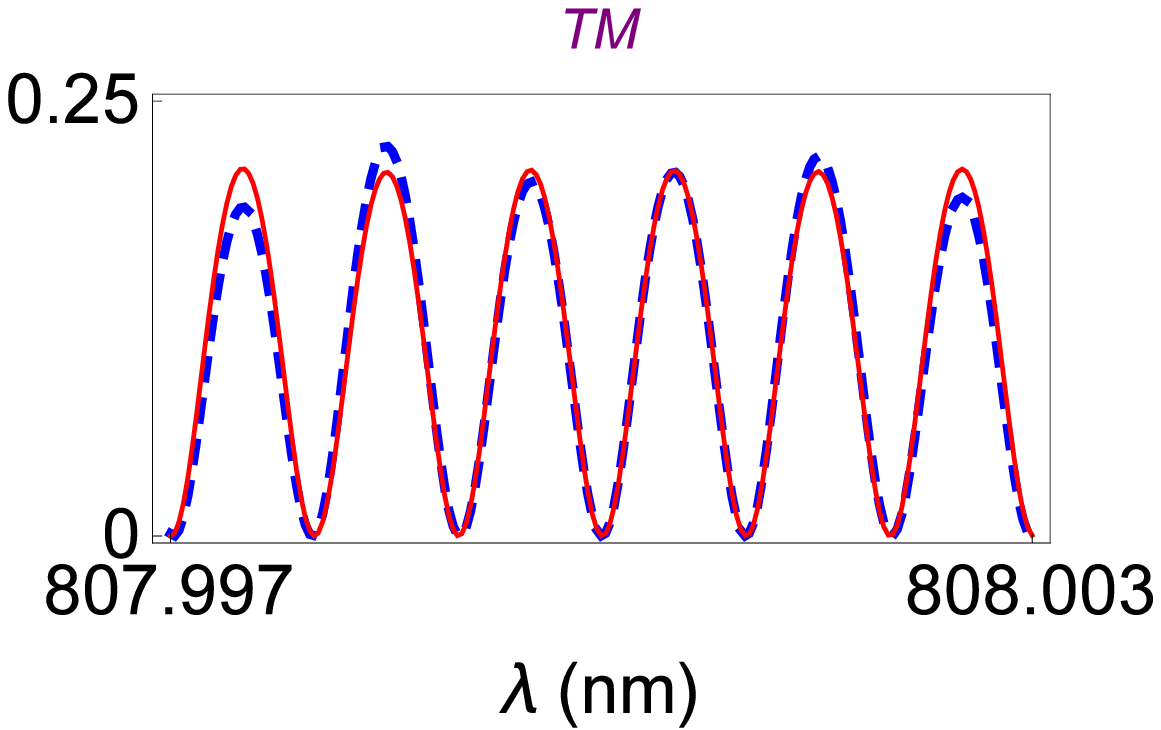}\\
    \includegraphics[scale=0.4]{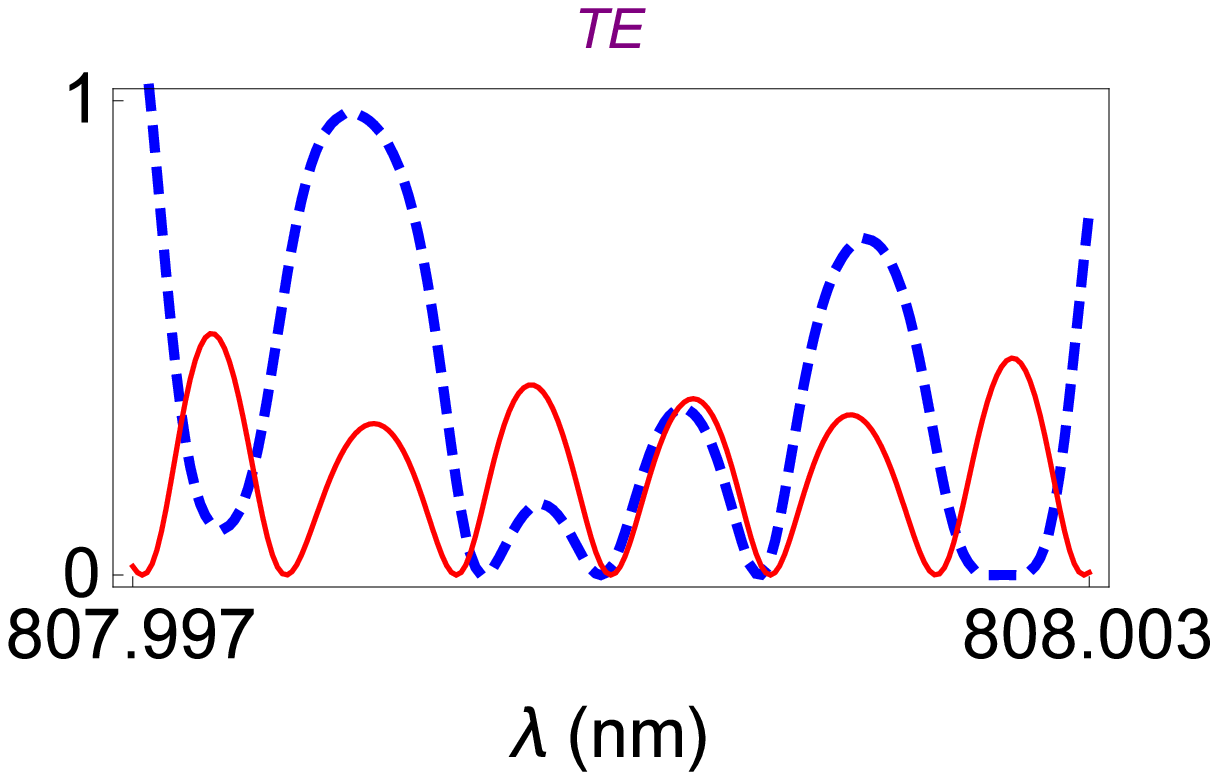}~~~
    \includegraphics[scale=0.4]{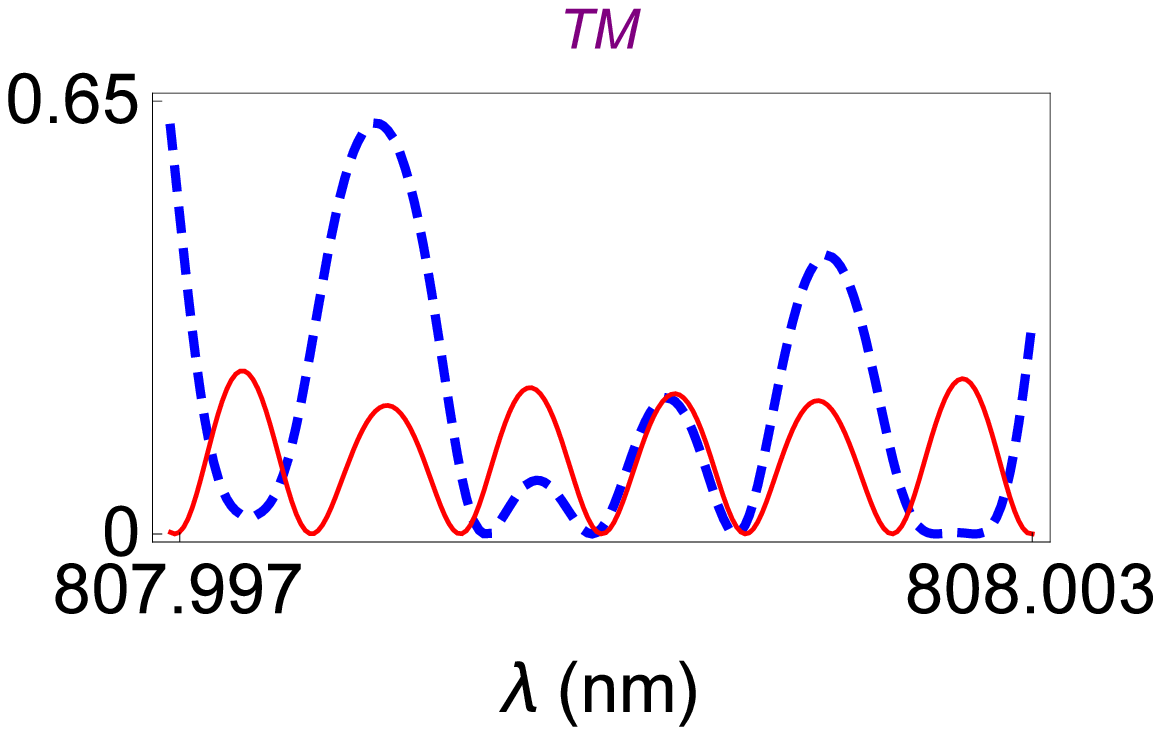}\\
    \includegraphics[scale=0.4]{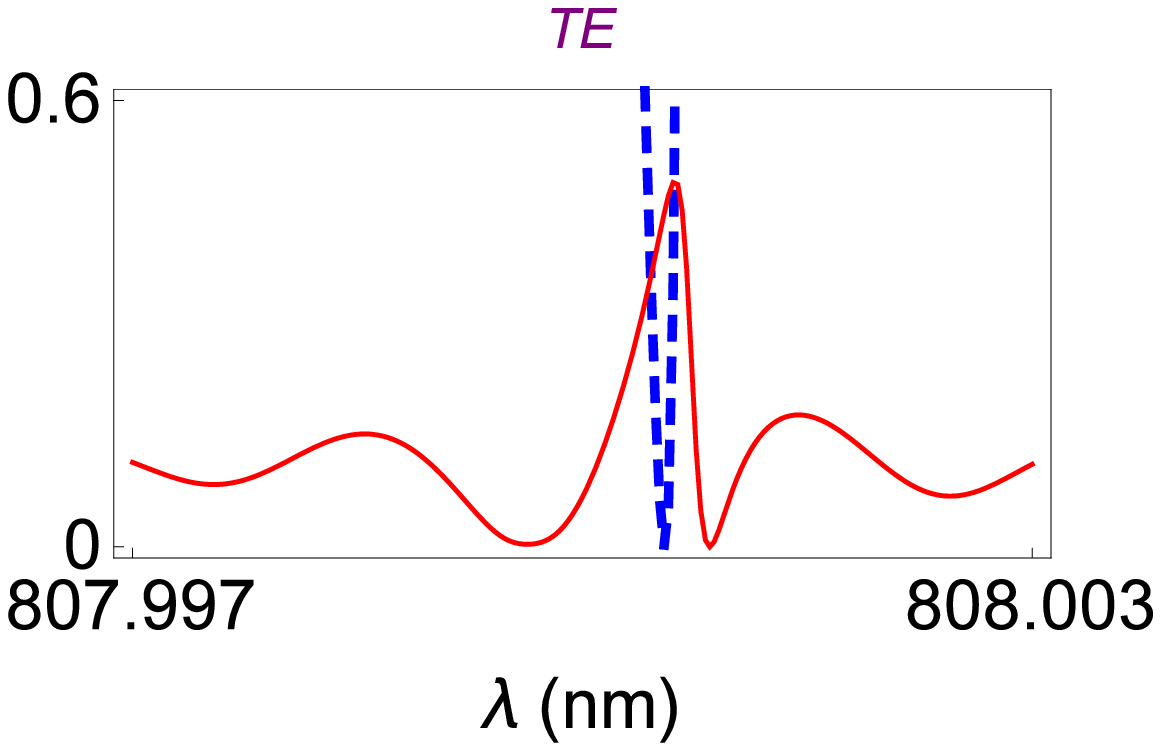}~~~
    \includegraphics[scale=0.4]{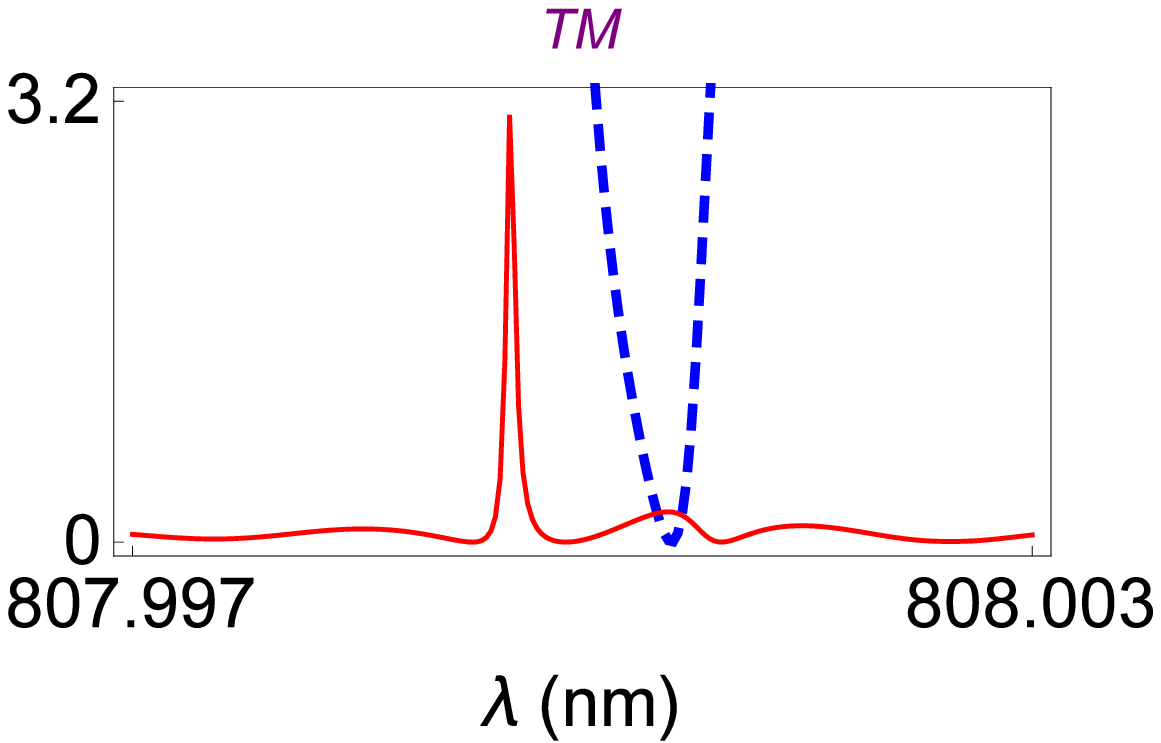}\\
    \includegraphics[scale=0.4]{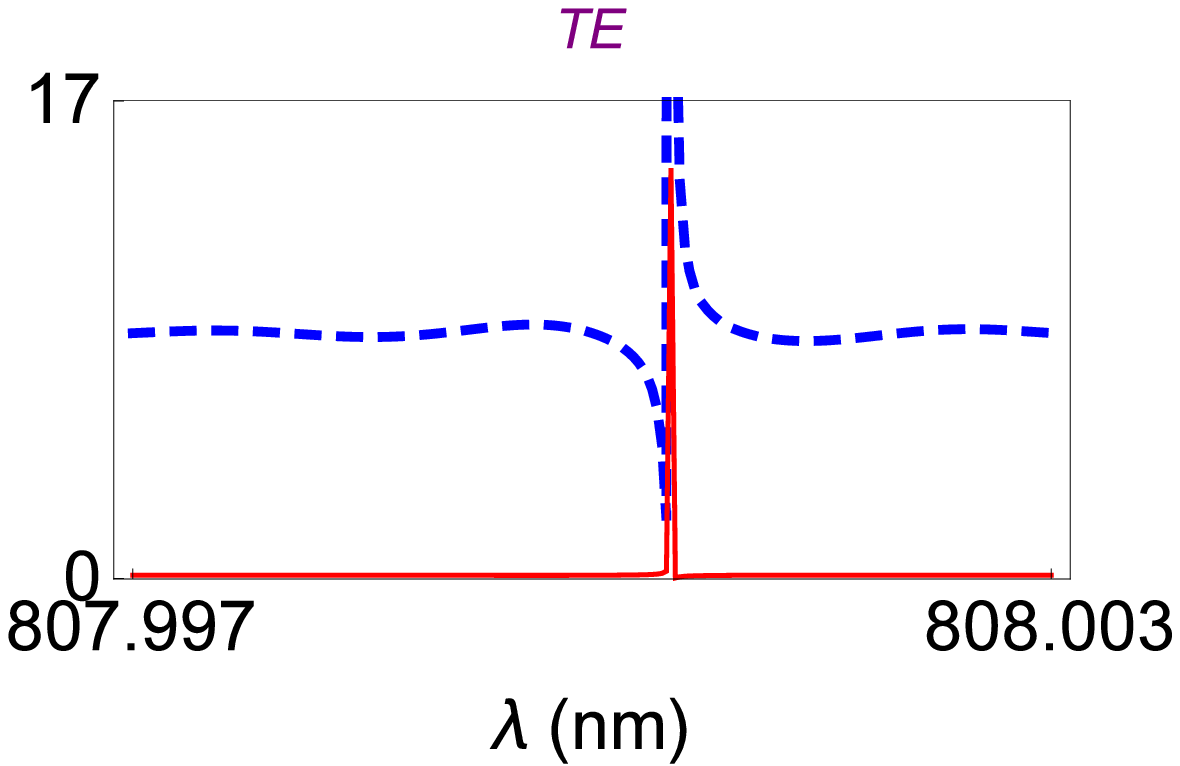}~~~
    \includegraphics[scale=0.4]{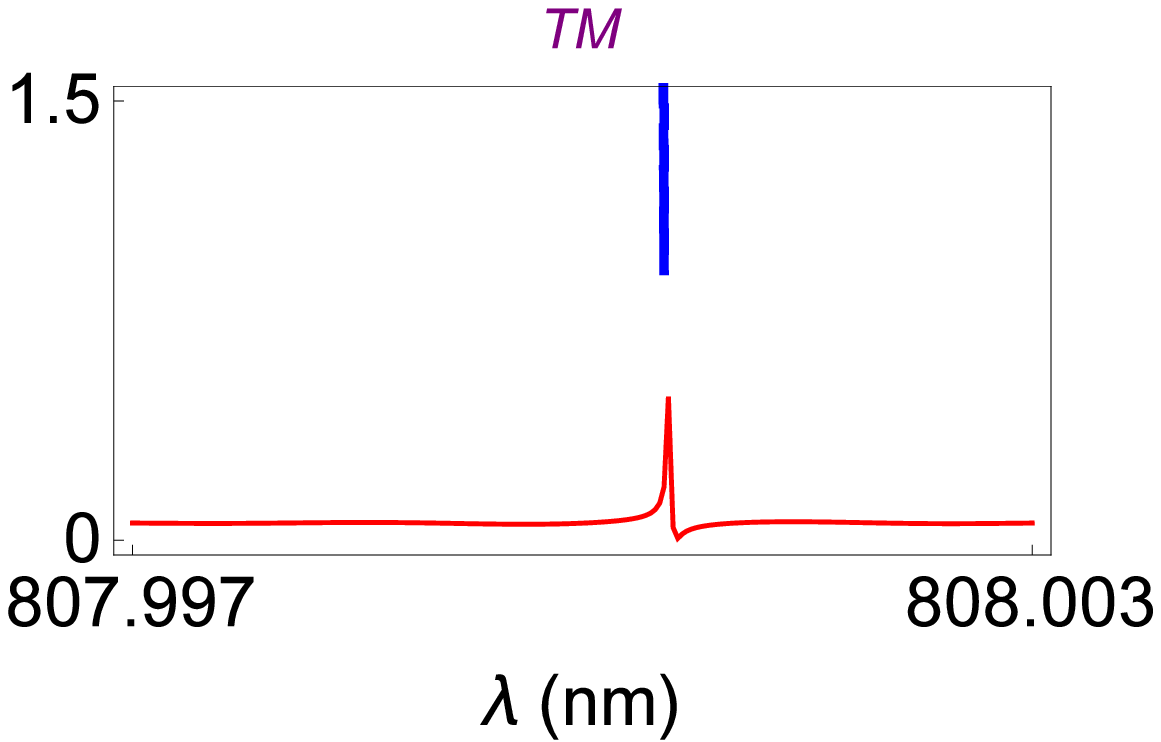}
	\caption{(Color online) Plots of $\left|R^{l}\right|^2$ (thick dashed blue curve) and $\left|R^{r}\right|^2$ (thin solid red curve) as a function of wavelength $\lambda$ corresponding to TE and TM wave solutions for the case of $\cP\cT$-symmetric Nd:YAG layers with a gap and thickness $L = 10~\textrm{cm}$ at incident angle $\theta = 30^{\circ}$. Graphs are plotted for gain values $g \approx 0.35~\textrm{cm}^{-1}$, $g \approx 0.60~\textrm{cm}^{-1}$, $g \approx 0.90~\textrm{cm}^{-1}$, $g \approx 1.2~\textrm{cm}^{-1}$ and $g \approx 1.5~\textrm{cm}^{-1}$ from top to down manner. It is clear that unidirectional reflectionlessness occur for gain values greater than some certain gain values (graphs in last two row), below which one observes bidirectional reflectionlessness. }
    \label{reflectionlessnessTEM1}
    \end{center}
    \end{figure}

Based on the guiding light of the perturbative analysis, we achieved the required gain values with the corresponding wavelength range for the left and right reflectionlessness. We can use this information to verify that out $\cP\cT$-symmetric system is indeed unidirectionally reflectionless. Besides, It is also quite natural to extract information directly from the components of transfer matrix since they give rise to deduce the quantities $\left|R^{l}\right|^2$ and $\left|R^{r}\right|^2$ in the light of consequences of the last subsection. In Fig.~\ref{reflectionlessnessTEM1} one observes the graphs of $\left|R^{l}\right|^2$ (thick dashed blue curve) and $\left|R^{r}\right|^2$ (thin solid red curve) as a function of wavelength $\lambda$ for the $\cP\cT$-symmetric Nd:YAG crystals with a gap possessing a constructive configuration with $s/s_0 = 20$ and $L = 10~\textrm{cm}$ at incidence angle of $\theta = 30^{\circ}$ for various gain values ranging from $g \approx 0.35~\textrm{cm}^{-1}$ to $g \approx 1.5~\textrm{cm}^{-1}$. These graphs clarify that below the gain values $g \approx 0.35~\textrm{cm}^{-1}$ only bidirectional reflectionlessness is observed, and above this gain value, unidirectional reflectionlessness originates, which verifies the results found in~Fig.~\ref{g01TEs}. If we still increase the amount of gain unidirectional reflectionlessness disappears and no reflectionlessness is observed, which verifies Fig.~\ref{g01TEs}. Notice that the best degree of reflectionlessness is guaranteed just above the intersection points of right and left zero-reflection amplitude curves. It is also worth to express that the reflectionless range of wavelength can be widened considerably if the gain value is well-adjusted (see the graphs in third row of Fig.~\ref{reflectionlessnessTEM1}).

\begin{figure}
	\begin{center}
    \includegraphics[scale=0.5]{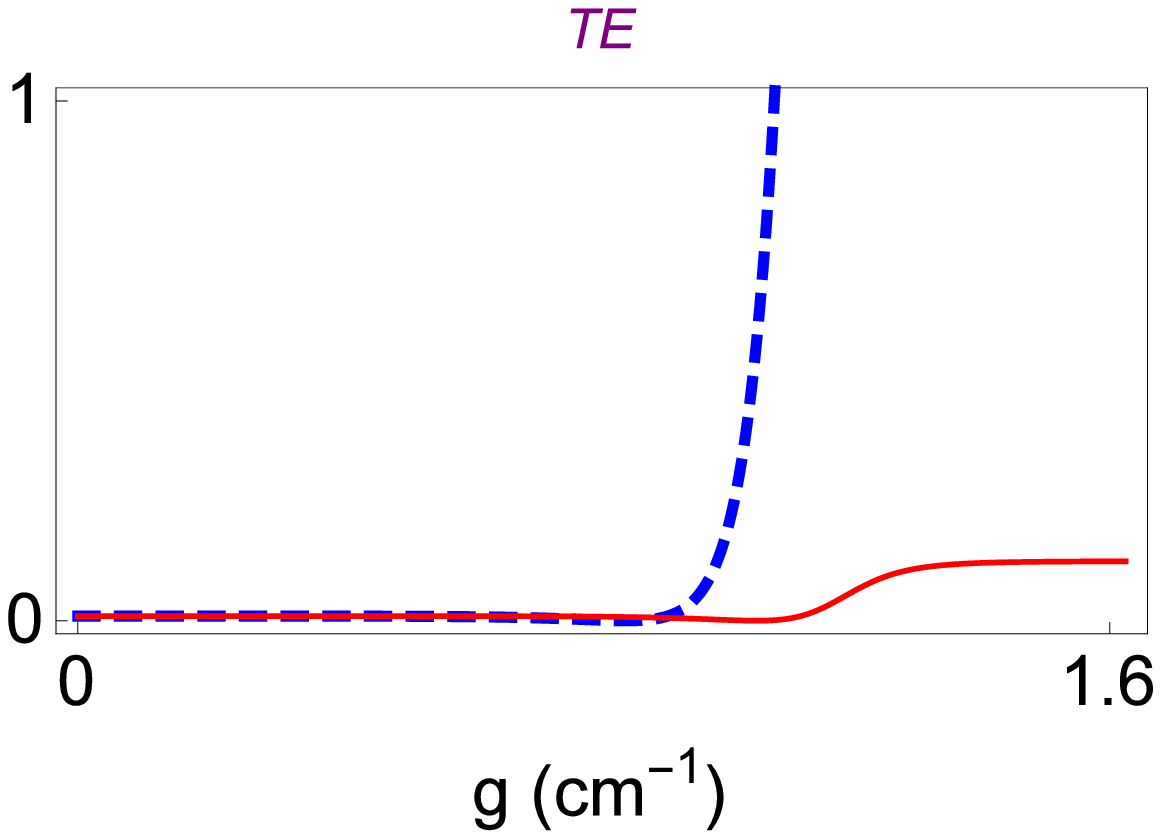}~~~
    \includegraphics[scale=0.5]{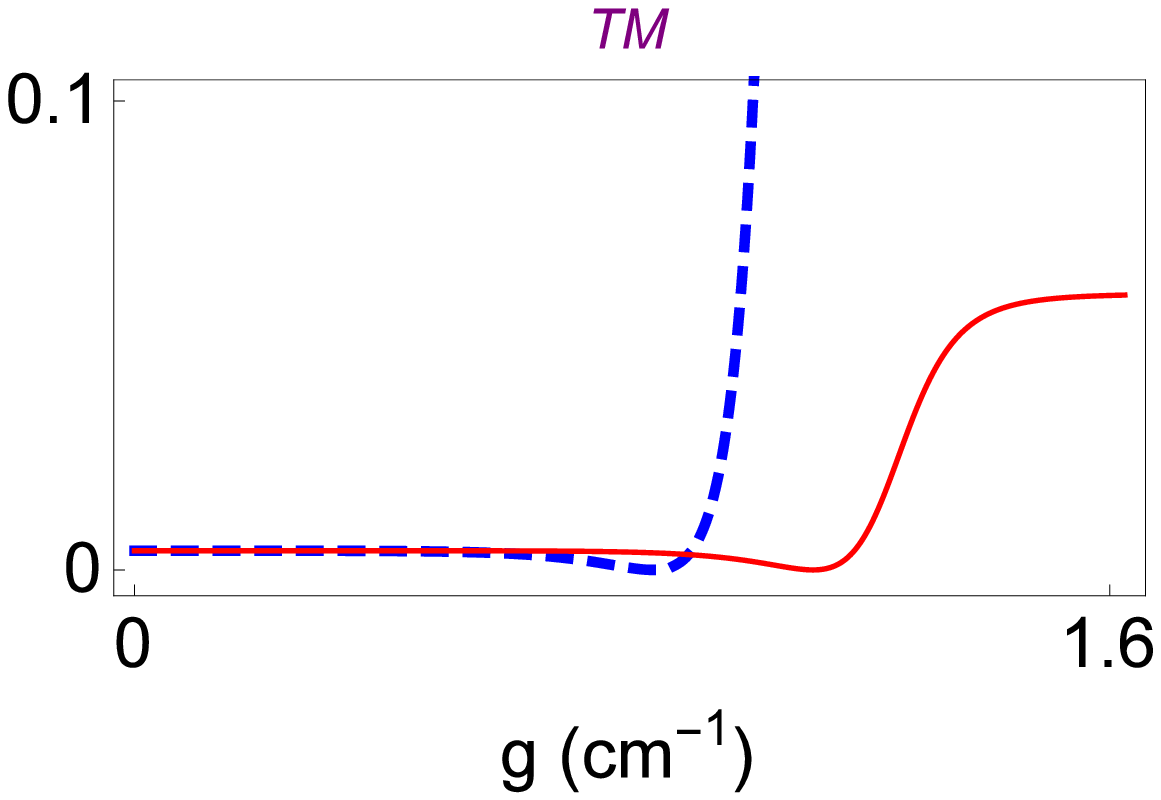}\\
    \includegraphics[scale=0.5]{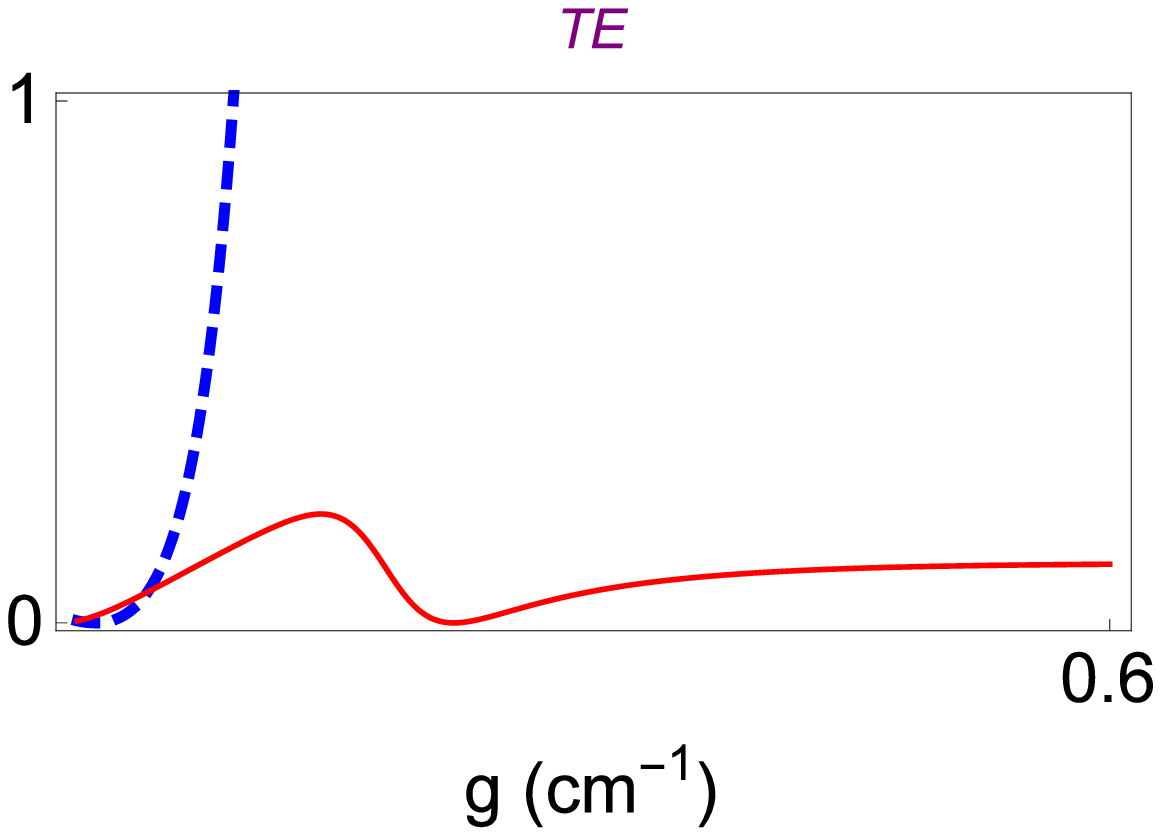}~~~
    \includegraphics[scale=0.5]{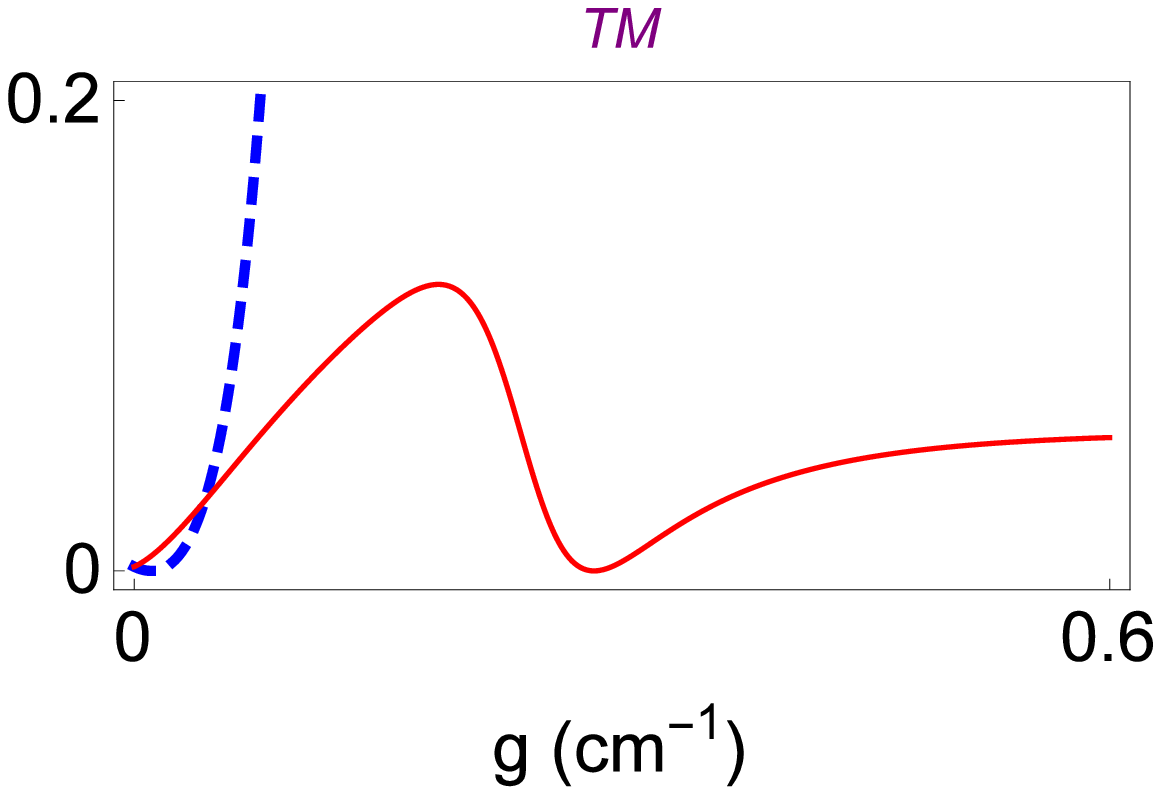}\\
    \includegraphics[scale=0.5]{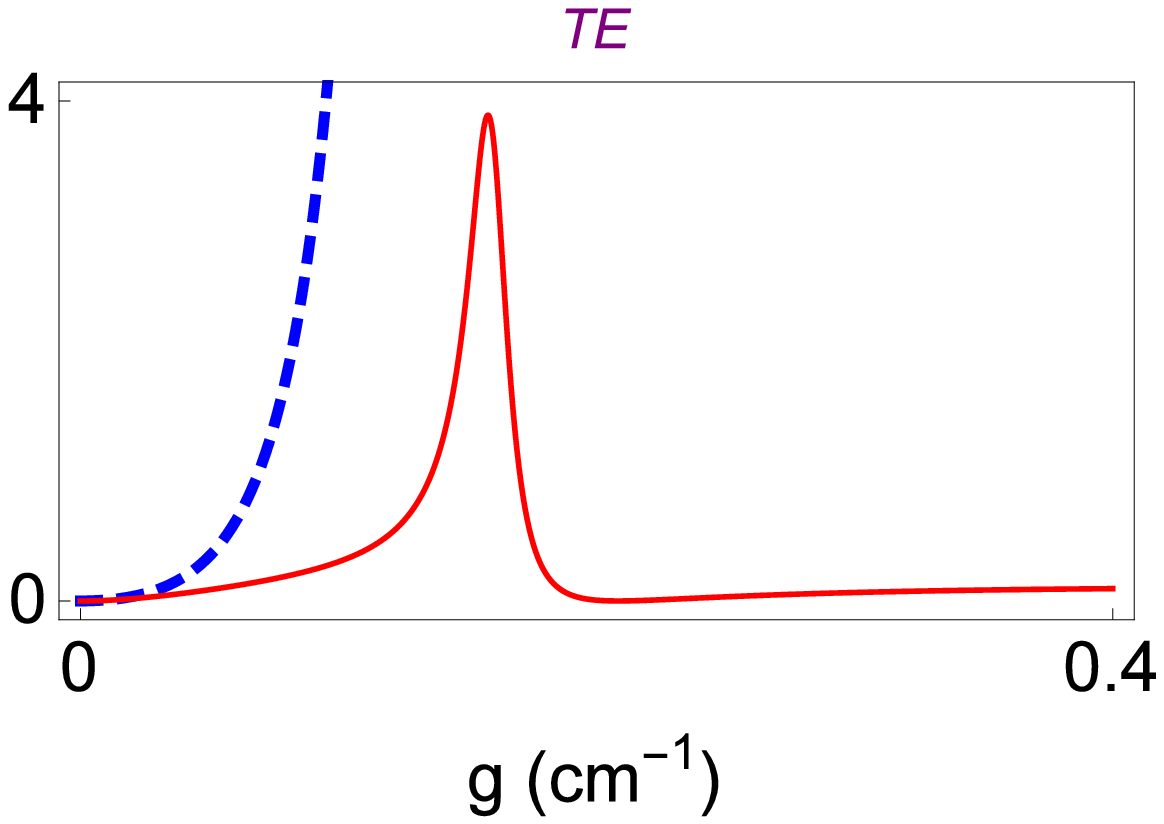}~~~
    \includegraphics[scale=0.5]{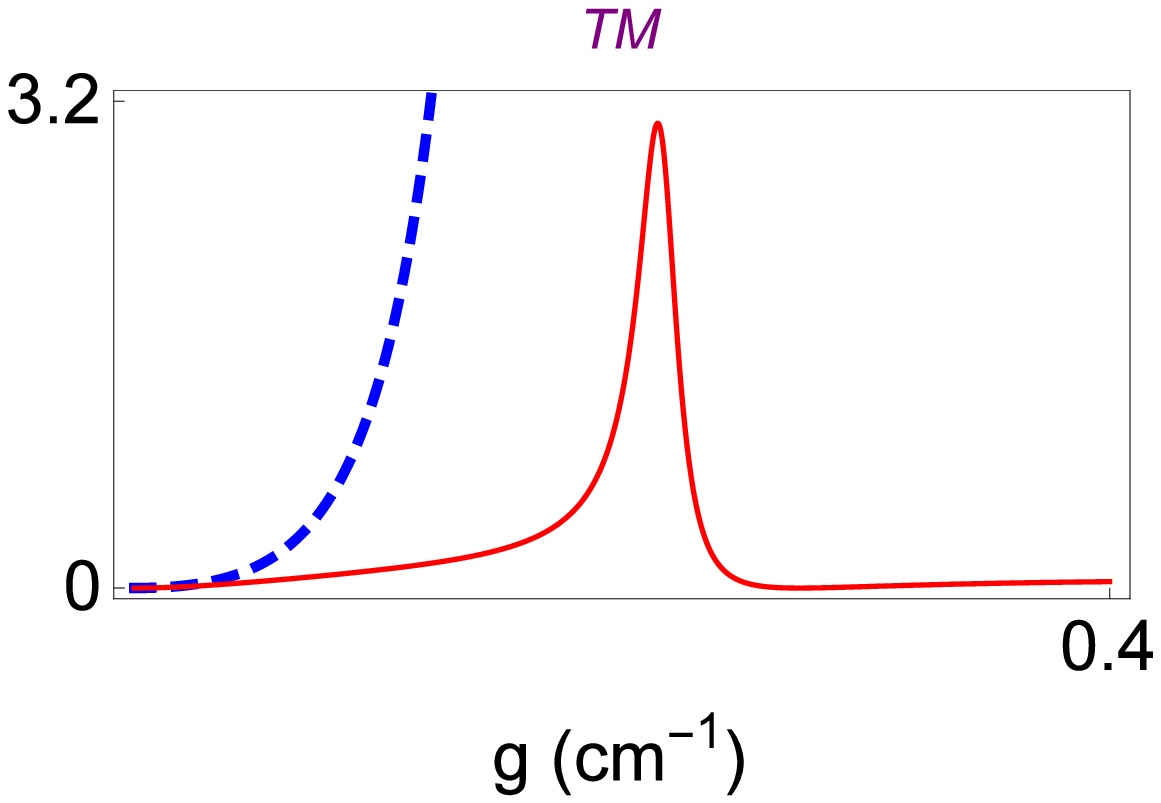}
	\caption{(Color online) Plots of $\left|R^{l}\right|^2$ (thick dashed blue curve) and $\left|R^{r}\right|^2$ (thin solid red curve) as a function of gain coefficient $g$ corresponding to TE and TM wave solutions for the case of $\cP\cT$-symmetric Nd:YAG layers with a gap.  Graphs are plotted for gain-loss separations comprising the constructive configuration with $s/s_0 = 20$ (top row), the generic configuration with $s/s_0 = 20.5$ (middle row) and almost destructive configuration with $s/s_0 = 20.99$ (bottom row).}
    \label{reflectionlessnessTEM2}
    \end{center}
    \end{figure}

In Fig.~\ref{reflectionlessnessTEM2}, the effect of gain coefficient on graphs of $\left|R^{l}\right|^2$ and $\left|R^{r}\right|^2$ is displayed for constructive, generic and destructive configuration cases for $\cP\cT$-symmetric gain-loss system with a gap, which is made out of Nd:YAG crystals with thickness $L = 10~\textrm{cm}$, and wave is sent out at $\lambda = 807.9997~\textrm{nm}$ at angle of incidence $\theta = 30^{\circ}$. It is obvious that required amount of gain for unidirectional reflectionlessness is lowered in passing from constructive to destructive cases. Left reflectionlessness is easier to achieve compared to the right one. Above certain amount of gain value both reflectionless situations clear away.

\begin{figure}
	\begin{center}
    \includegraphics[scale=0.5]{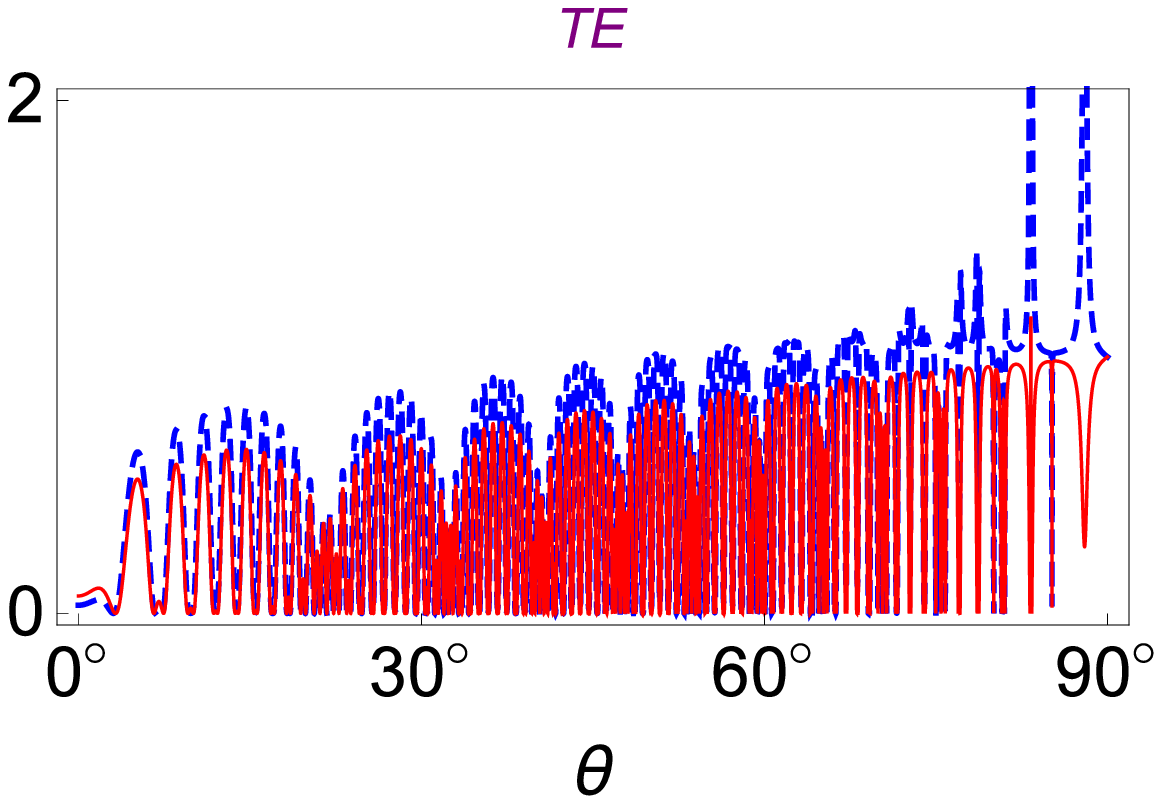}~~~
    \includegraphics[scale=0.5]{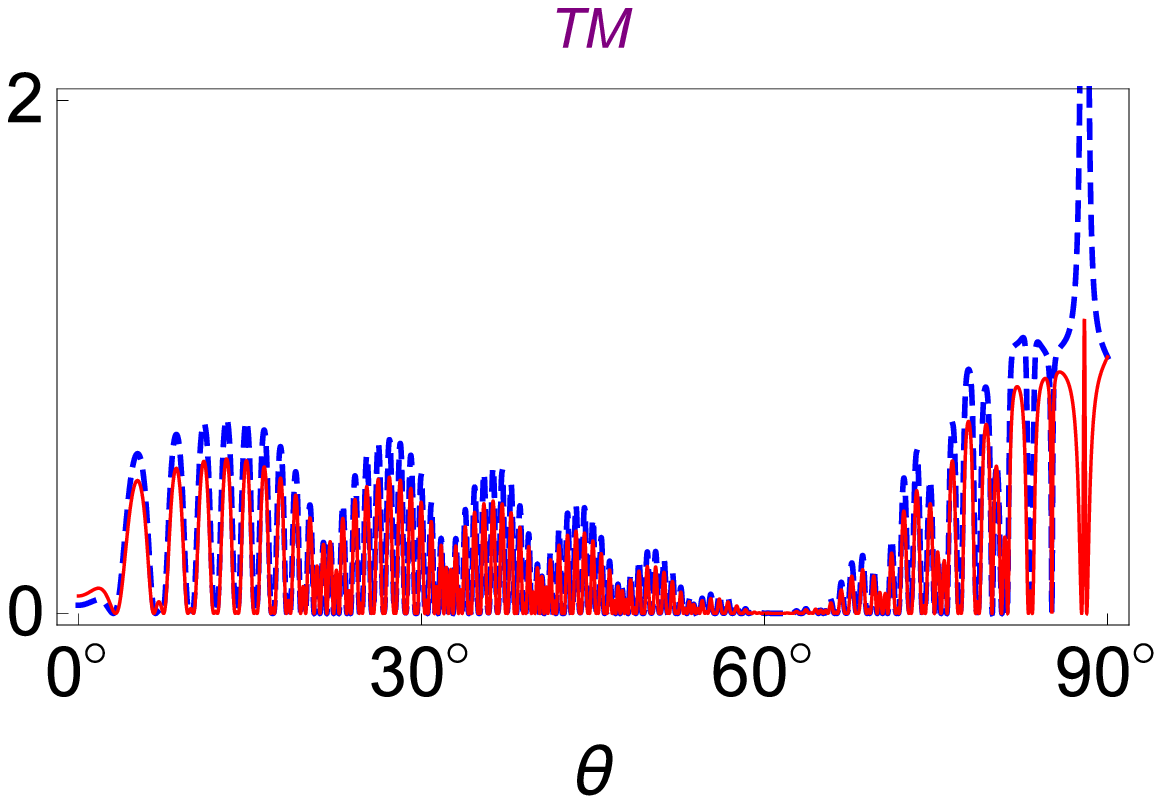}
	\caption{(Color online) Plots of $\left|R^{l}\right|^2$ (thick dashed blue curve) and $\left|R^{r}\right|^2$ (thin solid red curve) as a function of incident angle $\theta$ corresponding to TE and TM wave solutions for the case of $\cP\cT$-symmetric layers with a gap. }
    \label{reflectionlessnessTEM3}
    \end{center}
    \end{figure}

In Fig.~\ref{reflectionlessnessTEM3}, the dependence of $\left|R^{l}\right|^2$ and $\left|R^{r}\right|^2$ on incidence angle is observed for the $\cP\cT$-symmetric Nd:YAG layers with a gap . Gap width is taken to be $s \approx 4.782~\textrm{nm}$ and wave is sent out at $\lambda = 807.9997~\textrm{nm}$ for the gain value $g = 13.44~\textrm{cm}^{-1}$. To clarify the situation, we employ gain (loss) thickness of $L = 100~\mu\textrm{m}$. We see that not all incident angles leads to a uni/bi-directional reflectionless situation, but some discrete angles give rise to it. The best alternate is obtained at angles around valleys of big patterns. Also, it is worth to see the effect of Brewster's angle in TM, which yield a perfect reflectionless situation.

\begin{figure}
	\begin{center}
    \includegraphics[scale=0.4]{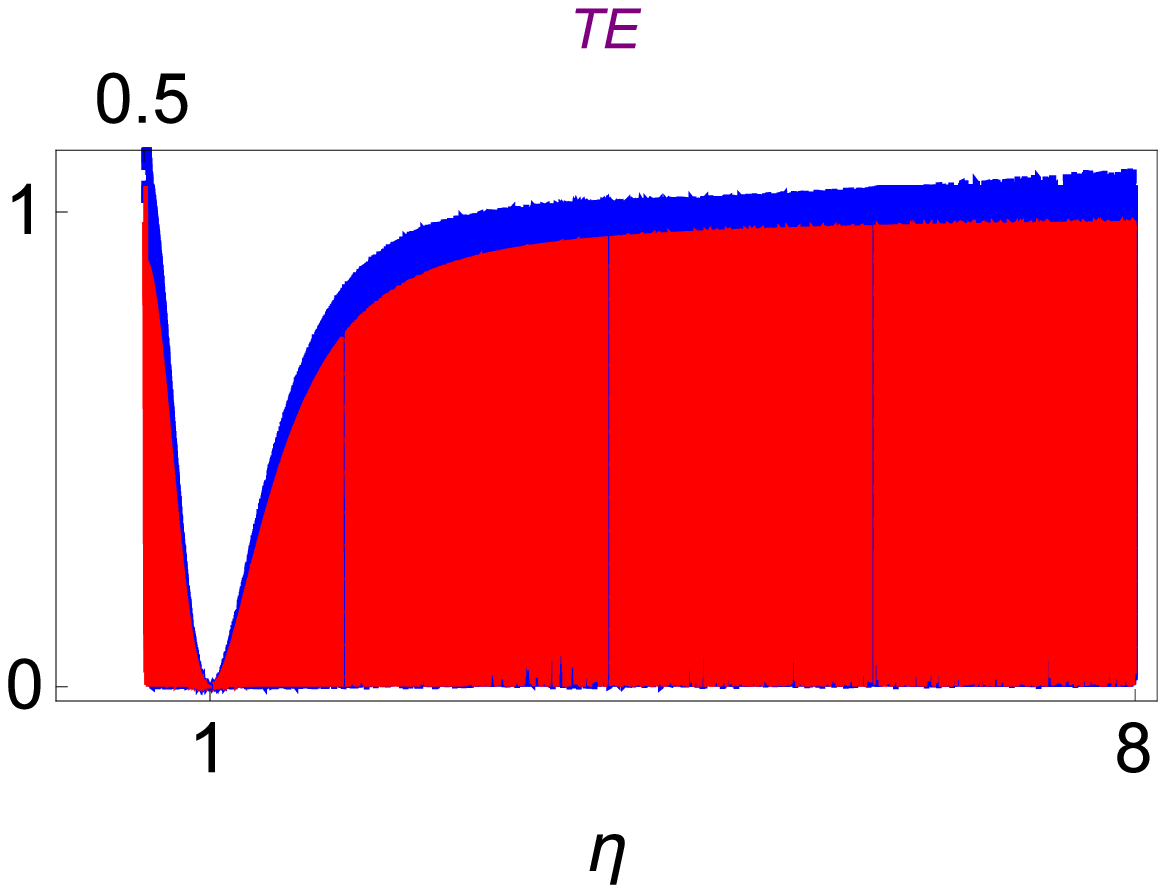}~~~
    \includegraphics[scale=0.4]{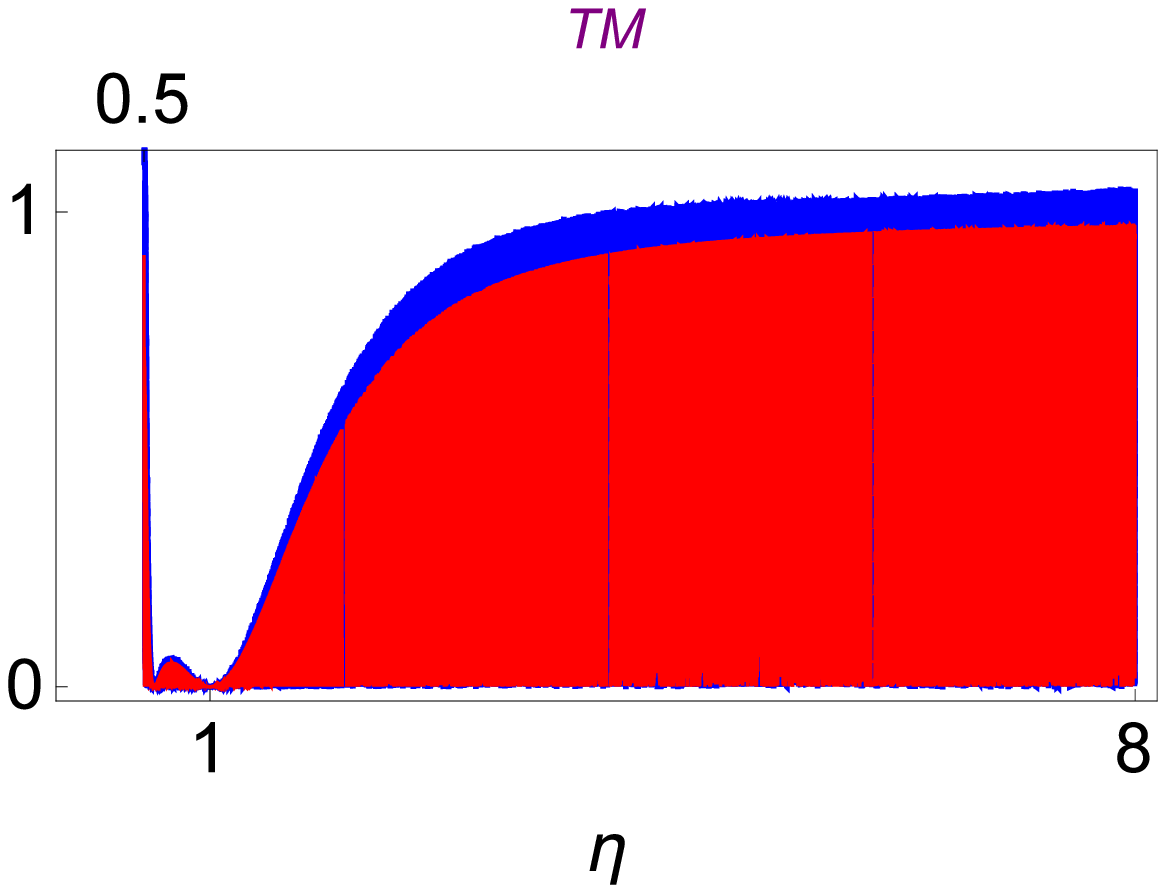}\\
    \includegraphics[scale=0.4]{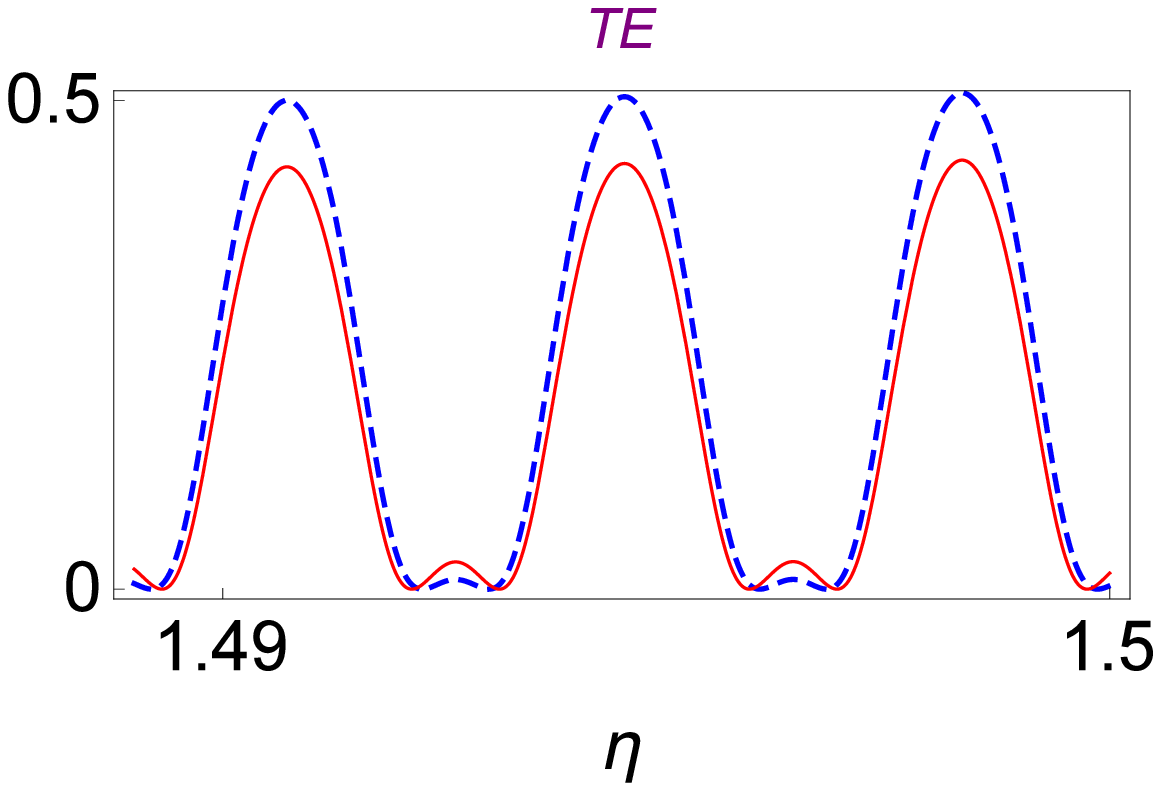}~~~
    \includegraphics[scale=0.4]{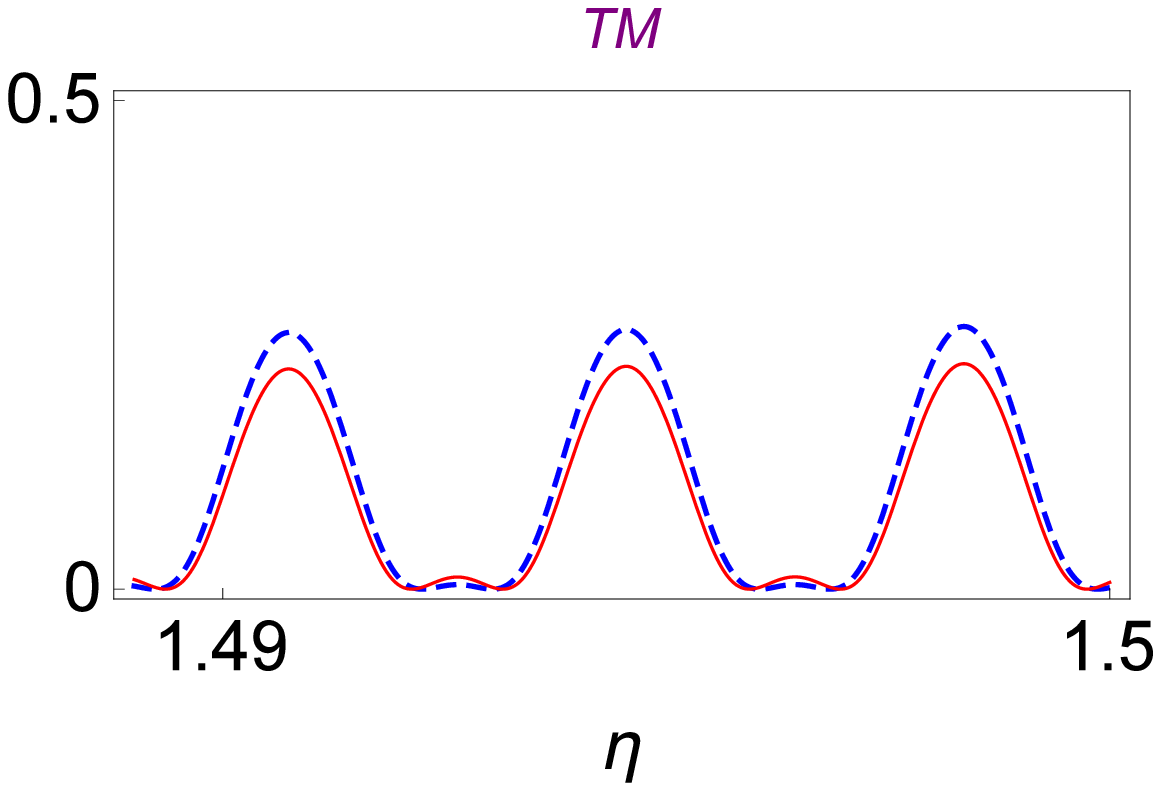}\\
    \includegraphics[scale=0.4]{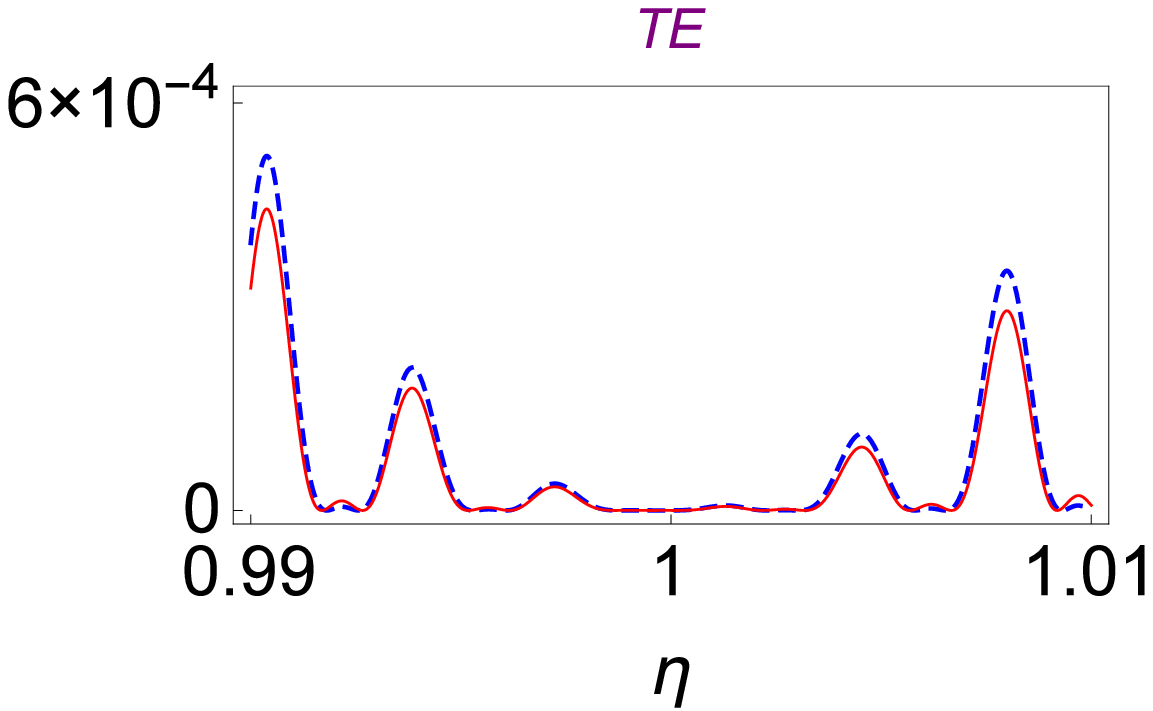}~~~
    \includegraphics[scale=0.4]{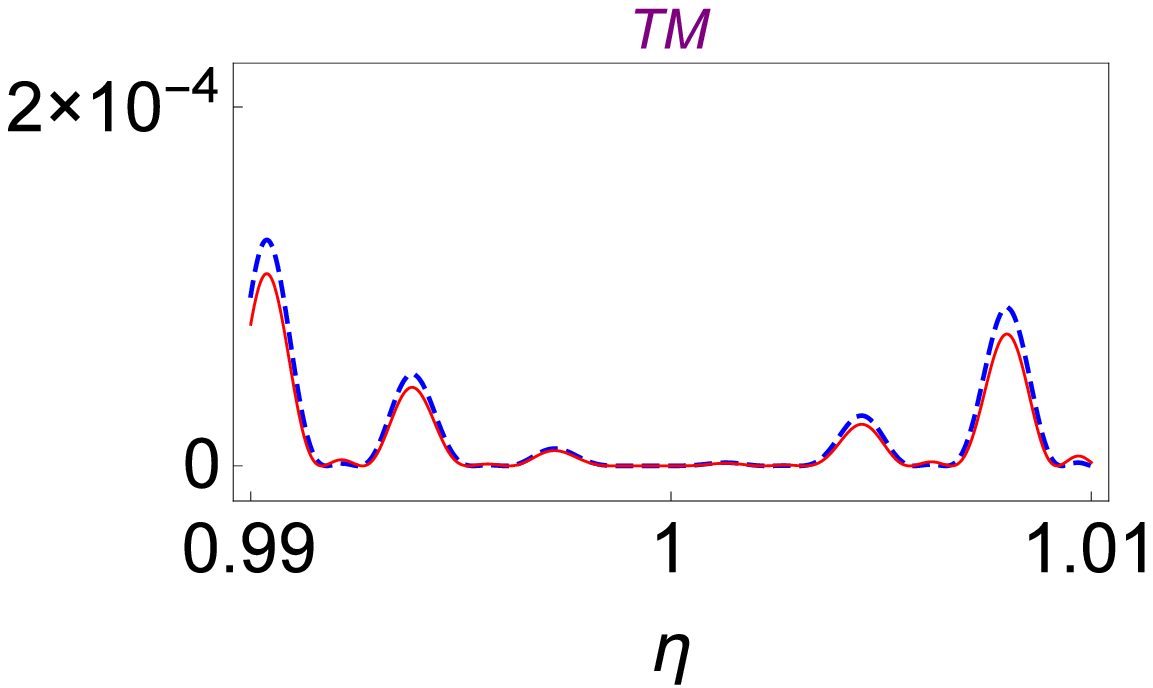}\\
    \includegraphics[scale=0.4]{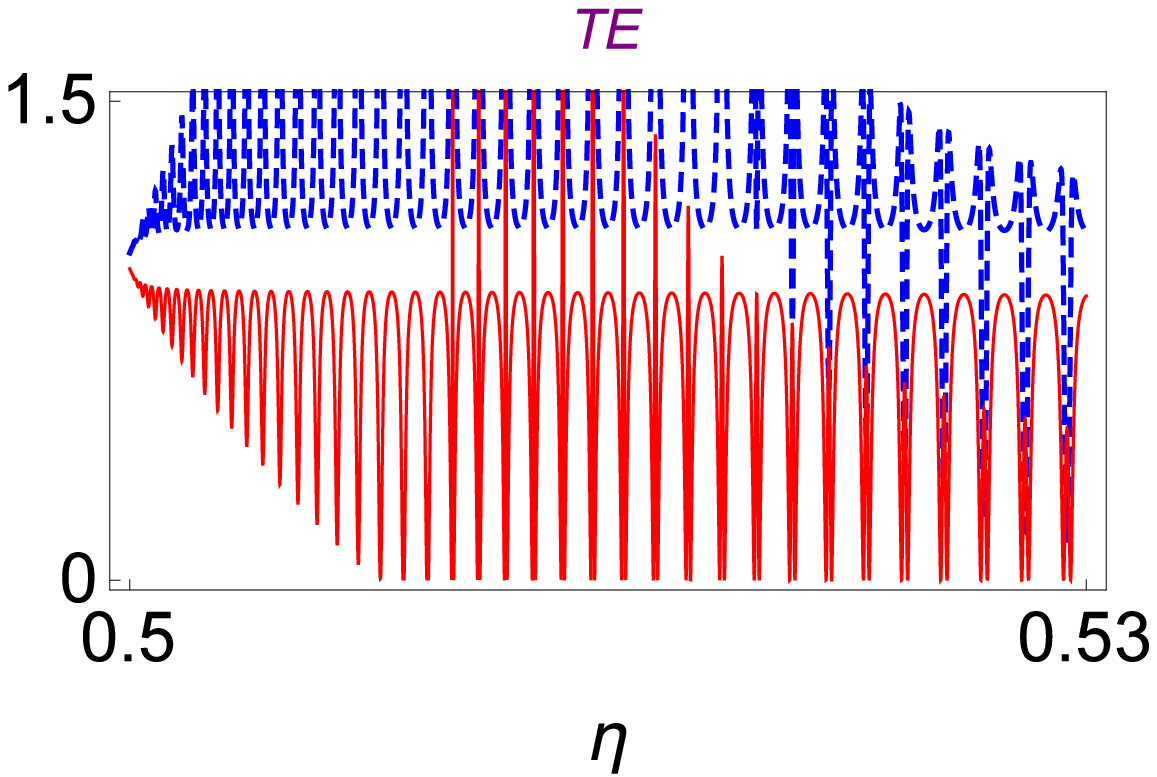}~~~
    \includegraphics[scale=0.4]{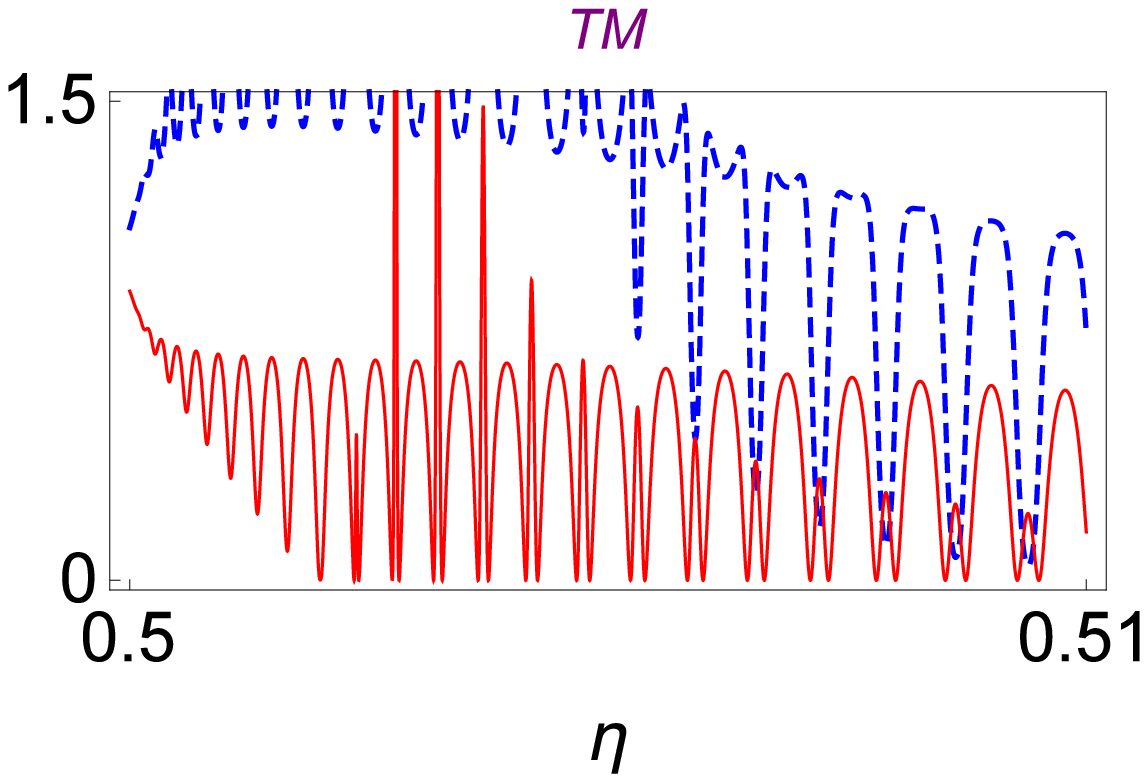}\\
    \includegraphics[scale=0.4]{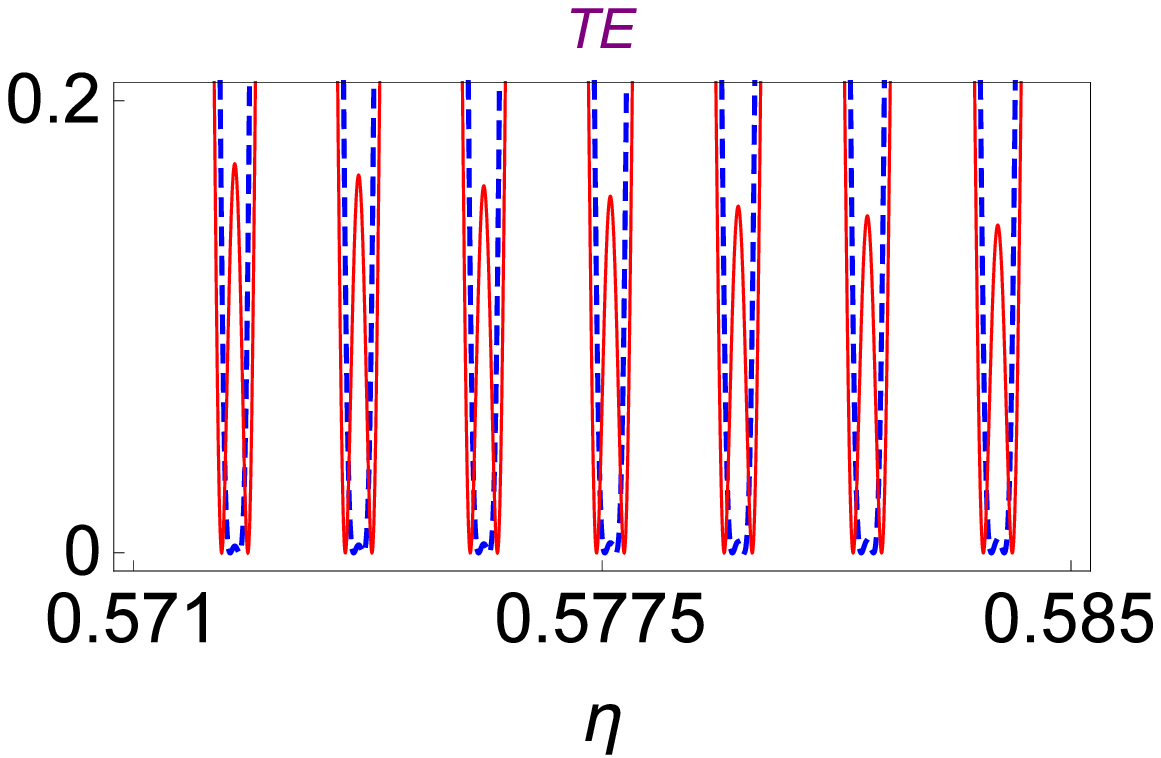}~~~
    \includegraphics[scale=0.4]{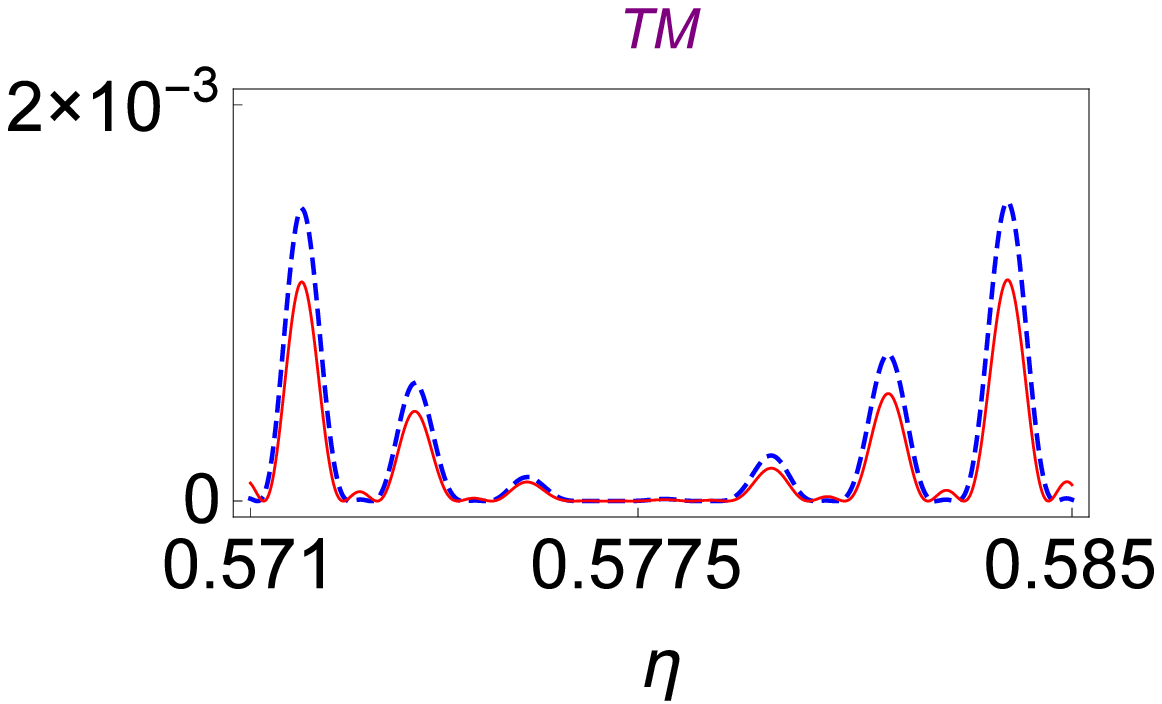}
	\caption{(Color online) Plots of $\left|R^{l}\right|^2$ (thick dashed blue curve) and $\left|R^{r}\right|^2$ (thin solid red curve) as a function of refractive index $\eta$ corresponding to TE and TM wave solutions for the case of $\cP\cT$-symmetric layers with a gap. }
    \label{reflectionlessnessTEM4}
    \end{center}
    \end{figure}

In Fig.~\ref{reflectionlessnessTEM4}, we use different materials by changing the refractive index $\eta$ to see how $\left|R^{l}\right|^2$ and $\left|R^{r}\right|^2$ are affected. For clarity, we use $\cP\cT$-symmetric gain-loss system with slab thickness $L = 100~\mu\textrm{m}$ and gain-loss separation distance $s = 4.782~\mu\textrm{m}$, and wave is sent out at the angle $\theta = 30^{\circ}$ and $\lambda = 807.9997~\textrm{nm}$ with the gain value $g = 9~\textrm{cm}^{-1}$. We observe that no natural material with refractive indice $\eta < 0.88$ can be found to yield a reflectionless situation. Again refractive indices form a discrete values and although there could be found good reflectionless situations at numerous refractive index values, the best option is to use materials with refractive indices nearby $\eta \approx 1$ in TE case, and $\eta \approx 0.578$ and $\eta \approx 1$ in TM case.

\section{Unidirectional Invisibility}

\begin{figure}
	\begin{center}
    \includegraphics[scale=0.5]{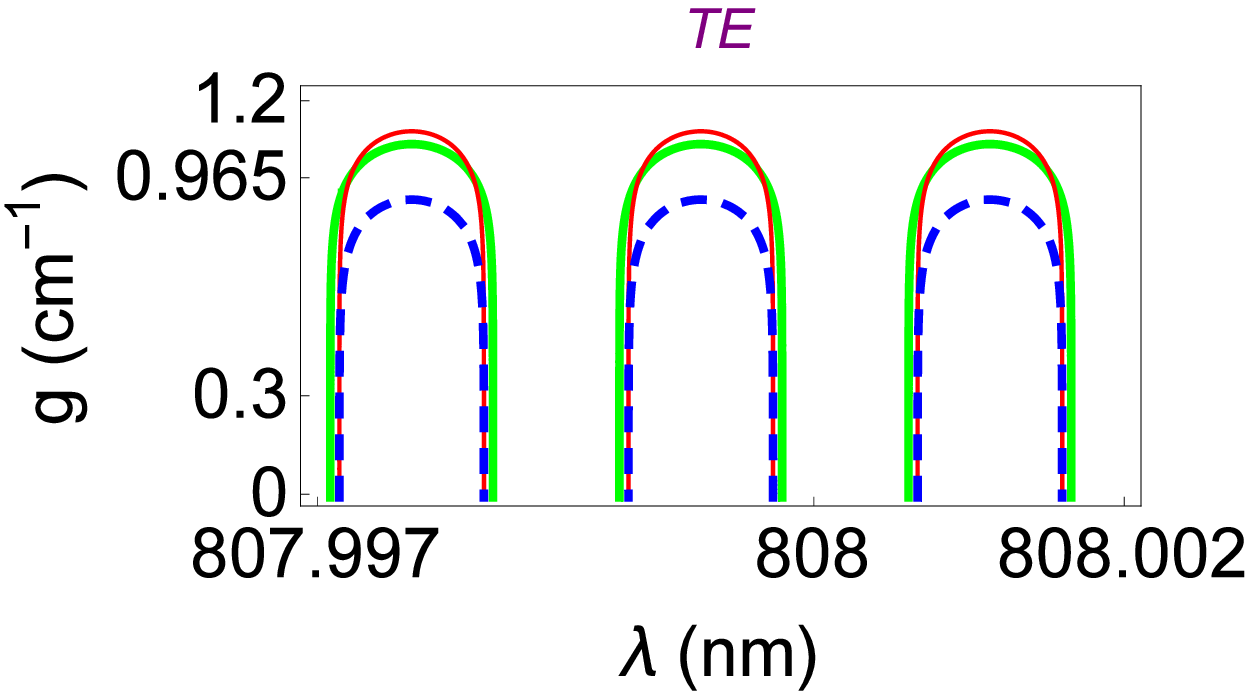}~~~
    \includegraphics[scale=0.5]{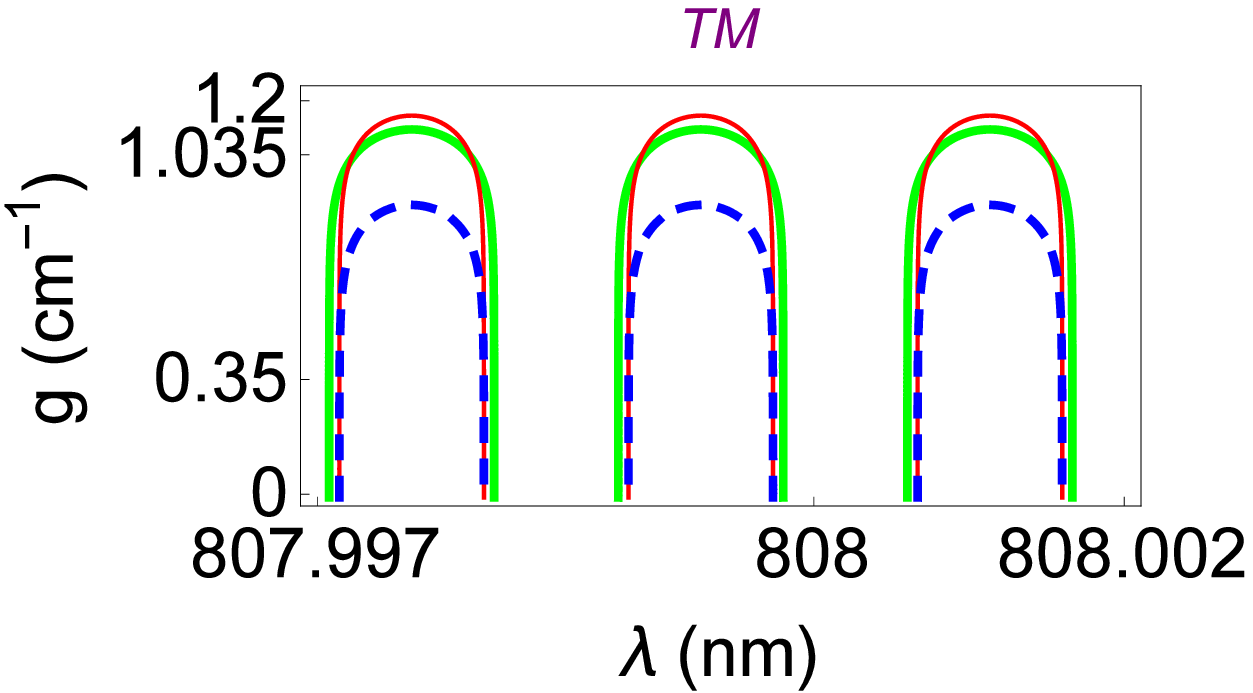}\\
    \includegraphics[scale=0.5]{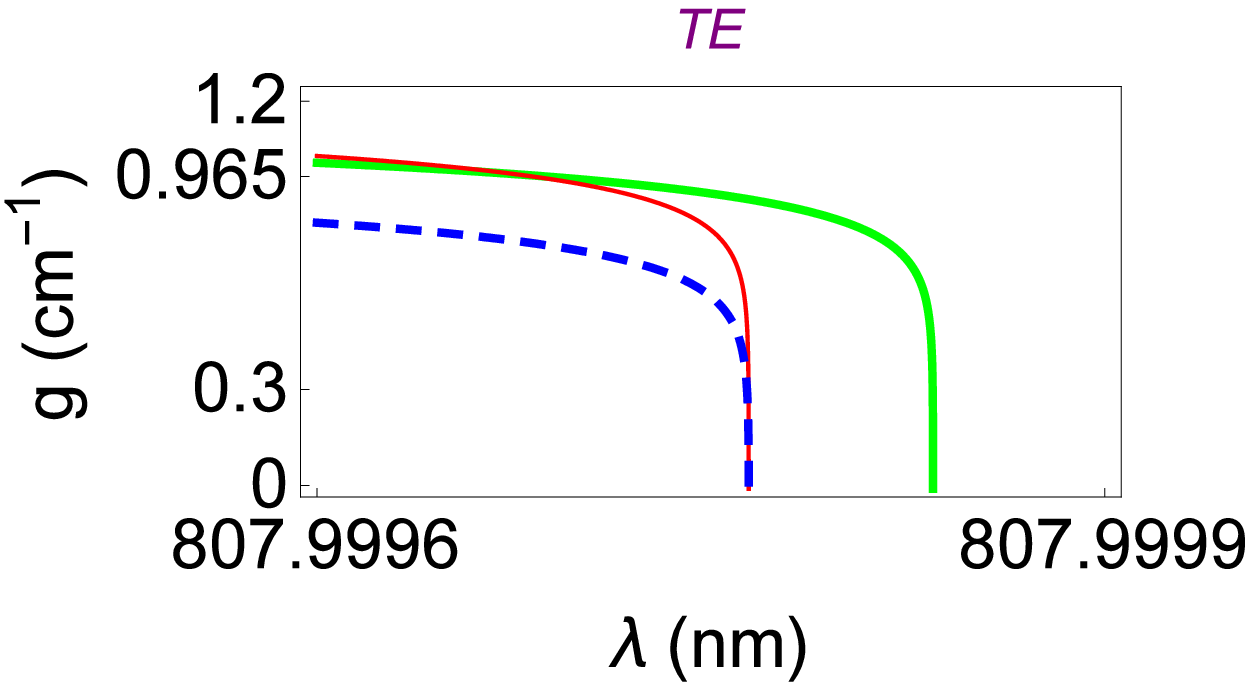}~~~
    \includegraphics[scale=0.5]{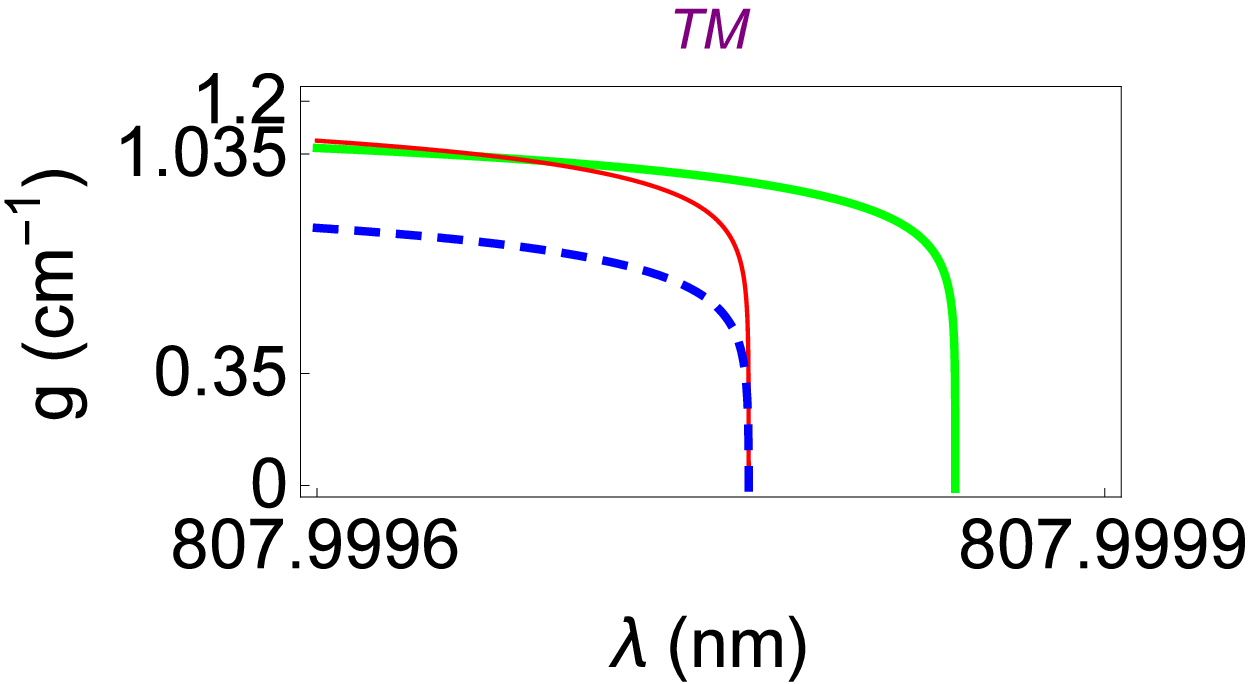}
	\caption{(Color online) Plots of gain coefficient $g$ as a function of wavelength $\lambda$ corresponding to Invisible TE and TM wave solutions of the constructive configuration for the case of $\cP\cT$-symmetric layers with a gap at angle of incidence $\theta = 30^{\circ}$. In these plots, thin solid red curves represent the right reflectionless configurations, dashed blue curves the left reflectionless one, and thick green solid curves the conditions for $M_{11} = M_{22} = 1$ as given in (\ref{transmission1}) and (\ref{transmission2}).}
    \label{invgain1}
    \end{center}
    \end{figure}

\begin{figure}
	\begin{center}
    \includegraphics[scale=0.5]{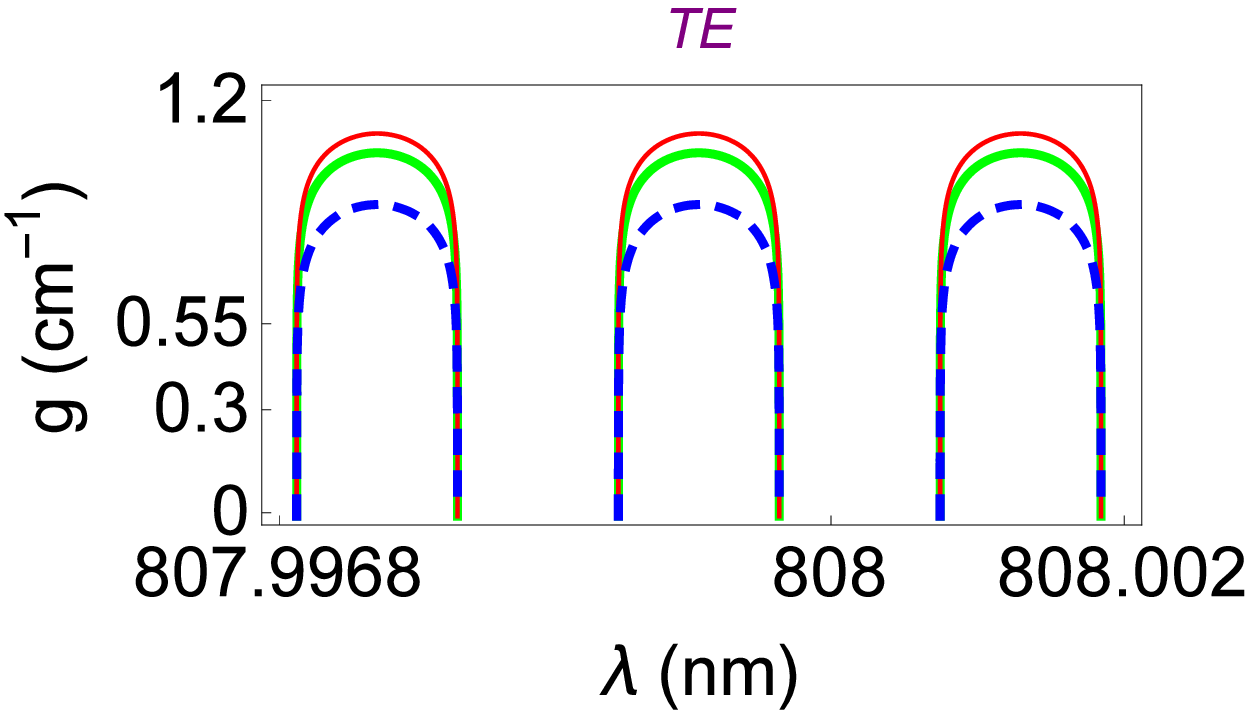}~~~
    \includegraphics[scale=0.5]{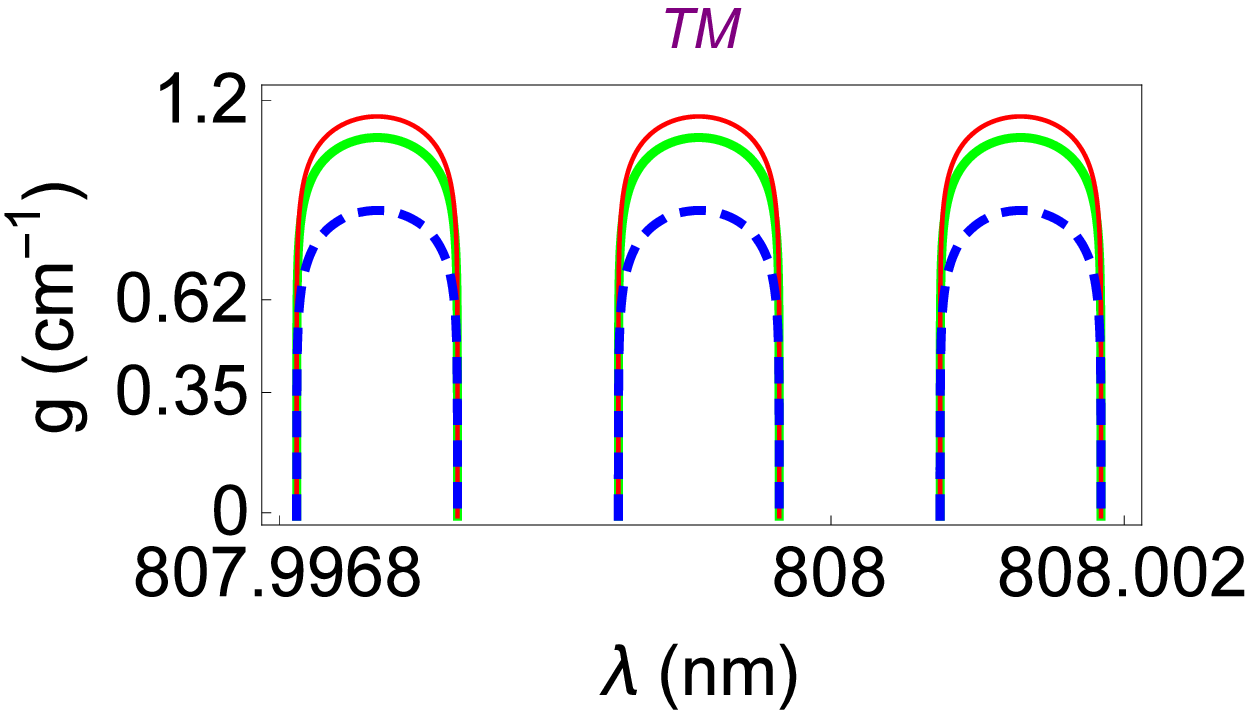}\\
    \includegraphics[scale=0.5]{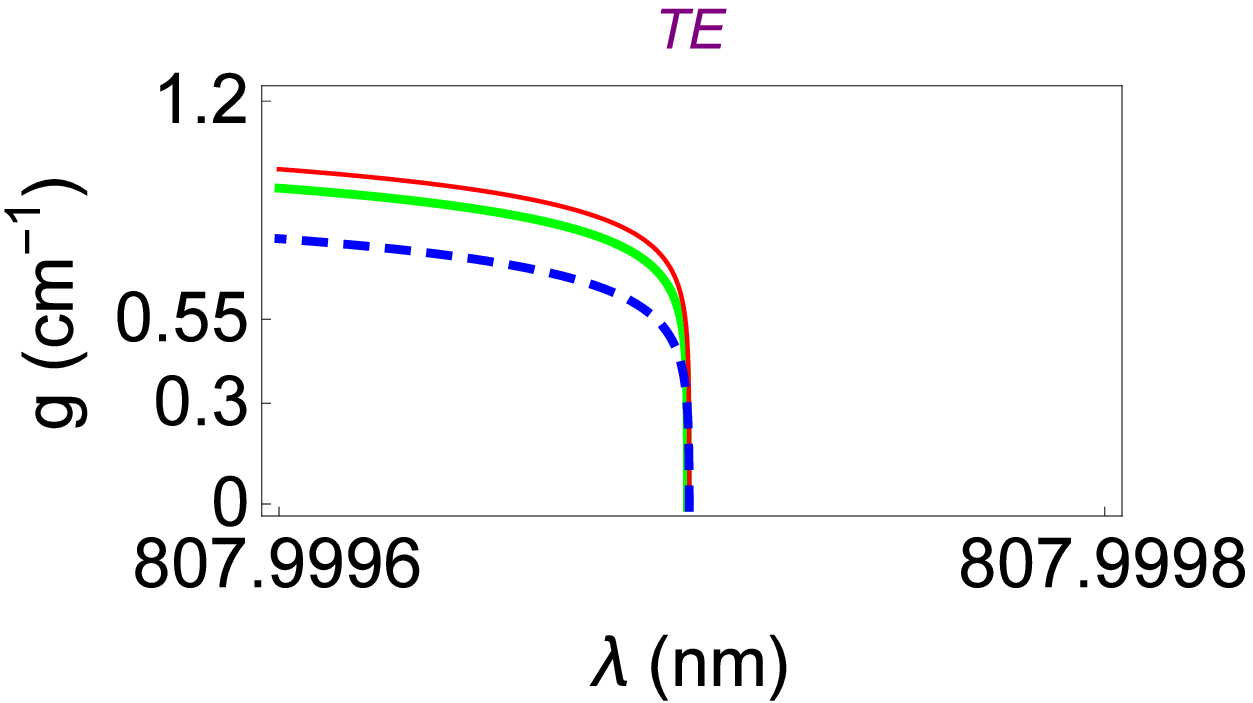}~~~
    \includegraphics[scale=0.5]{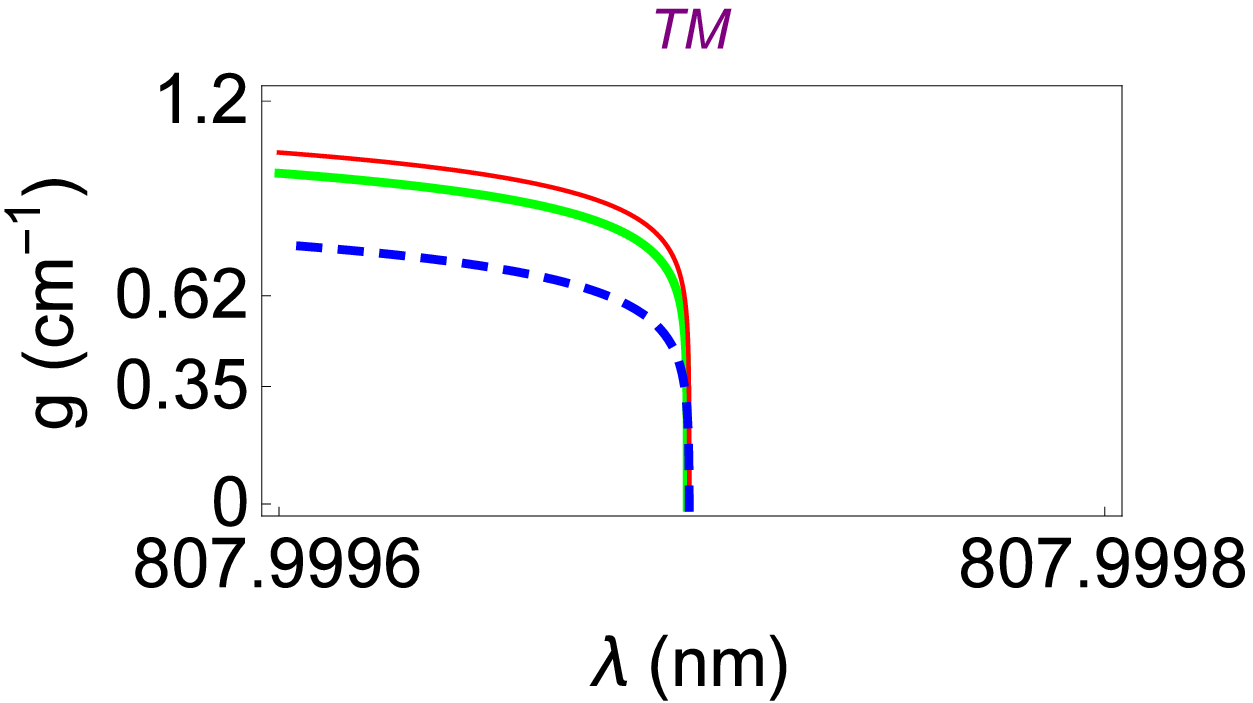}
	\caption{(Color online) Plots of gain coefficient $g$ as a function of wavelength $\lambda$ corresponding to Invisible TE and TM wave solutions of the constructive configuration for the case of $\cP\cT$-symmetric layers with a gap. In these graphs, thin solid red curves represent the right reflectionless configurations, dashed blue curves the left reflectionless one and thick green solid curves the conditions for $M_{11} = M_{22} = 1$ as given in (\ref{transmission1}) and (\ref{transmission2}).}
    \label{invgain2}
    \end{center}
    \end{figure}

\begin{figure}
	\begin{center}
    \includegraphics[scale=0.5]{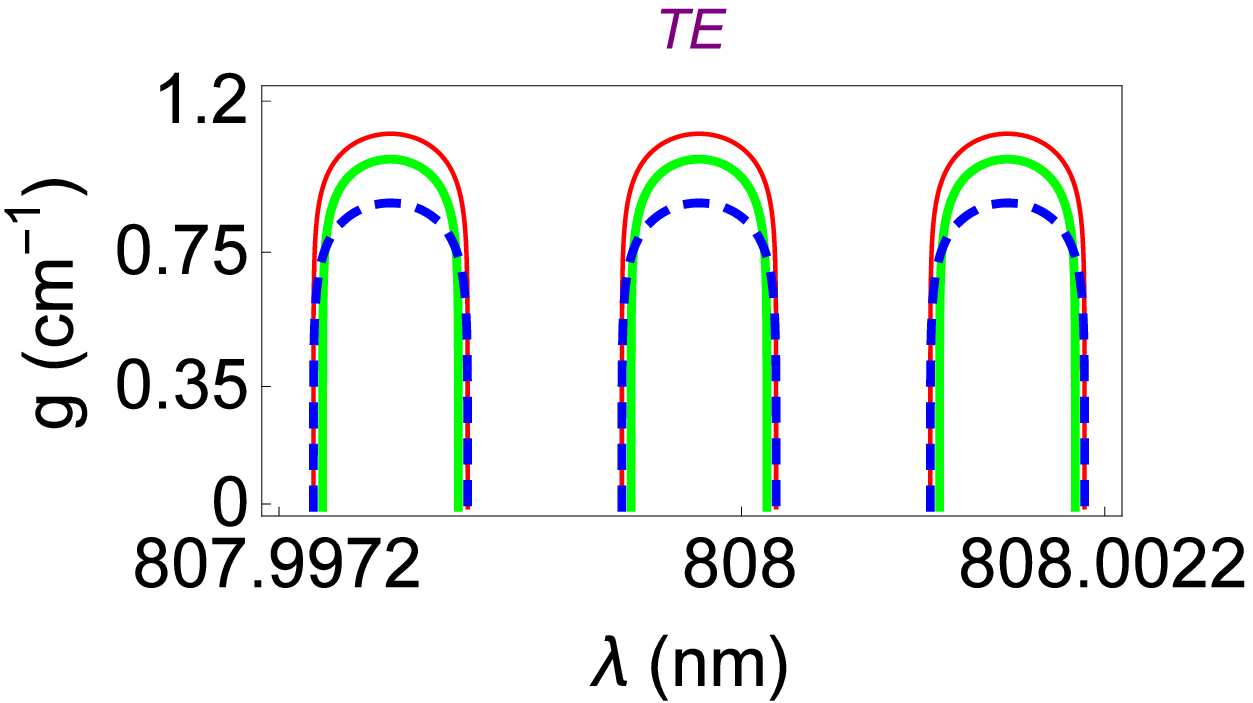}~~~
    \includegraphics[scale=0.5]{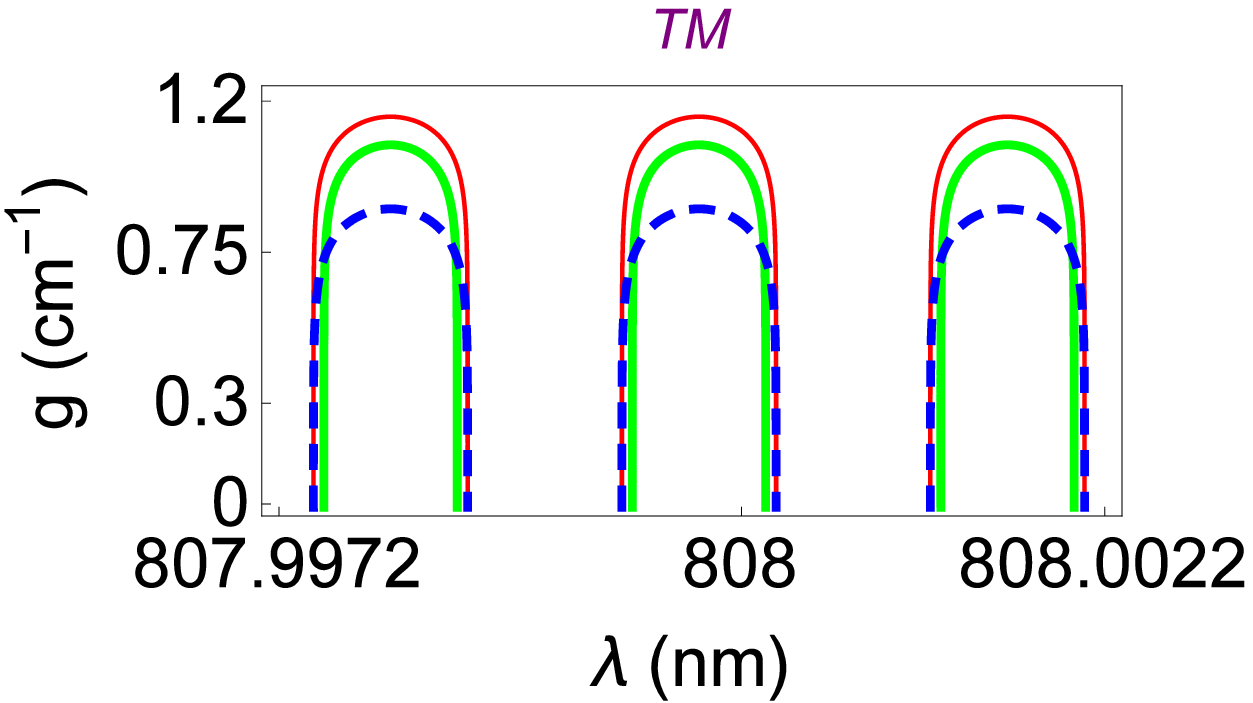}\\
    \includegraphics[scale=0.5]{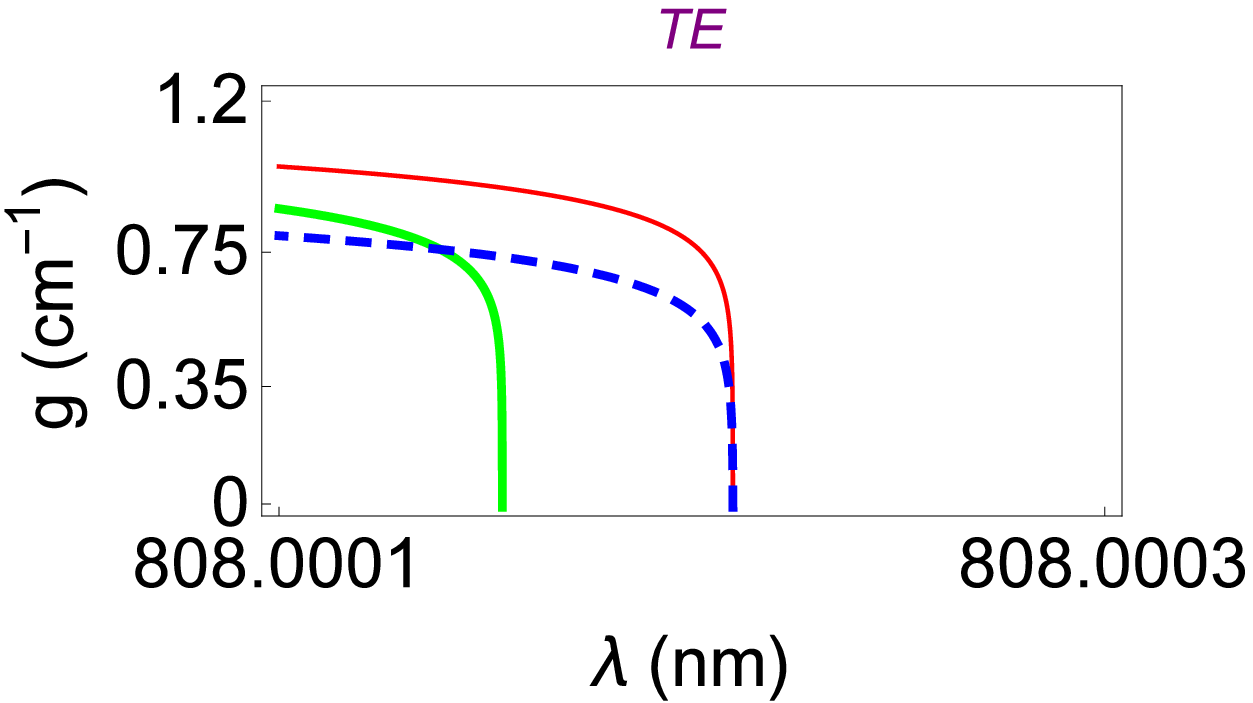}~~~
    \includegraphics[scale=0.5]{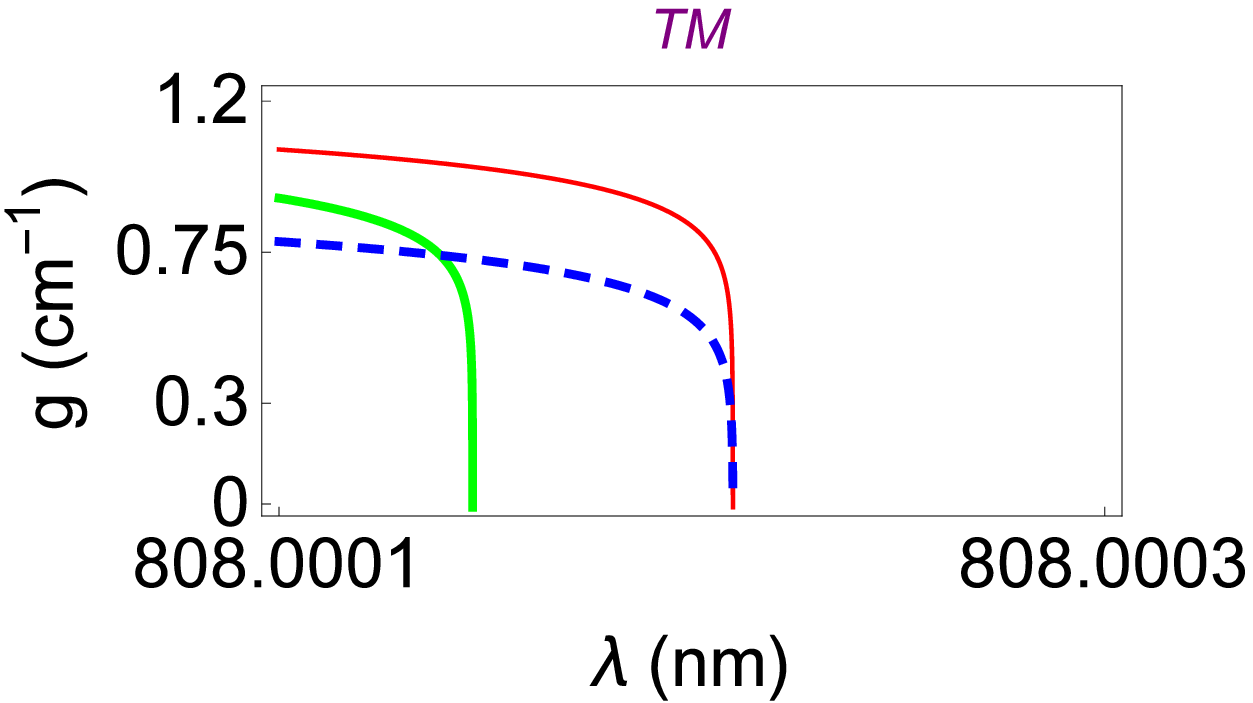}
	\caption{(Color online) Plots of gain coefficient $g$ as a function of wavelength $\lambda$ corresponding to Invisible TE and TM wave solutions for the case of $\cP\cT$-symmetric layers with a gap. In these plots, thin solid red curves represent the right reflectionless configurations, dashed blue curves the left reflectionless configurations and thick green solid curves the conditions for $M_{11} = M_{22} = 1$.}
    \label{invgain3}
    \end{center}
    \end{figure}

\begin{figure}
	\begin{center}
    \includegraphics[scale=0.5]{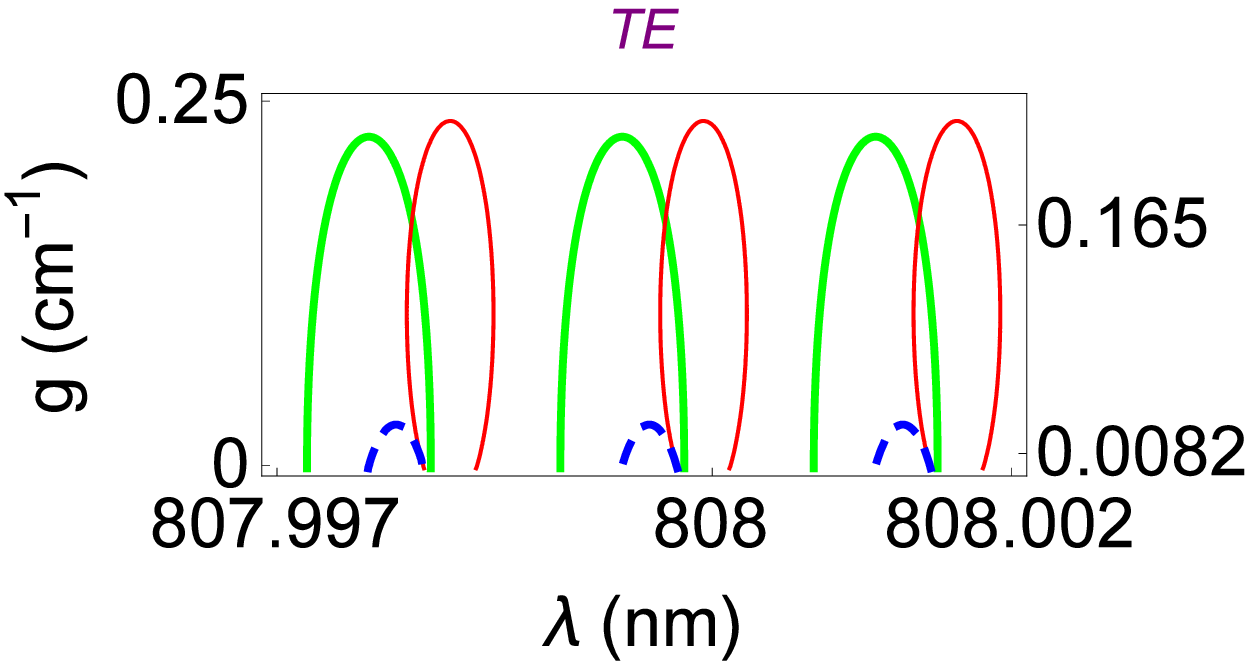}~~~
    \includegraphics[scale=0.5]{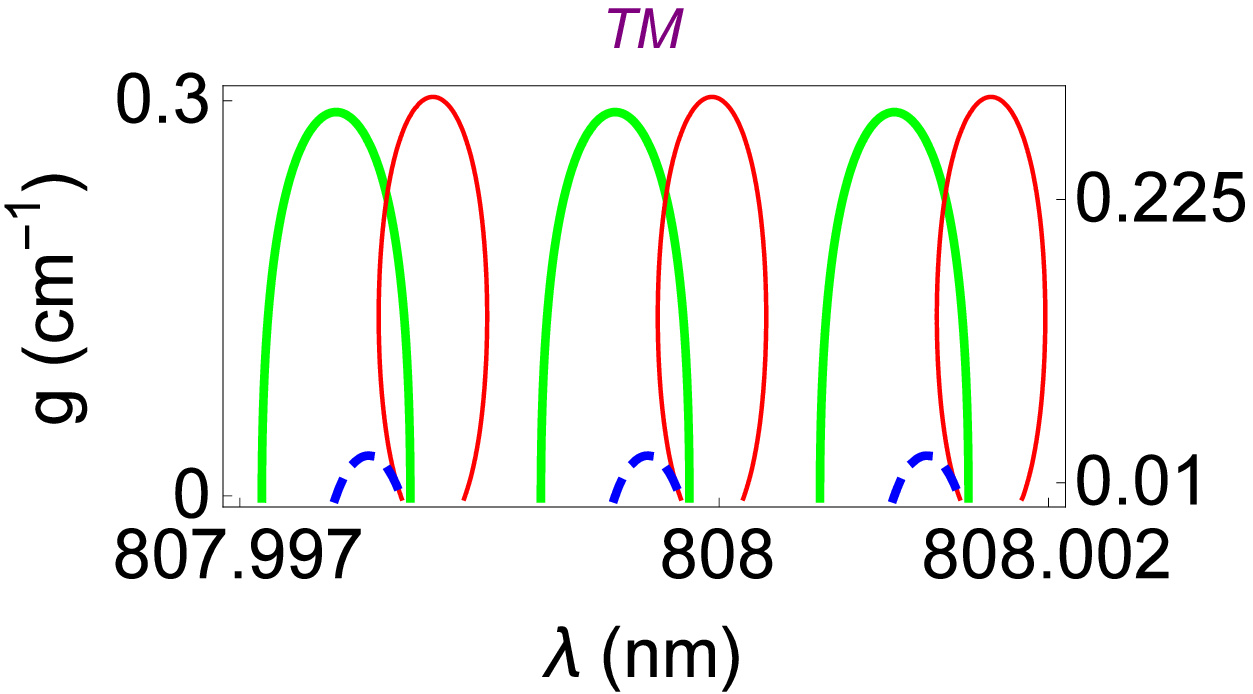}\\
    \includegraphics[scale=0.5]{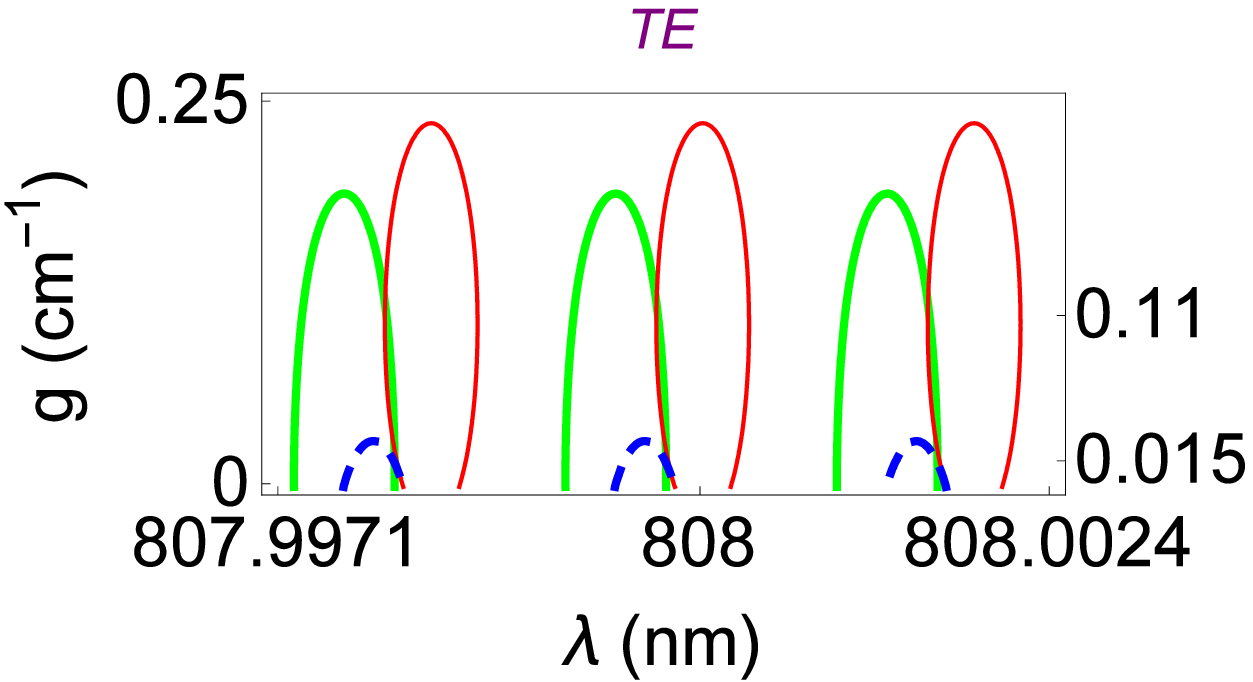}~~~
    \includegraphics[scale=0.5]{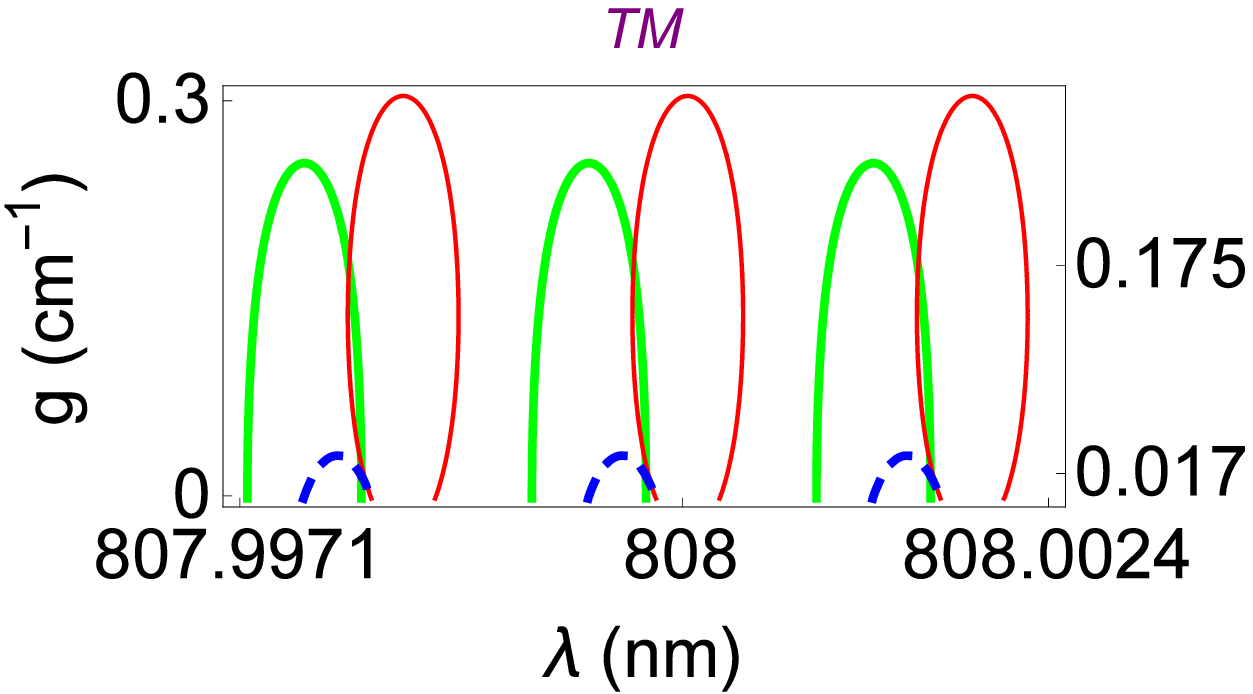}
	\caption{(Color online)  Plots of gain coefficient $g$ as a function of wavelength $\lambda$ corresponding to Invisible TE and TM wave solutions of the arbitrary configuration for the case of $\cP\cT$-symmetric layers with a gap ($s/s_0 = 20.5$). In these plots, thin solid red curves represent the right reflectionless configurations, dashed blue curves the left reflectionless ones and thick green solid curves the conditions for $M_{11} = M_{22} = 1$ as given in (\ref{transmission1}) and (\ref{transmission2}).  }
    \label{invgain4}
    \end{center}
    \end{figure}

\begin{figure}
	\begin{center}
    \includegraphics[scale=0.5]{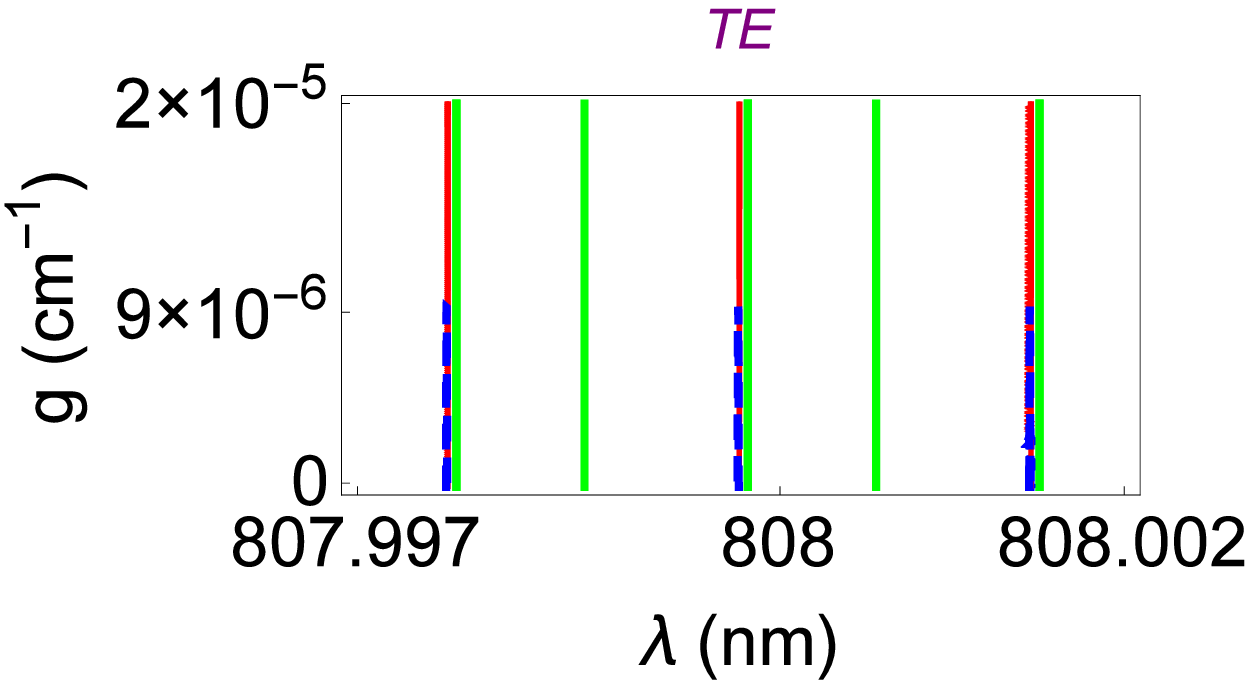}~~~
    \includegraphics[scale=0.5]{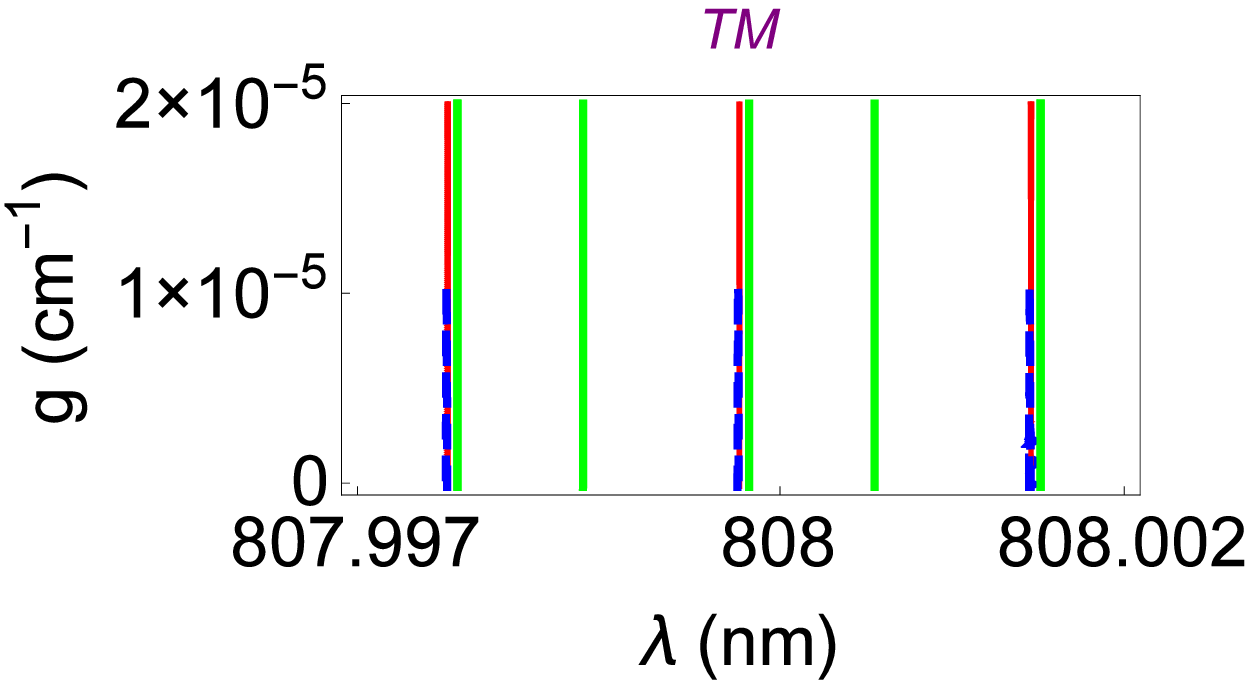}\\
    \includegraphics[scale=0.5]{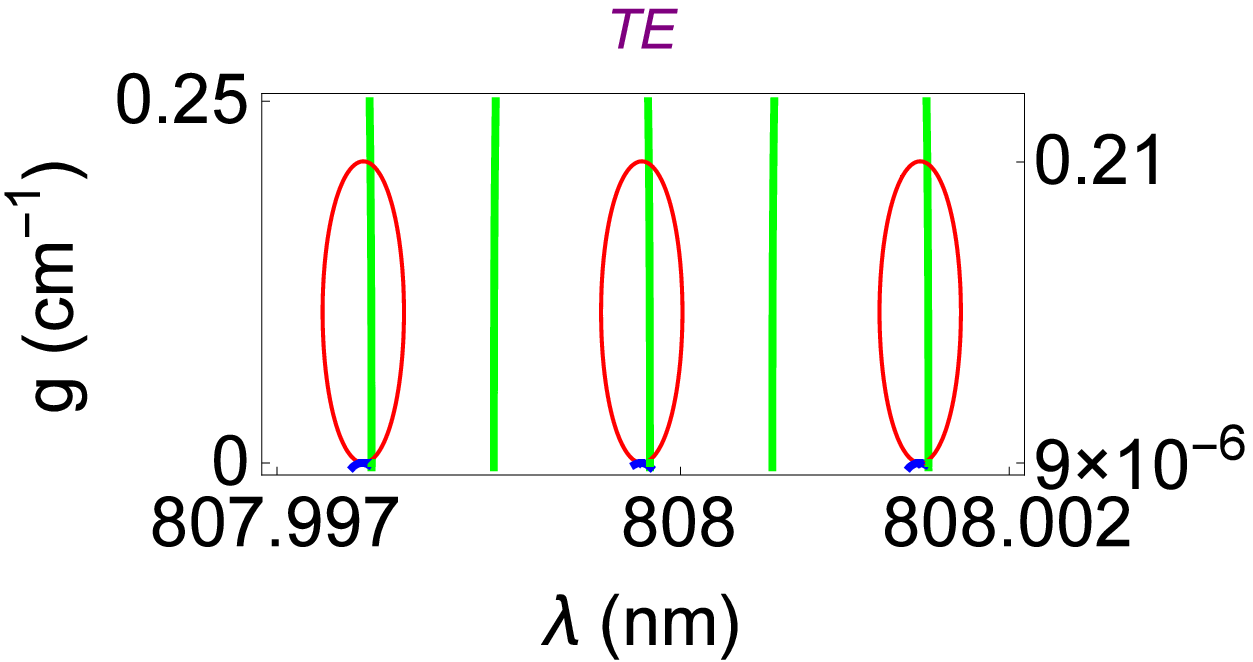}~~~
    \includegraphics[scale=0.5]{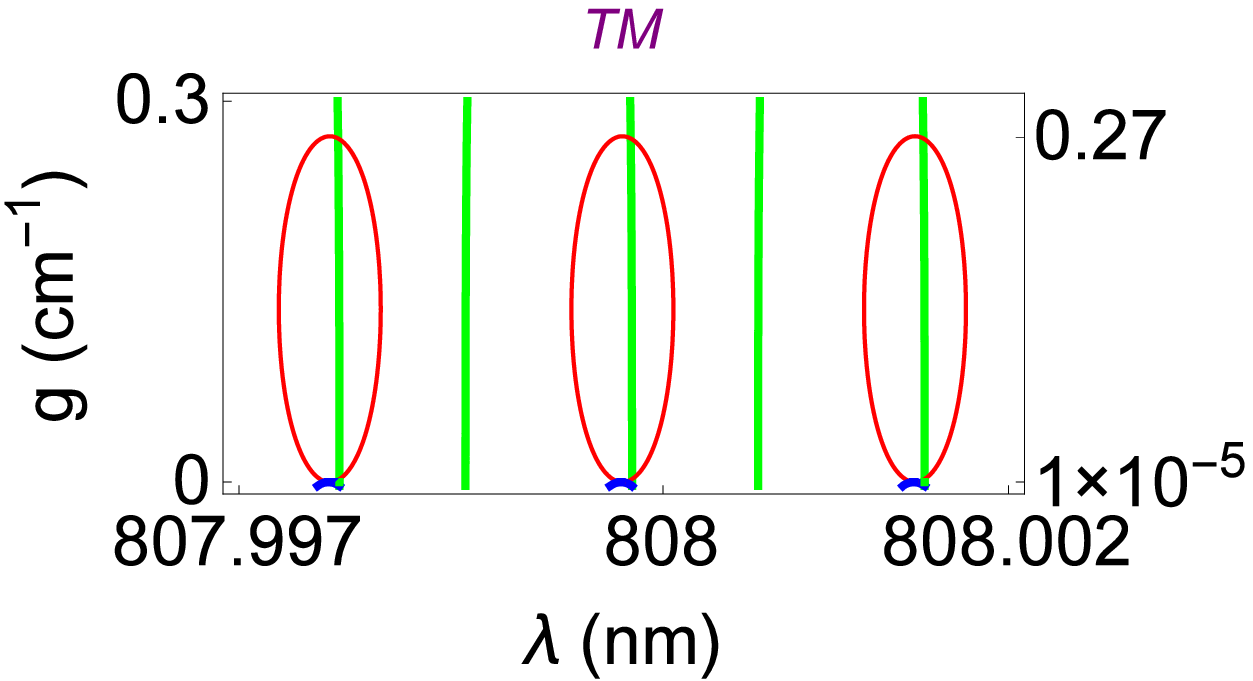}
	\caption{(Color online)  Plots of gain coefficient $g$ as a function of wavelength $\lambda$ corresponding to Invisible TE and TM wave solutions of the destructive configuration for the case of $\cP\cT$-symmetric layers with a gap. In these plots, thin solid red curves represent the right reflectionless configurations, dashed blue curves the left reflectionless ones and thick green solid curves the conditions for $M_{11} = M_{22} = 1$ in (\ref{transmission1}) and (\ref{transmission2}). }
    \label{invgain5}
    \end{center}
    \end{figure}

\begin{figure}
	\begin{center}
    \includegraphics[scale=0.5]{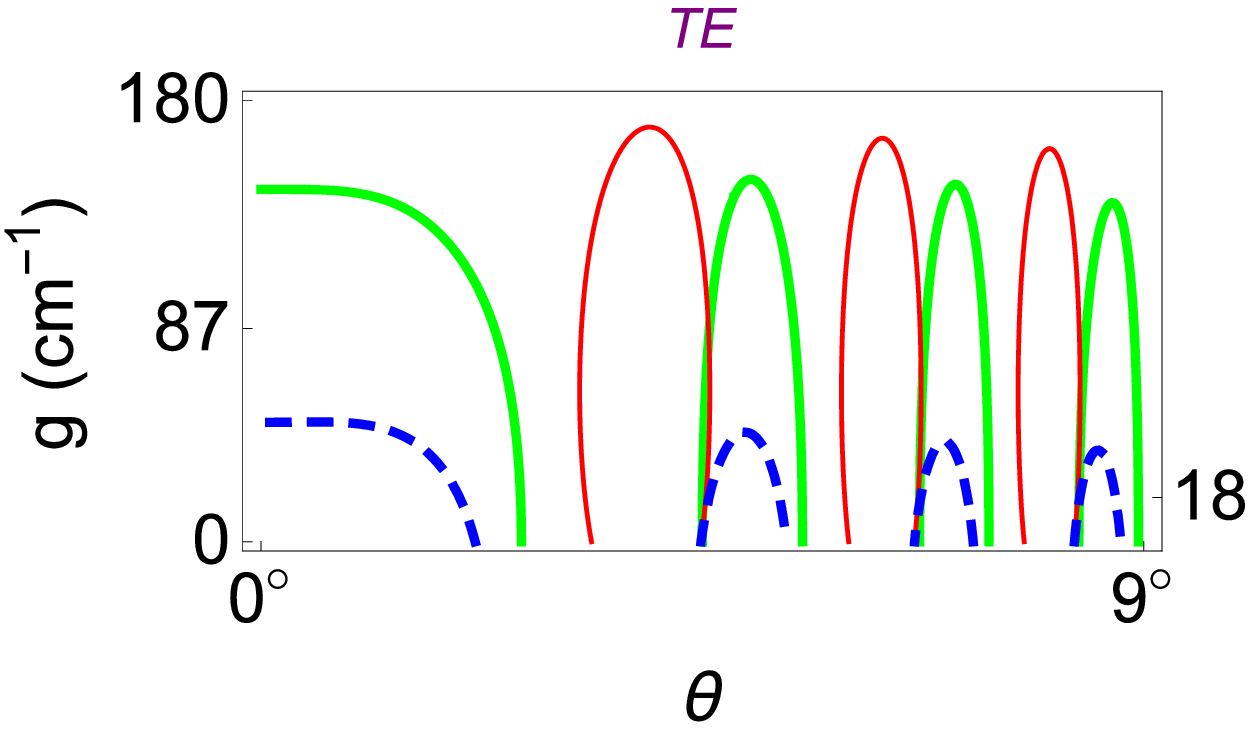}~~~
    \includegraphics[scale=0.5]{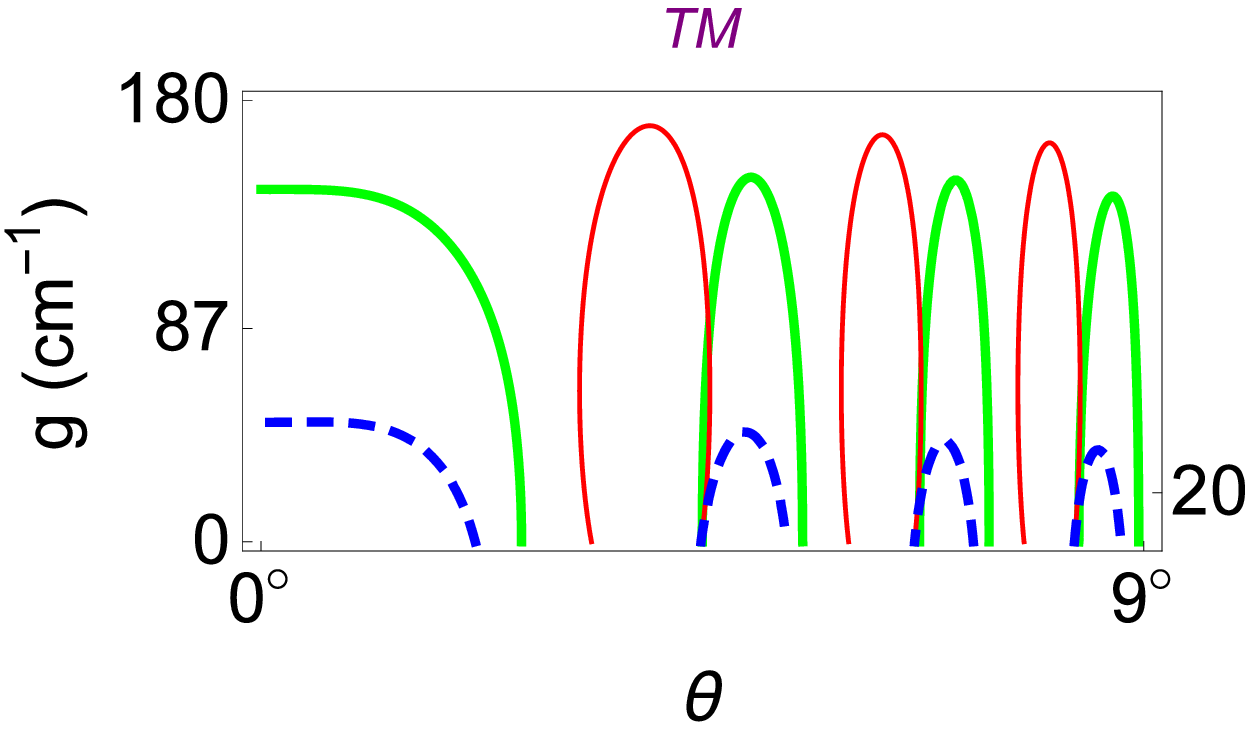}\\
    \includegraphics[scale=0.5]{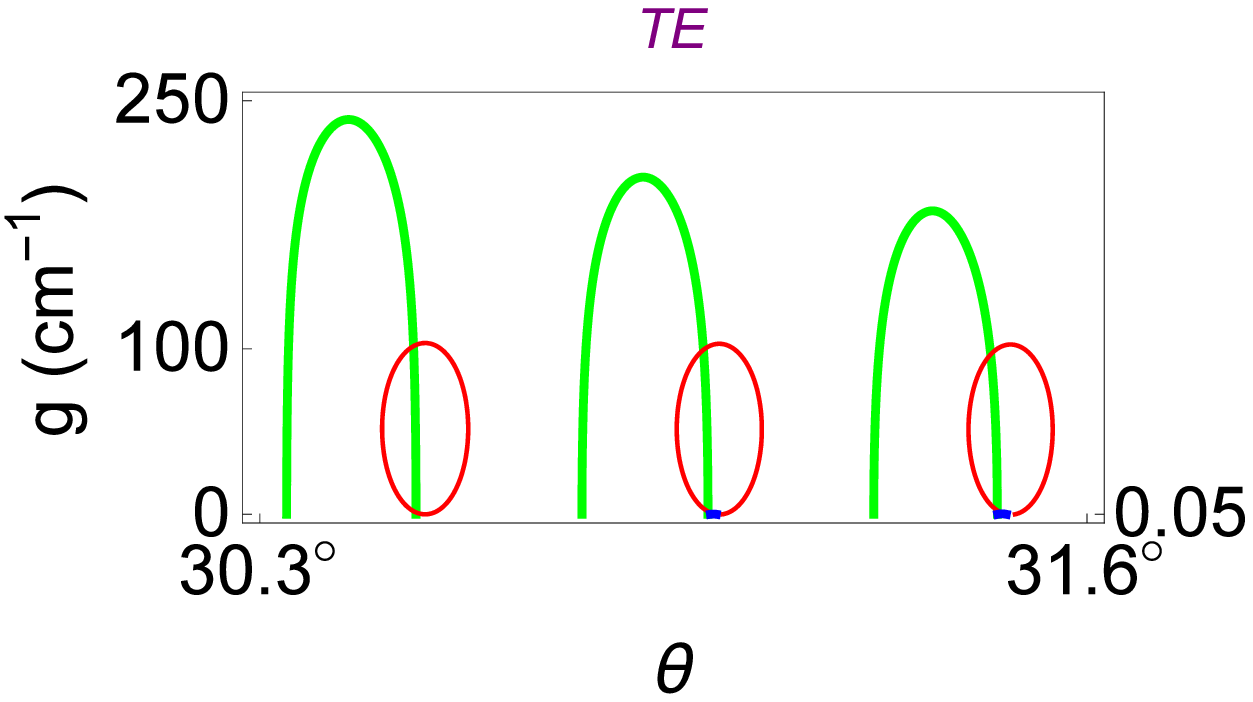}~~~
    \includegraphics[scale=0.5]{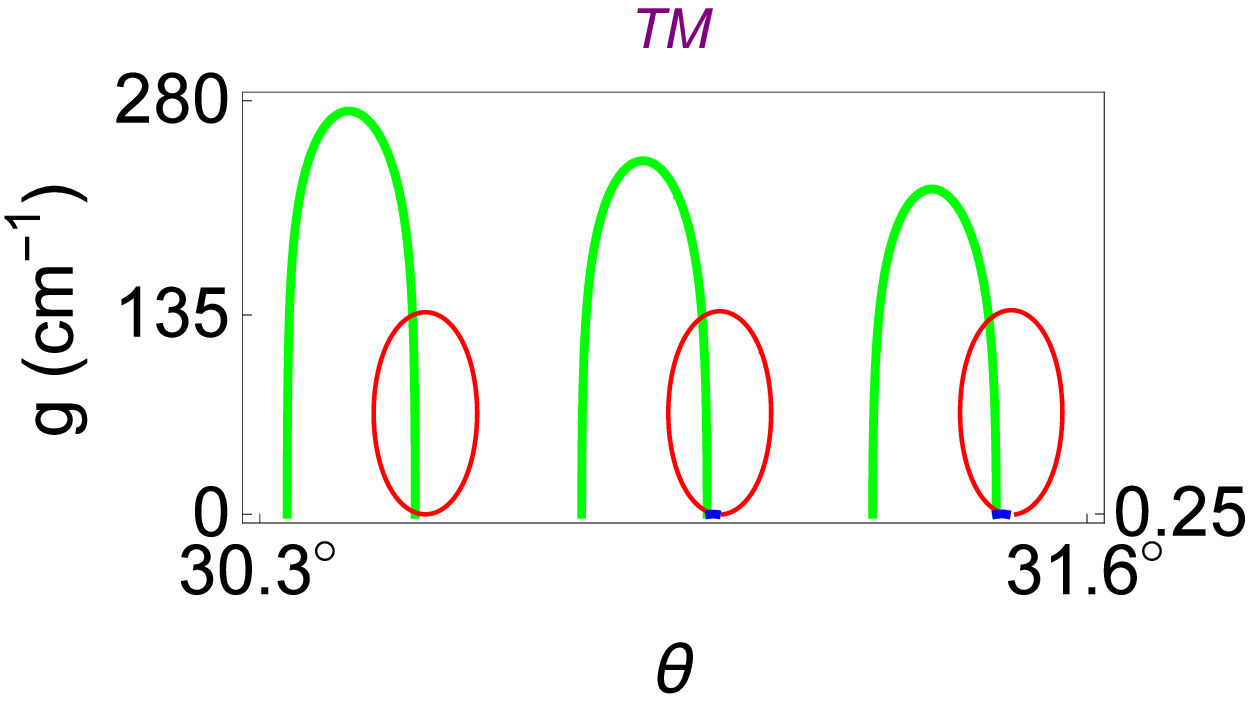}\\
    \includegraphics[scale=0.5]{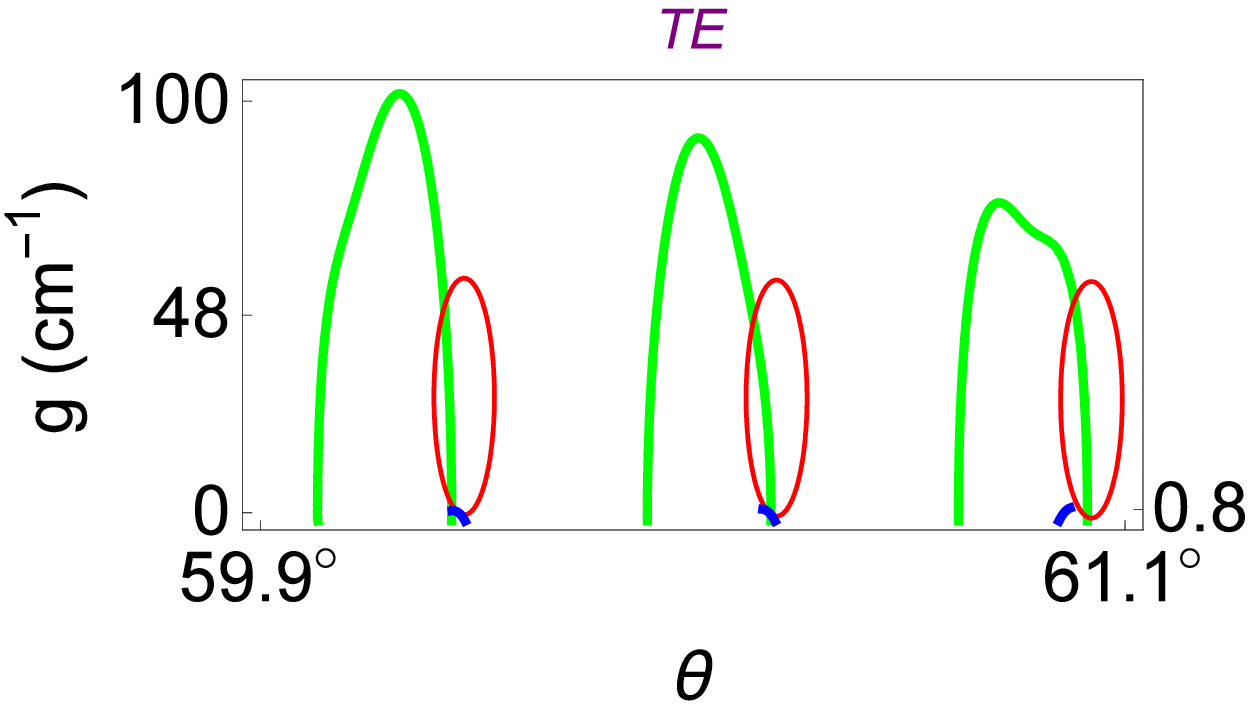}~~~
    \includegraphics[scale=0.5]{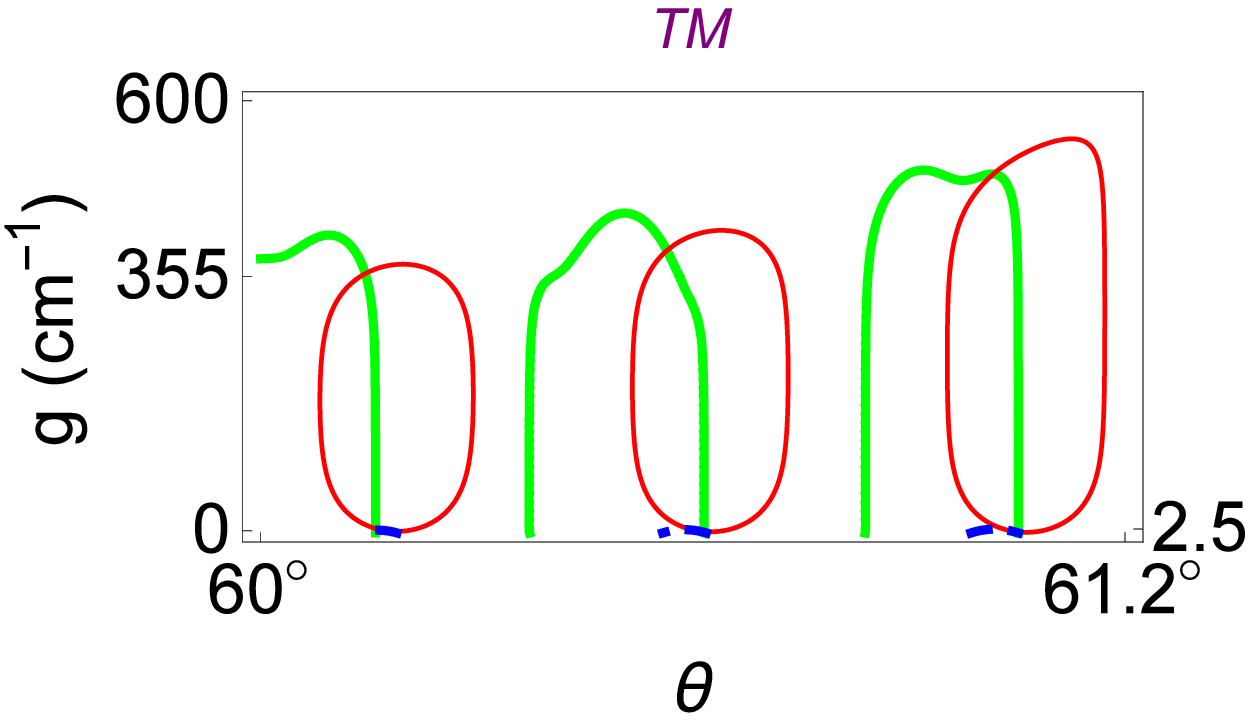}\\
    \includegraphics[scale=0.5]{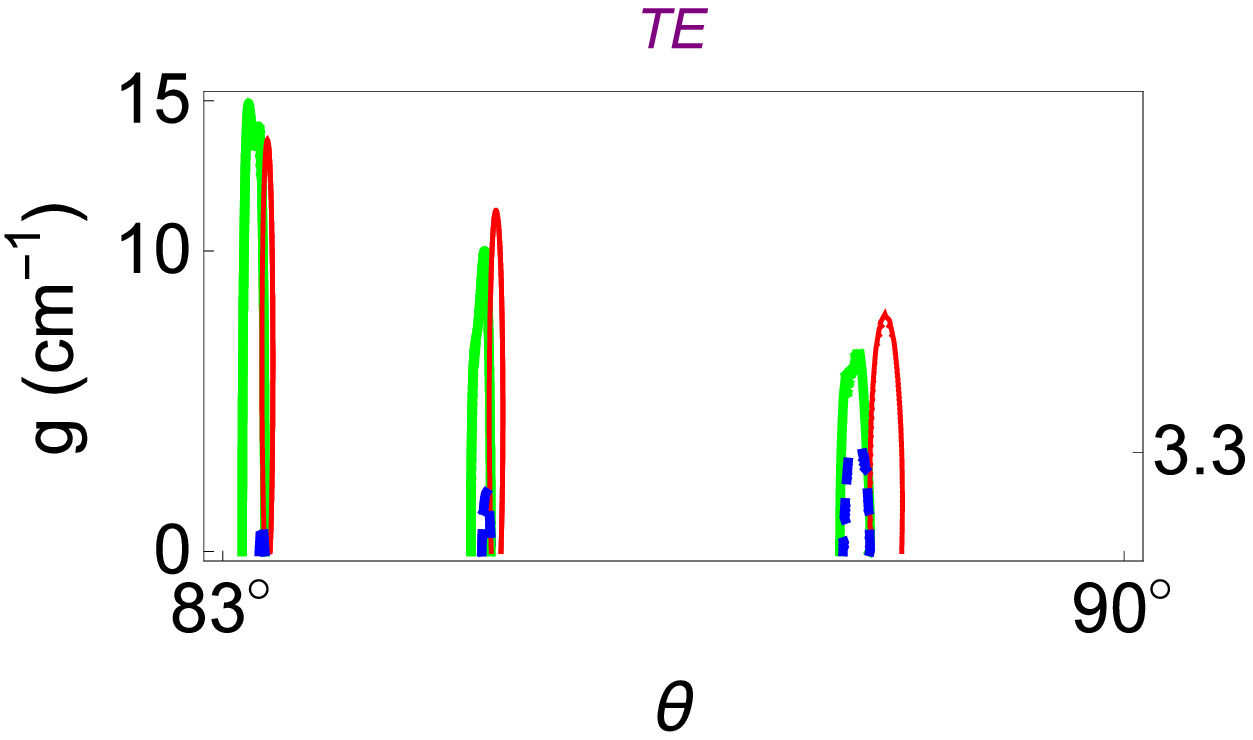}~~~
    \includegraphics[scale=0.5]{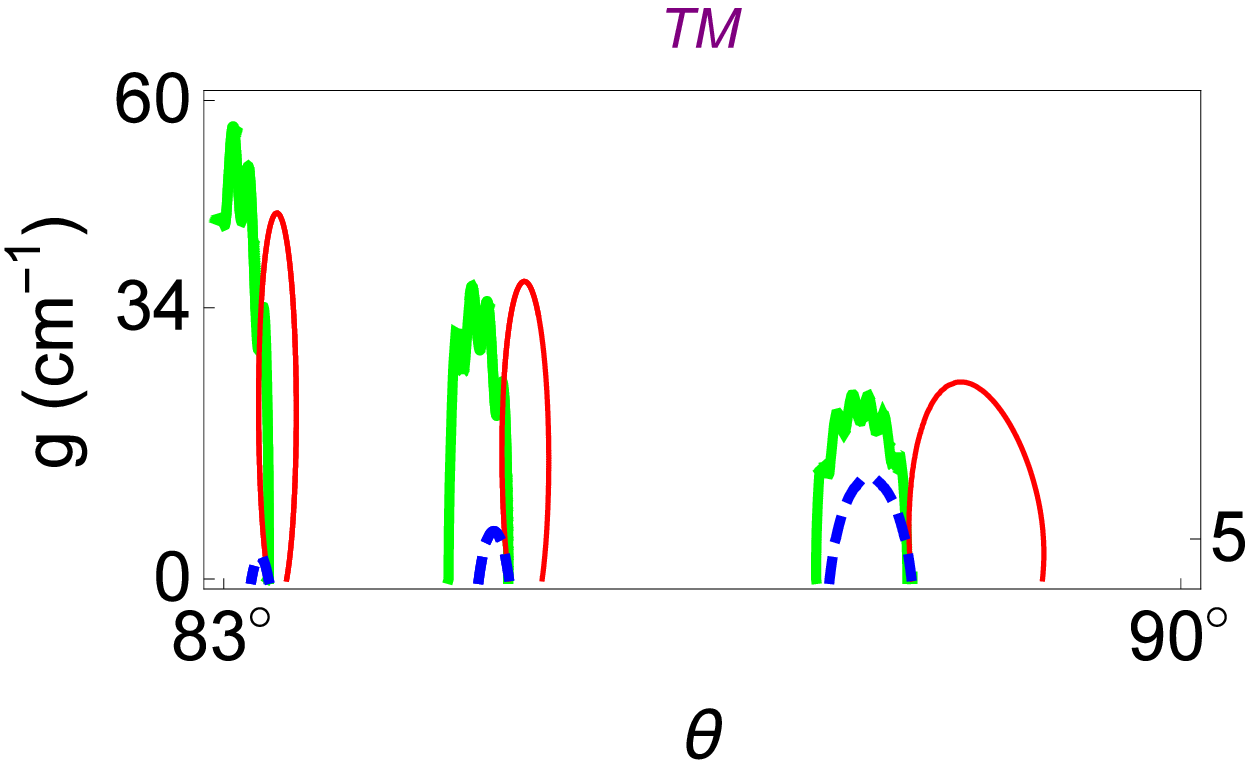}
	\caption{(Color online) Plots of gain coefficient $g$ as a function of incidence angle $\theta$ corresponding to invisible TE and TM wave solutions for the case of $\cP\cT$-symmetric layers with a gap of $s \approx 1.167~\mu\textrm{m}$. In these plots, thin solid red curves represent the right reflectionless configurations, dashed blue curves the left reflectionless ones and thick green solid curves the conditions for $M_{11} = M_{22} = 1$  in (\ref{transmission1}) and (\ref{transmission2}).}
    \label{invgain6}
    \end{center}
    \end{figure}

It is explicitly revealed that unidirectional reflectionlessness imposes some constraints on the gain coefficient which are restricted to lie within a certain wavelength range in a predetermined angle of incidence. This is in fact an obligation arising from the condition of unidirectional reflectionless potentials. Thus, gain coefficient is limited to take values between some minimum and maximum points. In addition, one needs a further condition, and in turn constraint on the gain coefficient if one desires invisibility. This case constricts the required wavelength interval and reduces the range of gain coefficient as apart from reflectionless configuration. The necessary condition of $M_{11} = M_{22} =1$ for invisibility gives rise to
\be
\cos \fa_4\left\{\cos \fa_2 - i\fu_2^{+}\sin \fa_2 \right\} + \sin \fa_4\left\{\left[\fu_2^{-}\fu_4^{-}e^{2ik_zs} - \fu_2^{+}\fu_4^{+}\right]\sin \fa_2 - i \fu_4^{+} \cos \fa_2\right\} = e^{-2ia_0}\notag
\ee
This can be expressed in an expanded form as follows
\begin{align}
e^{2i\fa_4} = \frac{4\fu_2\fu_4 e^{i(\fa_2 + \fa_4 -2k_z L)} + (\fu_4 + 1)^2 \left[(\fu_2 - 1)^2 e^{2i\fa_2} - (\fu_2 + 1)^2\right] - (\fu_2^2 - 1)(\fu_4^2 - 1)e^{2ik_zs}\left(e^{2i\fa_2} - 1\right)}{(\fu_4 - 1)^2 \left[(\fu_2 - 1)^2 e^{2i\fa_2} - (\fu_2 + 1)^2\right] - (\fu_2^2 - 1)(\fu_4^2 - 1)e^{2ik_zs}\left(e^{2i\fa_2} - 1\right)} \notag
\end{align}
This complex relation can be split into real and imaginary parts in a perturbative manner as performed in above section to yield
\begin{align}
&\left[1 - \cos 2k_z s + \gamma_{\ell}^2 (1 + \cos 2k_z s)\right]\cos 2a_0\tilde{\eta}
- (1 - \gamma_{\ell}^2)(1 - \cos 2k_z s)\cosh (\tilde{g}L) = \frac{\gamma_{\ell}^2}{2} \cos (2k_z L)&\label{transmission1}\\
&(1 - \gamma_{\ell}^2)\sin 2k_z s \left[\cos 2a_0\tilde{\eta} - \cosh (\tilde{g}L)\right]
+ 2\gamma_{\ell}\sin 2a_0\tilde{\eta} + 2 \alpha_{\ell} (1 - \gamma_{\ell}^2)\sinh (\tilde{g}L) = \frac{\gamma_{\ell}^2}{2} \sin (2k_z L)& \label{transmission2}
\end{align}
 In Figs.~\ref{invgain1}, \ref{invgain2}, \ref{invgain3}, \ref{invgain4}, \ref{invgain5} and \ref{invgain6}, the behaviors of (\ref{transmission1}) and (\ref{transmission2}) on the invisibility phenomenon at various angles and wavelength ranges corresponding to constructive, destructive and arbitrary cases are explicitly displayed.
 In Figs~\ref{invgain1}, \ref{invgain2} and \ref{invgain3}, invisibility patterns corresponding to constructive configuration with $s/s_0 = 20$ for various angles are shown for the $\cP\cT$-symmetric system consisting of Nd:YAG crystals with slab thickness $L = 10~\textrm{cm}$. In Figure~\ref{invgain1}, incident angle is $\theta = 30^{\circ}$. In this case, no left invisibility is observed, and right invisibility is encountered at the gain value of $g \approx 0.965~\textrm{cm}^{-1}$ in TE case and $g \approx 1.035~\textrm{cm}^{-1}$ in TM case. Likewise, bidirectional reflectionlessness is observed up to $g \approx 0.3~\textrm{cm}^{-1}$ in TE case and $g \approx 0.35~\textrm{cm}^{-1}$ in TM case. Once the amounts of gain are increased from these values, unidirectional reflectionlessness is observed. Once the precision of measurements is increased, the ranges of invisibility come down to the specified points and only right and left reflectionless configurations stand explicitly, see bottom figures in Fig.~\ref{invgain1}. We notice that varying incident angle results in the curve of unity-transmission to move around the zero-reflection amplitude curves. In Figure~\ref{invgain2}, incident angle is slightly increased to the value of $\theta = 30.6^{\circ}$. This angle causes the right invisibility gain range to lie within $(0.3, 0.55)~\textrm{cm}^{-1}$ in TE case and $(0.35, 0.62)~\textrm{cm}^{-1}$ in TM case. Notice that bidirectional invisibility is observed below these gain values. Likewise, left reflectionlessness is seen at gain values higher than  $g \approx 0.3~\textrm{cm}^{-1}$ in TE case and  $g \approx 0.35~\textrm{cm}^{-1}$ in TM case, and right reflectionlessness at gain values higher than  $g \approx 0.55~\textrm{cm}^{-1}$ in TE case and  $g \approx 0.62~\textrm{cm}^{-1}$ in TM case. Again once the precision of measurements is increased, the left invisibility in TE case and right invisibility in TM case stand just at a single point in $(\lambda, g)$-plane. In Figure~\ref{invgain3}, incident angle is slightly increased to $\theta = 30.9^{\circ}$. This time right invisibility gives its place to left one. Left invisibility in encountered at the gain value of $g \approx 0.75~\textrm{cm}^{-1}$ for both TE and TM cases. Bidirectional and unidirectional reflectionlessness is observed in a similar manner.  Top and bottom figures clearly show that only a perfect right invisible configurations occur between gain values $(0.35, 0.75)~\textrm{cm}^{-1}$ in TE case and $(0.3, 0.75)~\textrm{cm}^{-1}$ in TM case even if the precision of measurement is increased.

 In Figure~\ref{invgain4}, invisibility patterns of generic configuration with $s/s_0 = 20.5$ are displayed for the incident angles of $\theta = 30^{\circ}$ (top row) and $\theta = 30.32^{\circ}$ (bottom row). We immediately notice that required gain values scales down considerably, and left zero-reflection amplitudes are always less than right zero-amplitudes. The curve for unity-transmission case locates evely as the right zero reflection amplitude. At angle $\theta = 30^{\circ}$, right invisibility is observed at gain values of $g \approx 0.165~\textrm{cm}^{-1}$ in TE case and $g \approx 0.225~\textrm{cm}^{-1}$ in TM case. The required gain values for bidirectional reflectionlessness are less than $g \approx 0.0082~\textrm{cm}^{-1}$ in TE case and $g \approx 0.01~\textrm{cm}^{-1}$ in TM case. But, the angle $\theta = 30.32^{\circ}$ lets the gain range of right invisibility increase up to the amounts of $g \approx 0.11~\textrm{cm}^{-1}$ in TE case and $g \approx 0.175~\textrm{cm}^{-1}$ in TM case. In this configuration, it is hard to get left invisibility. Consequently, we understand that right invisibility mostly occurs in case of TE mode in almost all angles. Once the precision of measurement is increased,  top figures do not yield invisibility, and bottom ones still holds the right invisibility.

 In Figure~\ref{invgain5}, invisibility pattern for the almost destructive case with $s/s_0 = 20.99$ corresponding to  $\cP\cT$-symmetric Nd:YAG layers with slab thickness of $L = 10~\textrm{cm}$ is observed at the incident angle of $\theta = 30^{\circ}$. We see that required gain for the zero-reflection amplitudes lowers significantly, especially left one. This results in achieving right invisibility range of gain to take values from $g \approx 9\times 10^{-6}~\textrm{cm}^{-1}$ in TE case and $g \approx \times 10^{-5}~\textrm{cm}^{-1}$ in TM case, up to an extended level. Also there are one more points of right invisibility for both cases at gain values of $g \approx 0.21~\textrm{cm}^{-1}$ in TE case and $g \approx 0.27~\textrm{cm}^{-1}$ in TM case. As a consequence, no left reflectionlessness and invisibilities are observed in this case. In the broad range of wavelength, this case leads all three distinct curves to coincide at the same wavelength values whereas producing only right invisible configurations above certain gain values. But when the precision is increased as in above figures, right and left invisibilities are distinguished.

 In Figure~\ref{invgain6}, Reflectionlessness and invisibility patterns are shown in the plane of incident angle and gain coefficient for the materials of $\cP\cT$-symmetric gain-loss system with a gap. For clarity, we make use of quite small slab thickness of $L = 200~\mu\textrm{m}$ at wavelength $\lambda = 808~\textrm{nm}$, which leads to a large amount of gain values. At fixed thickness $L$, gap value of $s$ and wavelength $\lambda$, not all angles, but discrete finite number of angles produce reflectionless and invisible situations. At the prescribed value of $s \approx 1.167~\mu m$, corresponding to $s/s_0 = 5$, small angles nearby $\theta = 0^{\circ}$ we have right invisibilities on large amount of gain. Also bidirectional invisibilities and reflectionlessness are observed at the displayed values of gain. Required gain values drop off incredibly when the incident angle is adjusted well so that the valleys of pattern at which the left zero-reflection curve forms up are favorable for this purpose, which happens around $\theta = 30^{\circ}$, $\theta = 60^{\circ}$ and $\theta = 90^{\circ}$ for our case.

We now investigate the behaviour of Reflection and Transmission amplitudes in view of these results. Following figures, Figs.~\ref{invgain7},~\ref{invgain8} and ~\ref{invgain9} demonstrate invisible wavelength ranges at various angle and gain values. In Fig.~\ref{invgain7}, plots of $\left|R^{l}\right|^2$, $\left|R^{r}\right|^2$ and $\left|T\right|^2 -1$ are shown as a function of  wavelength $\lambda$ for incident angles of $\theta = 0^{\circ}, 60^{\circ}$ and $\theta = 85^{\circ}$ for $\cP\cT$-symmetric Nd:YAG crystals with thickness $L = 1~\textrm{cm}$, $s = 0$, and gain coefficient $g = 46.66~\textrm{cm}^{-1}$. We clearly see that once the incident angle is small around $\theta = 0^{\circ}$, the range of wavelength producing invisibility increases, and unidirectional reflectionlessness slightly occurs. When the incident angle rises incredibly the width of wavelength range decreases for invisibility.

\begin{figure}
	\begin{center}
    \includegraphics[scale=0.5]{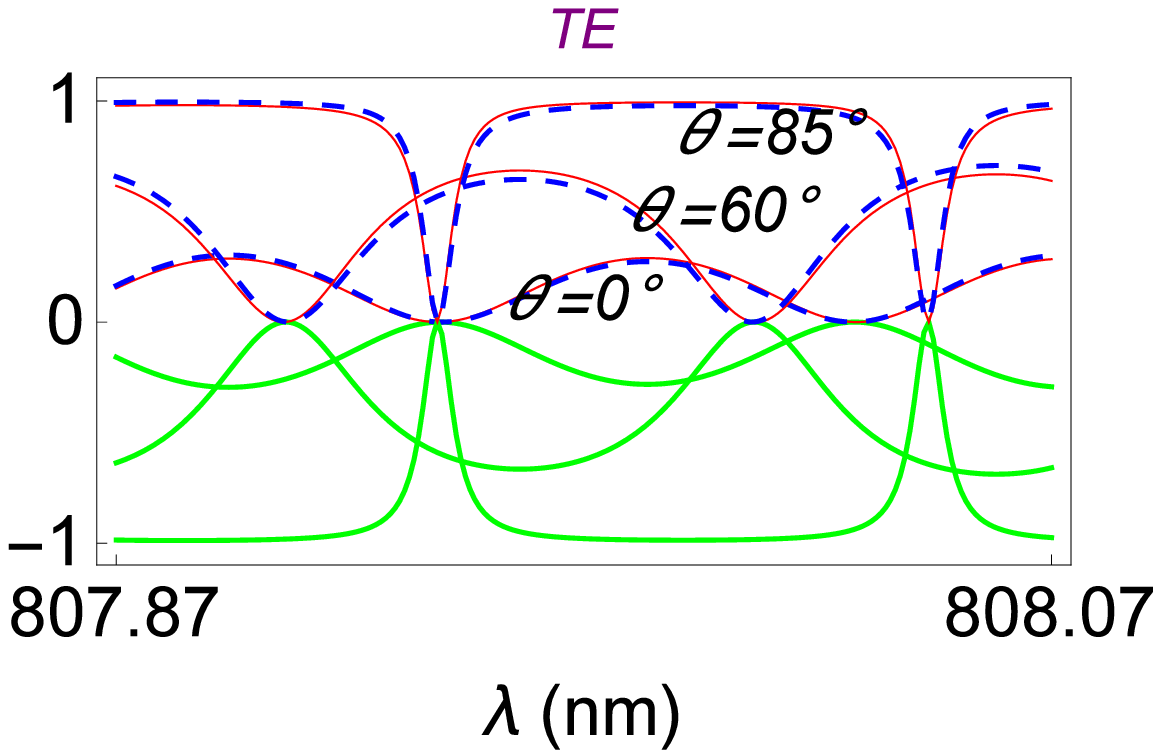}~~~
    \includegraphics[scale=0.5]{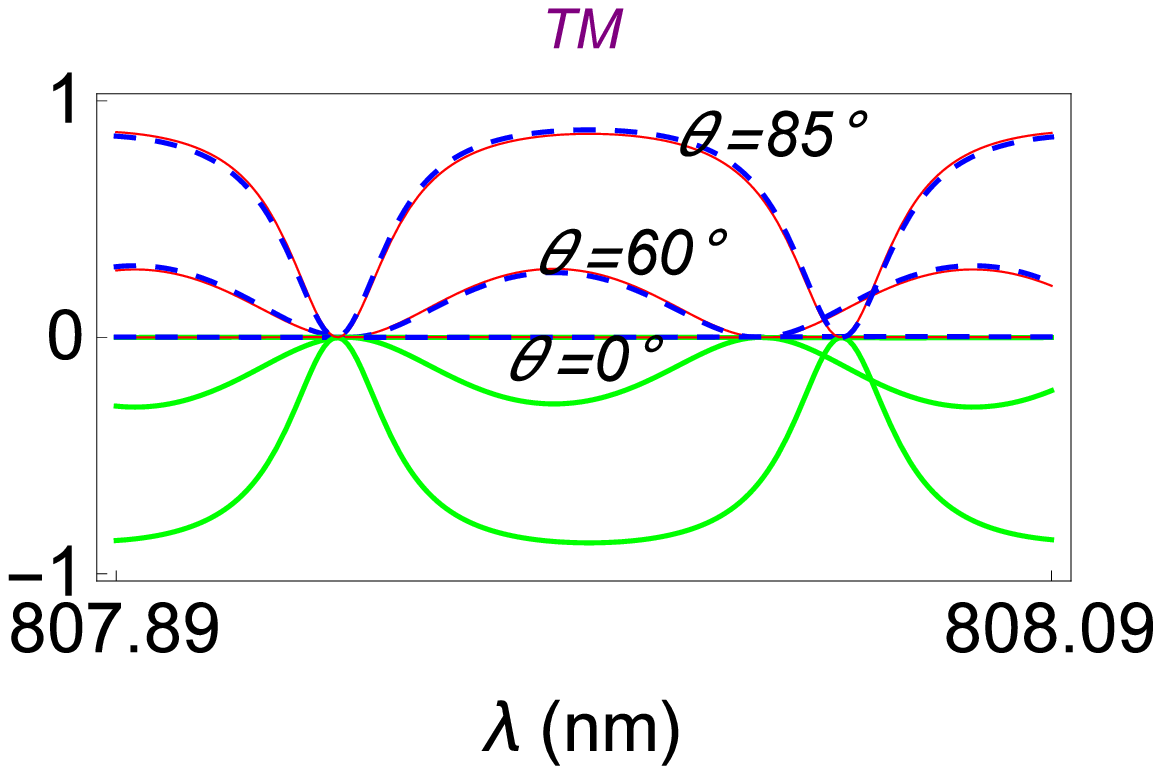}\\
    \includegraphics[scale=0.5]{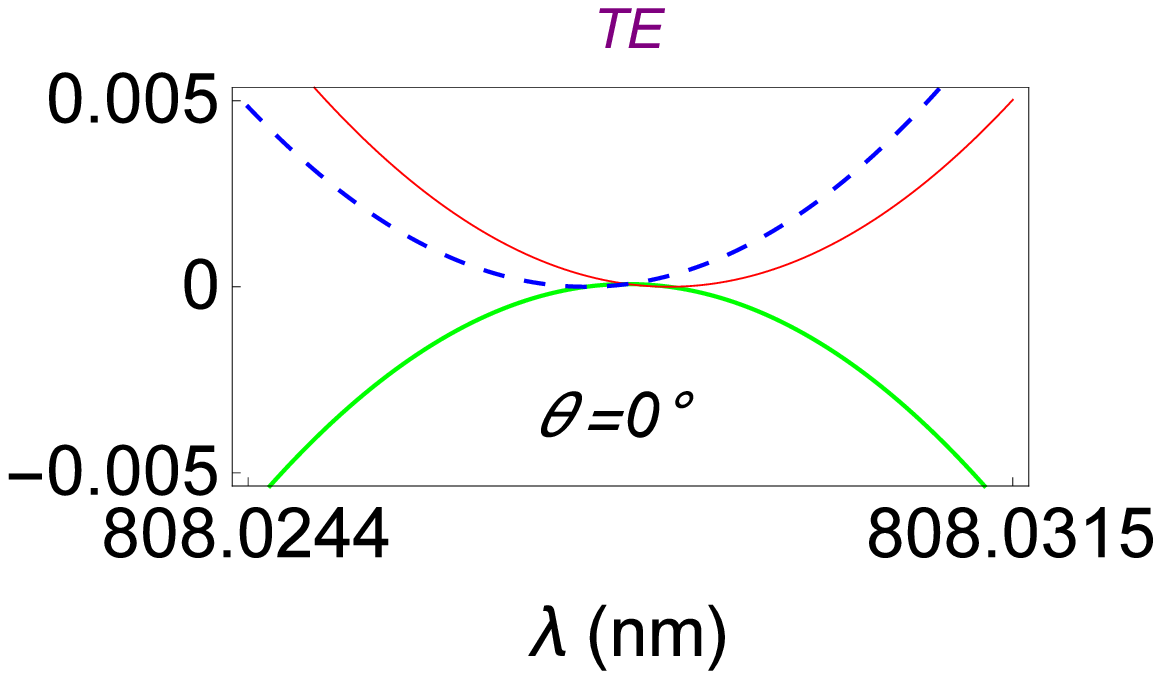}~~~
    \includegraphics[scale=0.5]{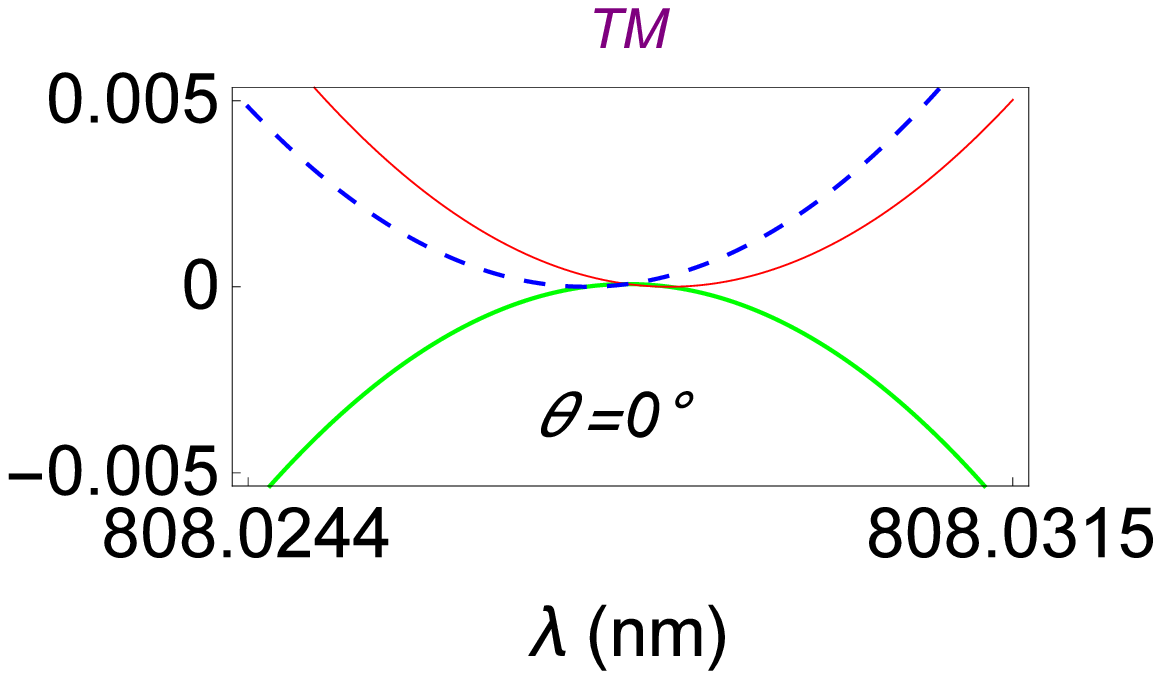}\\
    \includegraphics[scale=0.5]{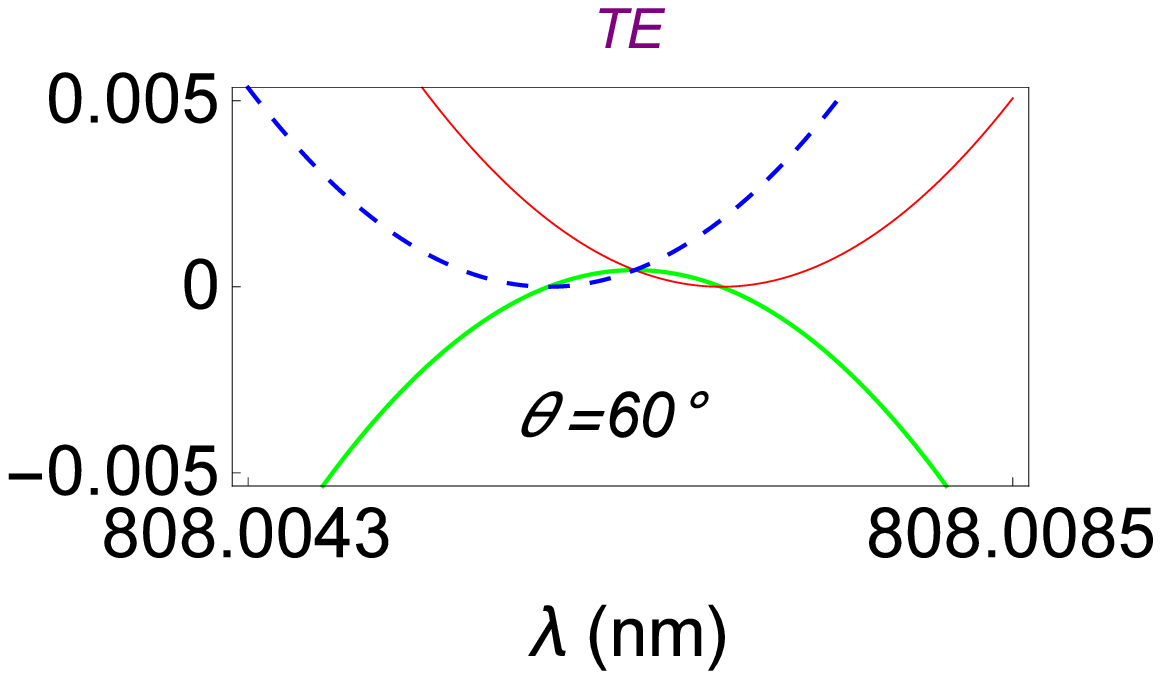}~~~
    \includegraphics[scale=0.5]{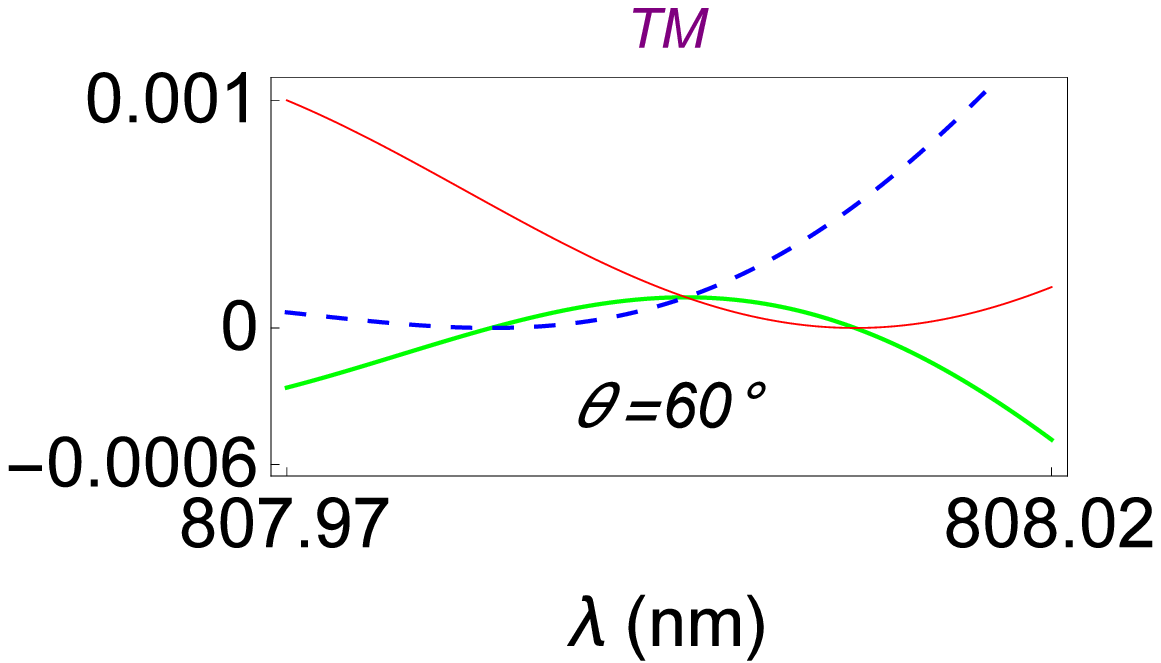}\\
    \includegraphics[scale=0.5]{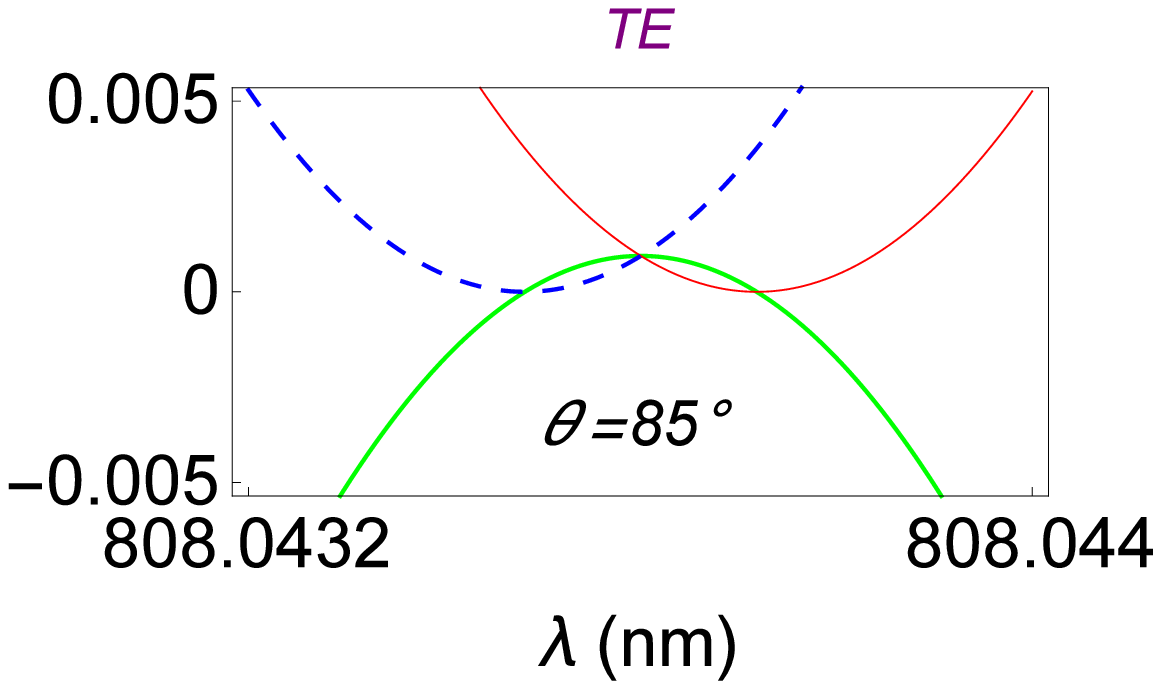}~~~
    \includegraphics[scale=0.5]{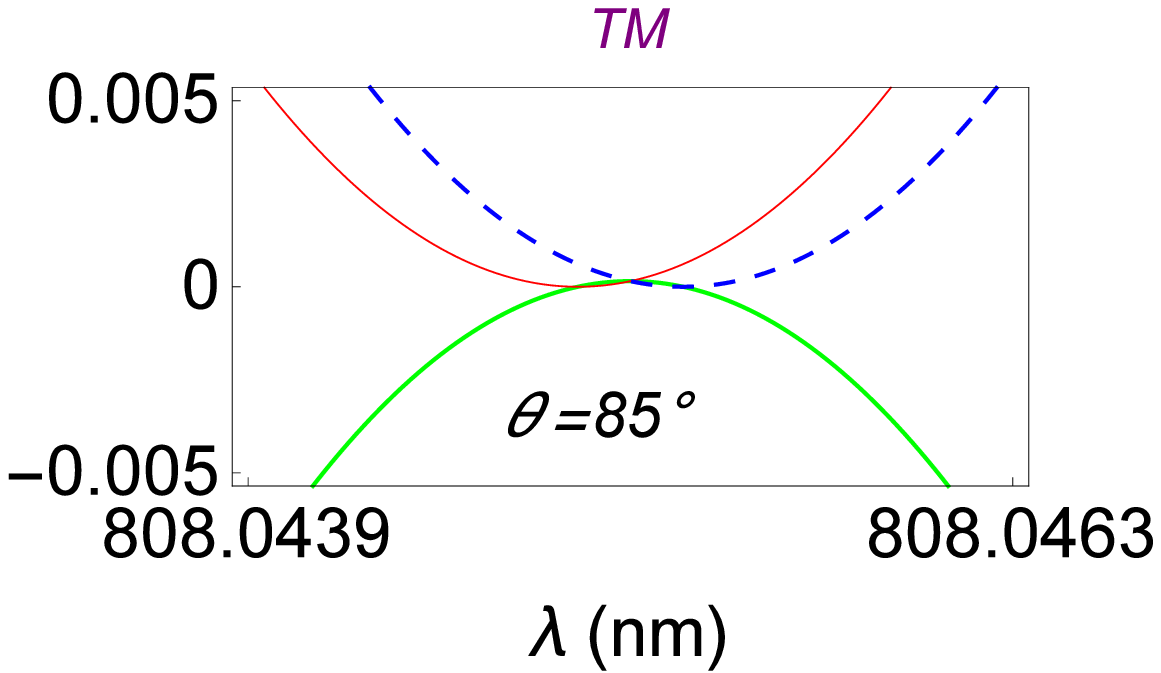}
	\caption{(Color online) Plots of $\left|R^{l}\right|^2$ (dashed blue curve), $\left|R^{r}\right|^2$ (solid thin red curve) and $\left|T\right|^2 -1$ (solid thick green curve) as a function of  wavelength $\lambda$ corresponding to TE and TM wave solutions at various angles for the case of $\cP\cT$-symmetric layers with a gap. }
    \label{invgain7}
    \end{center}
    \end{figure}

In Fig.~\ref{invgain8}, dependence of $\left|R^{l}\right|^2$, $\left|R^{r}\right|^2$ and $\left|T\right|^2 -1$ on wavelength is revealed for the cases of constructive, almost destructive and generic configurations. We again use Nd:YAG crystals with slab thickness $L = 10~\textrm{cm}$ for all configurations. Top figure corresponds to the constructive case with gain value of $g = 0.7~\textrm{cm}^{-1}$ and $s/s_0 = 20$. We see that point $a$ bidirectional invisibility point, point $b$ is left invisibility point and point $c$ is the right invisibility one if one explores invisibility within wavelength range of ten thousandths. For a wide range of wavelength, they all appear to be same and only one point leads to bidirectional invisibility. If we take a look at middle figure with gain value of $g = 0.0031~\textrm{cm}^{-1}$ and $s/s_0 = 20.5$ in the same range of wavelength we observe that both points $a$ and $b$ are bidirectionally invisible points, and around these points unidirectional reflectionlessness occurs. Finally, if we adjust the gap amount to be $s/s_0 = 20.99$ with the corresponding gain value $g = 0.00016~\textrm{cm}^{-1}$ so that almost destructive case arises, between points $a$ and $b$ appears to be bidirectionally invisible. Further away from these points leads to unidirectional reflectionlessness.

\begin{figure}
	\begin{center}
    \includegraphics[scale=0.5]{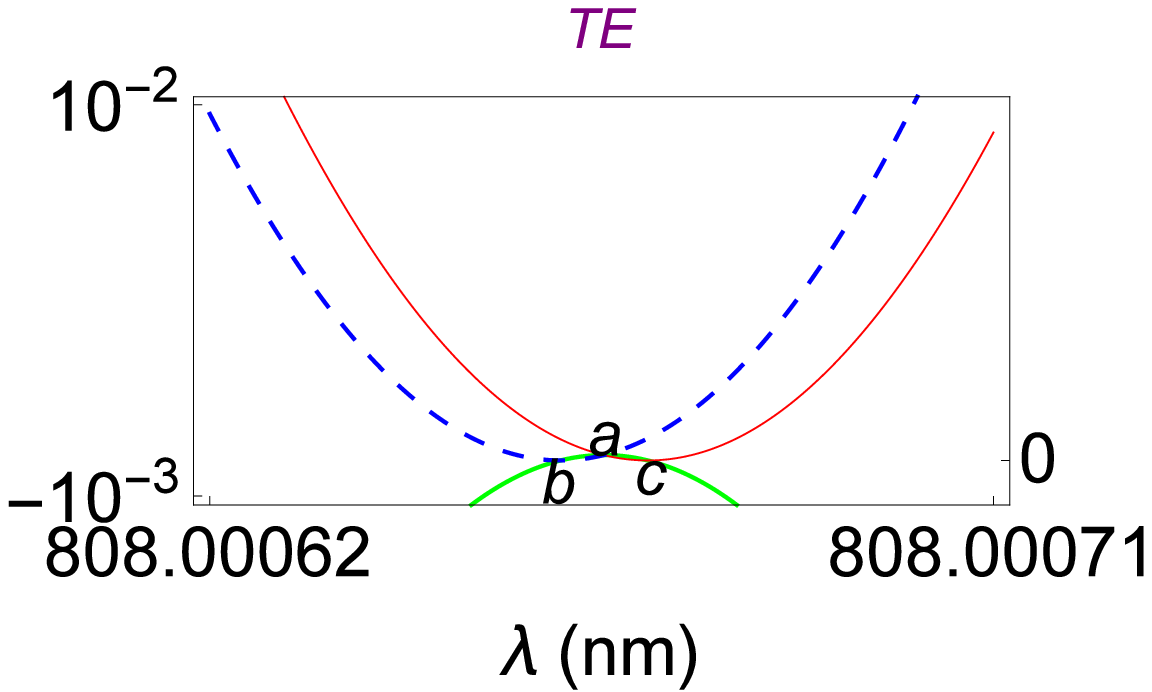}~~~
    \includegraphics[scale=0.5]{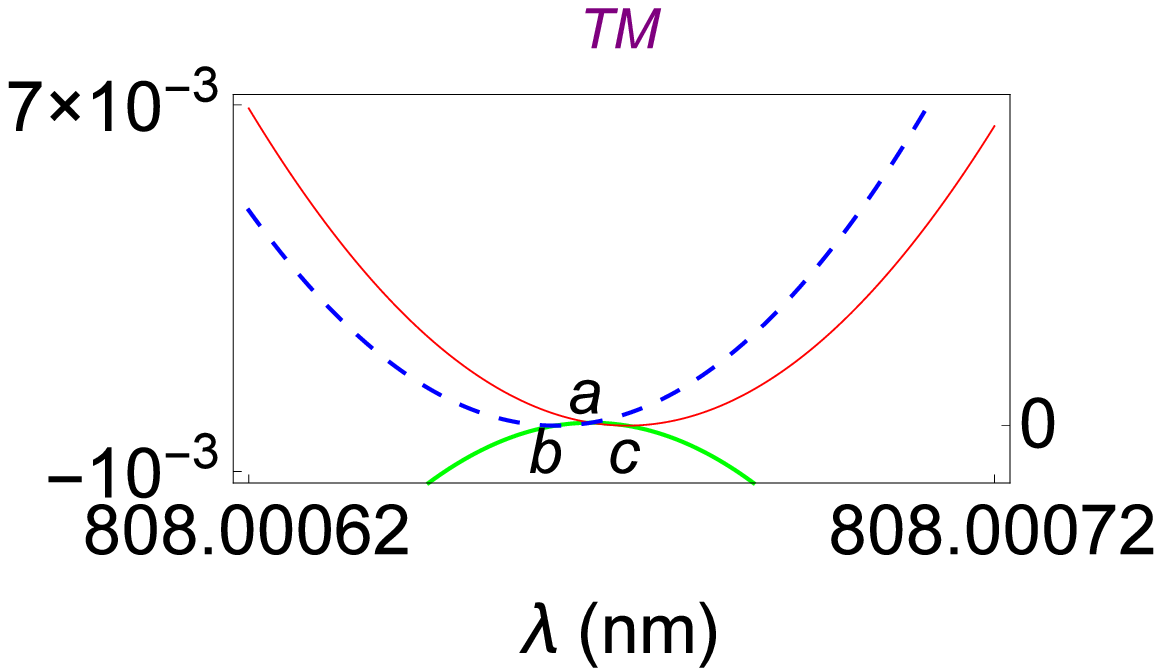}\\
    \includegraphics[scale=0.5]{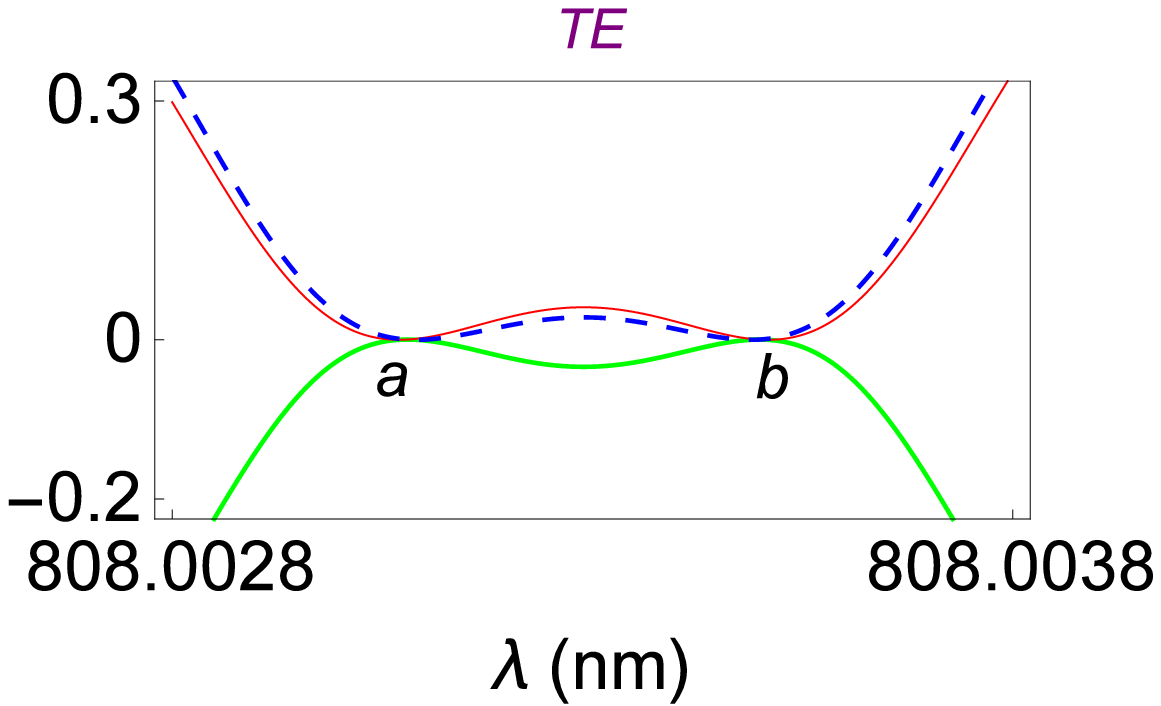}~~~
    \includegraphics[scale=0.5]{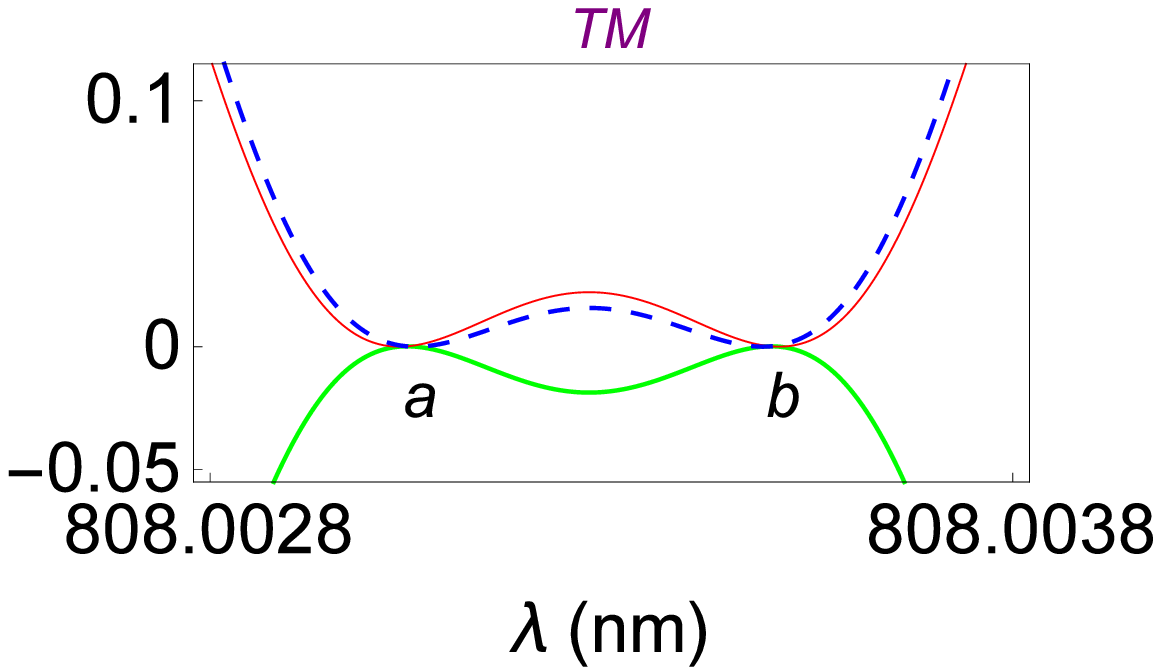}\\
    \includegraphics[scale=0.5]{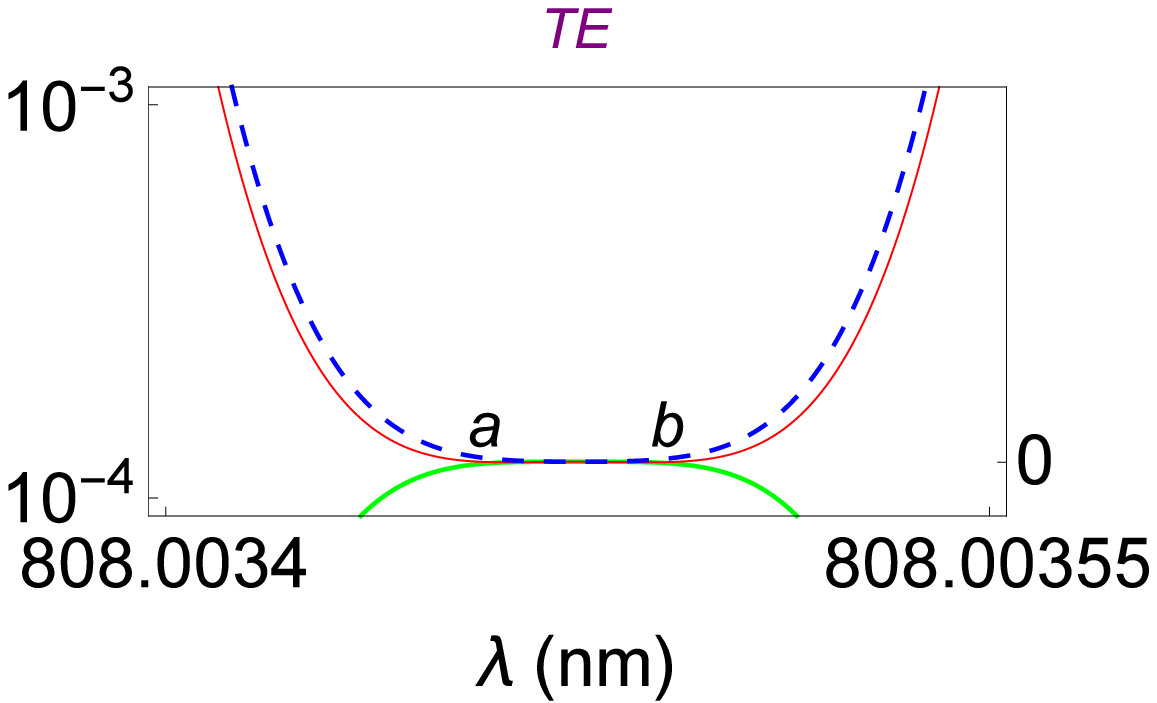}~~~
    \includegraphics[scale=0.5]{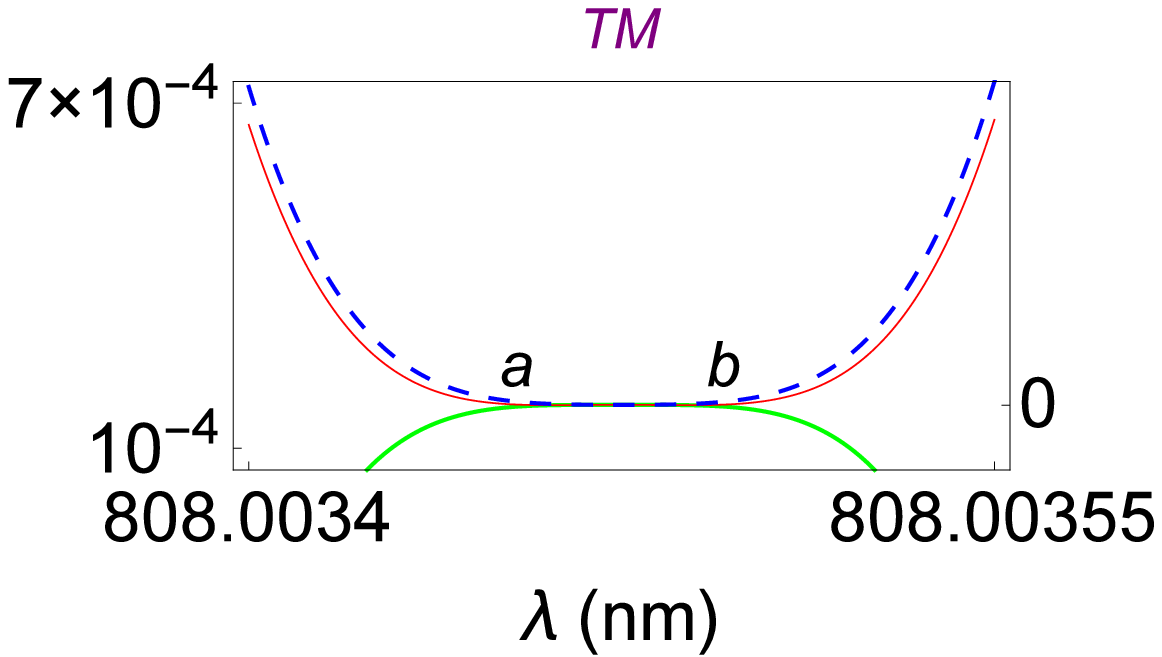}
	\caption{(Color online) Plots of $\left|R^{l}\right|^2$ (dashed blue curve), $\left|R^{r}\right|^2$ (solid thin red curve) and $\left|T\right|^2 -1$ (solid thich green curve) as a function of  wavelength $\lambda$ corresponding to constructive, generic and destructive TE and TM wave solutions for the case of $\cP\cT$-symmetric layers with a gap. The materials of $\cP\cT$-symmetric gain-loss system with a gap is made out of Nd:YAG crystals with $\eta = 1.8217$, thickness $L = 10~\textrm{cm}$, }
    \label{invgain8}
    \end{center}
    \end{figure}

In Fig.~\ref{invgain9}, plots of $\left|R^{l}\right|^2$, $\left|R^{r}\right|^2$ and $\left|T\right|^2 -1$ as a function of incidence angle $\theta$ are seen at parameter values of $L = 10~\textrm{cm}$, $g = 0.0078~\textrm{cm}^{-1}$ and wavelength $\lambda = 808~\textrm{nm}$. We notice that at fixed parameters, only certain prescribed angles give rise to reflectionless and invisible configurations. In TE case, angles around $\theta = 90^{\circ}$ do not yield any reflectionless and invisible patterns.

\begin{figure}
	\begin{center}
    \includegraphics[scale=0.5]{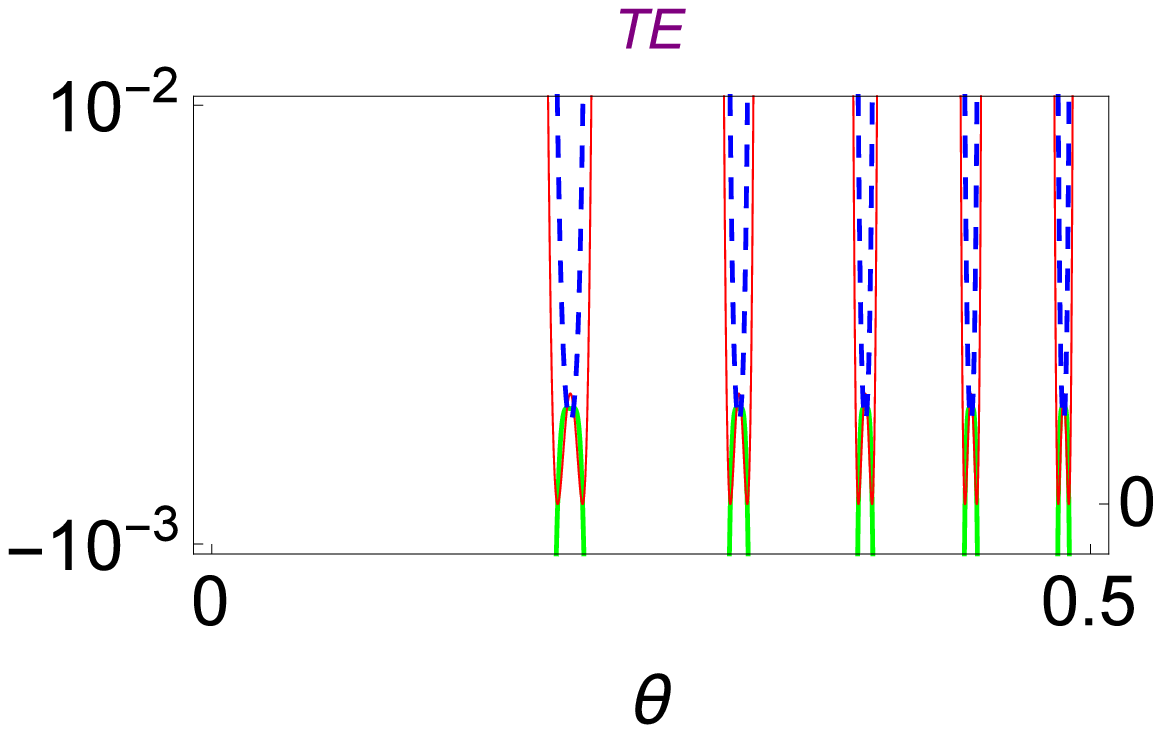}~~~
    \includegraphics[scale=0.5]{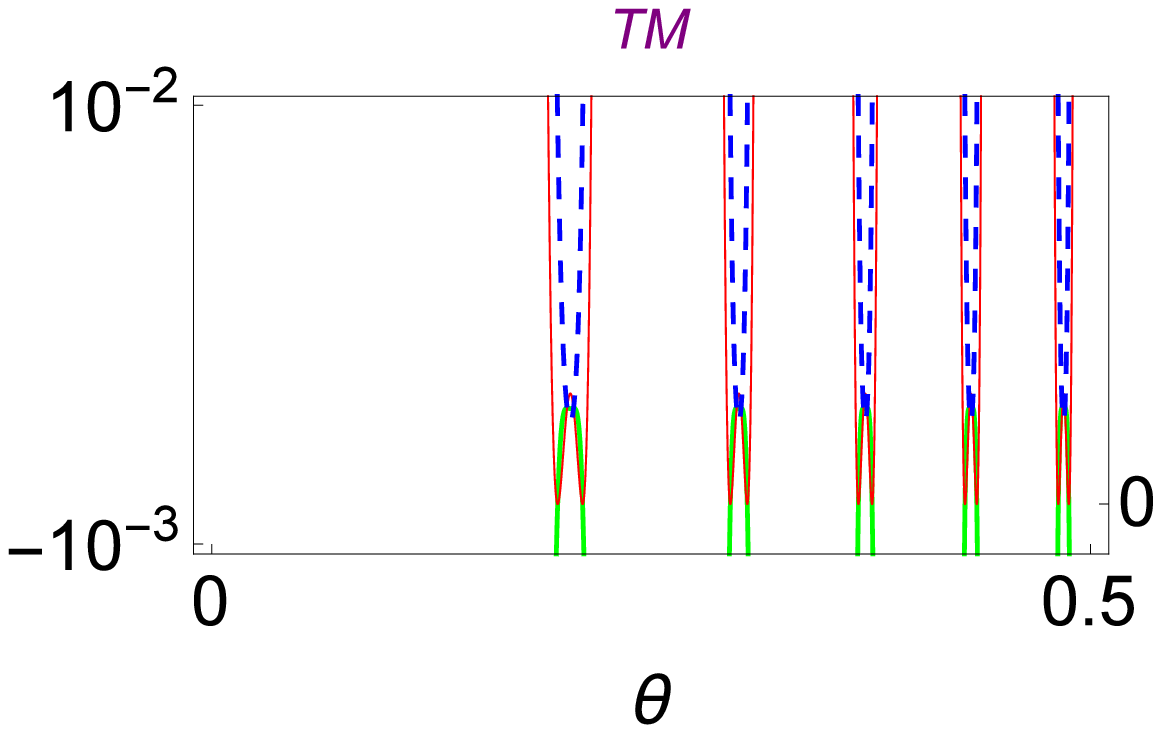}\\
    \includegraphics[scale=0.5]{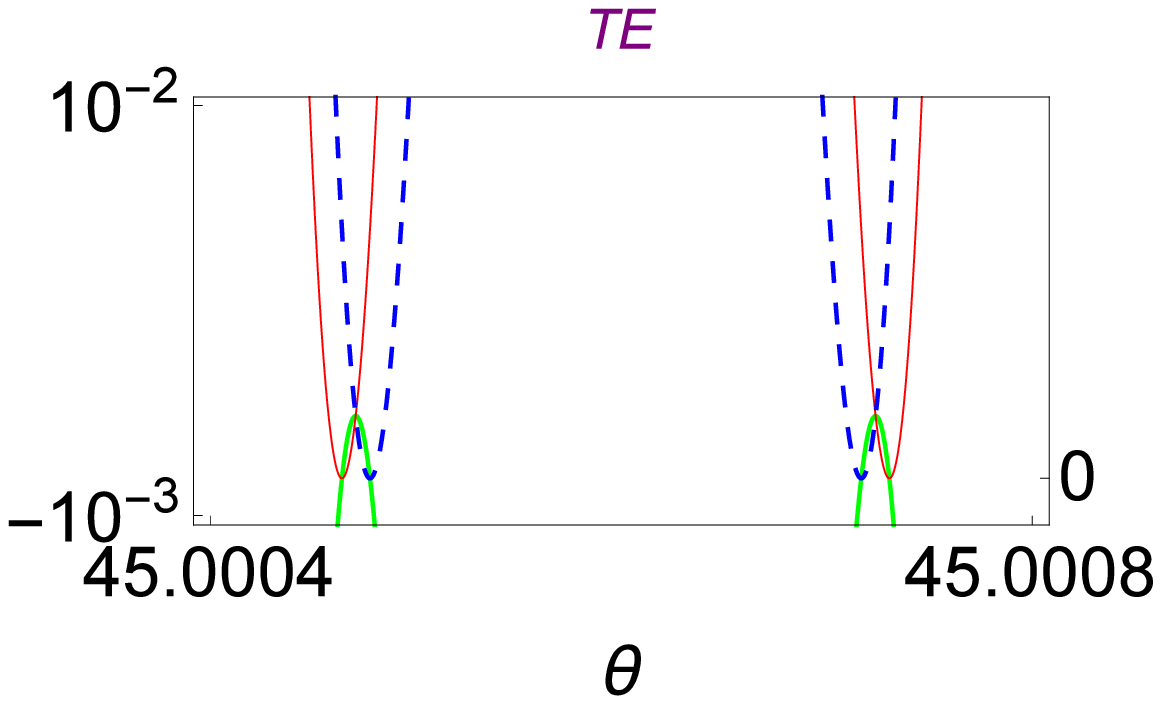}~~~
    \includegraphics[scale=0.5]{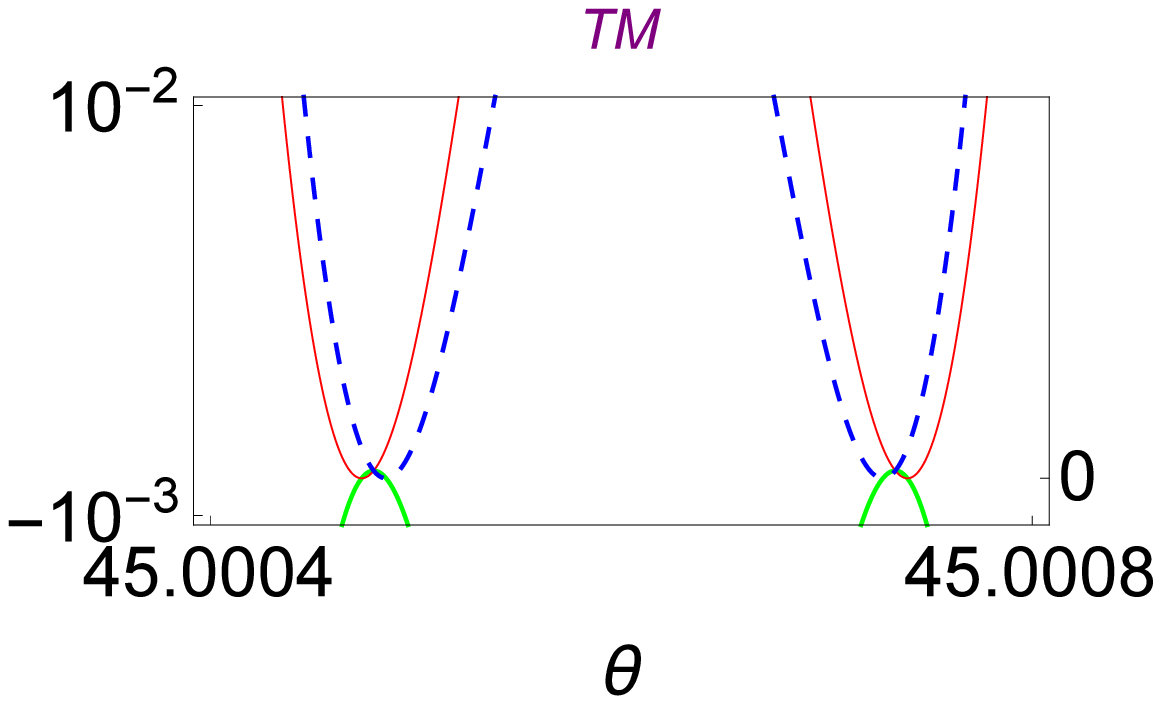}\\
    \includegraphics[scale=0.5]{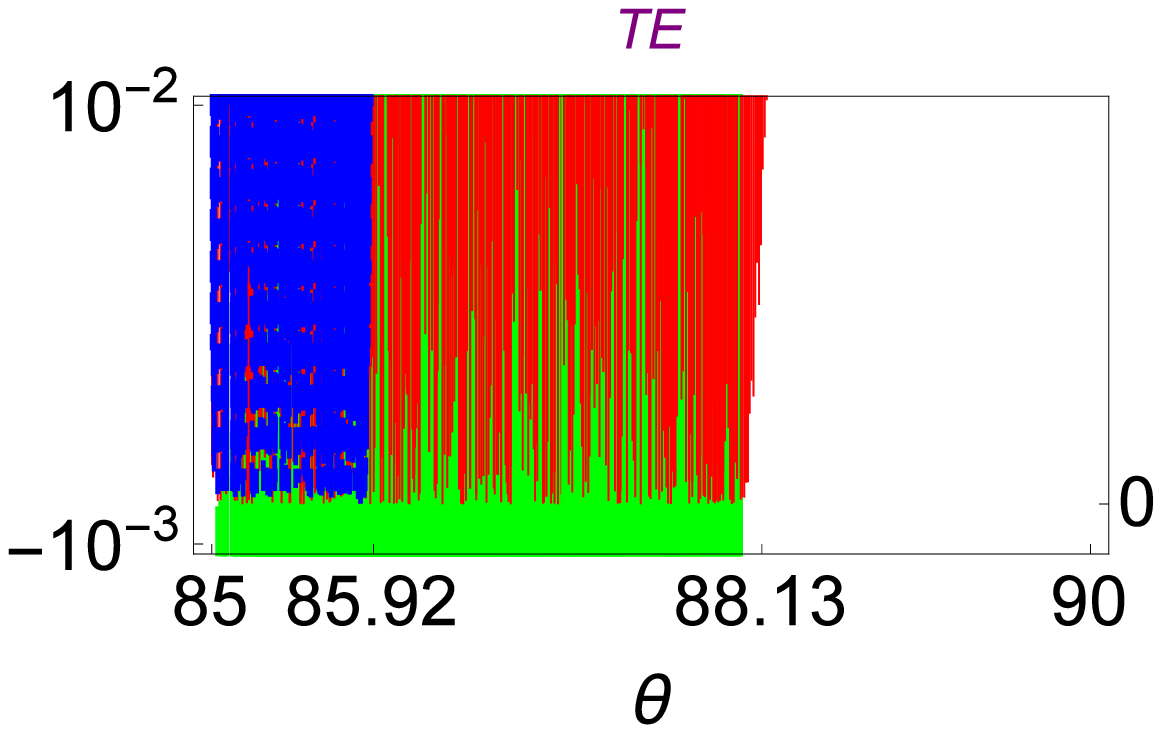}~~~
    \includegraphics[scale=0.5]{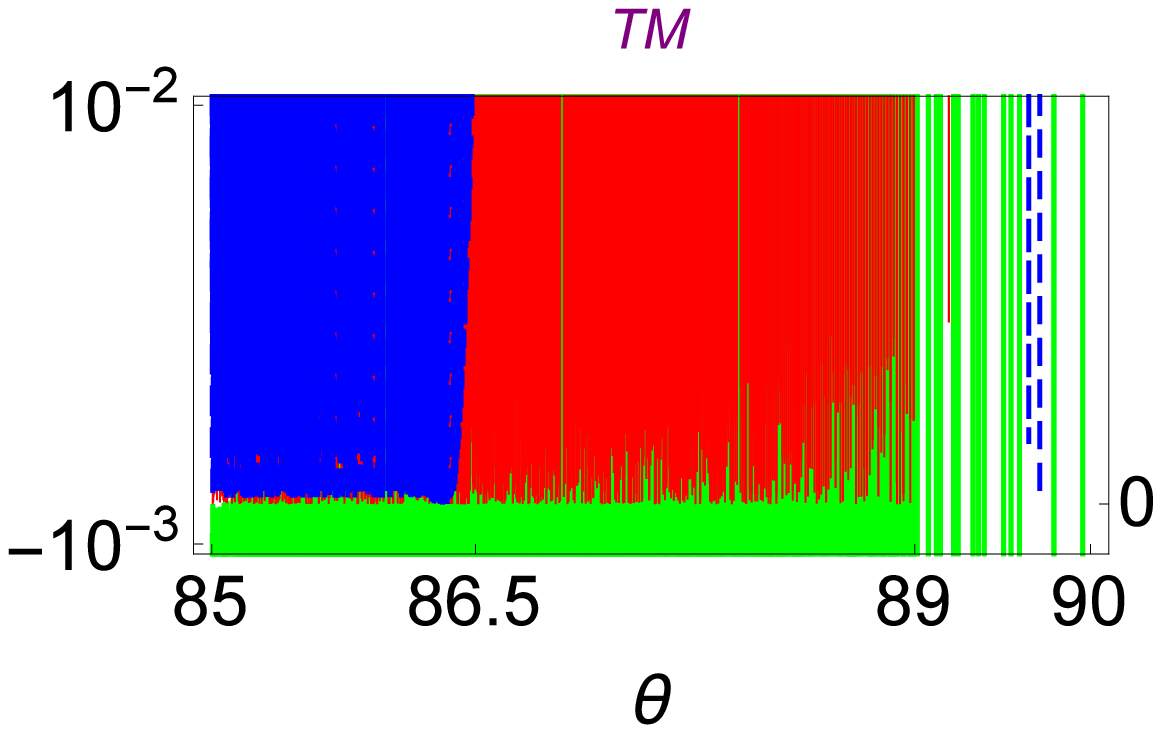}
	\caption{(Color online) Plots of $\left|R^{l}\right|^2$ (dashed blue curve), $\left|R^{r}\right|^2$ (solid thin red curve) and $\left|T\right|^2 -1$ (solid thick green curve) as a function of incident angle $\theta$ corresponding to TE and TM wave solutions for the case of $\cP\cT$-symmetric layers with a gap. }
    \label{invgain9}
    \end{center}
    \end{figure}

\section{Concluding Remarks}
\label{S9}

In this article, we analyzed the behavior of $\cP\cT$-symmetric bilayer and two-layer system in oblique TE and TM cases corresponding to unidirectional reflectionlessness and invisibility, and their optical realizations. We exploited the power of transfer matrix which emphasizes the validity of boundary conditions arising from the solutions directly coming from Maxwell's equations. It is a direct consequence of transfer matrix that a single layer consisting of just gain or loss can not produce an invisible configuration whereas gain-loss system constituting a two-layer pattern can do. In our analysis we developed a method which can yield invisibility at various angles for TE and TM wave solutions. We also obtained necessary and sufficient conditions leading to reflectionless and invisible solutions. We showed that the separation between gain and loss plays a crucial role in obtaining reflectionless and invisible patterns.

We obtained that reflectionless and invisible patterns are very sensitive to incident angle, and occur only at specific angles. Also, amount of gain and gap value between gain and loss regions determines the ascribed phenomena such that optimal values of these parameters should be adjusted in a given system if one desires reflectionless and invisible situations. Unidirectional invisibility requires a certain range of gain values at the predetermined system parameters. Precision of measurement plays an important role since it can split reflection and transmission amplitudes consistently whereas they were invisible before at far wavelength range, they may not be invisible at small ranges.\\[6pt]

\noindent{\em Acknowledgments:}  We are grateful to Ali Mostafazadeh for fruitful discussions and invaluable comments. This work has been supported by  the Scientific and Technological Research Council of Turkey (T\"UB\.{I}TAK) in the framework of the project no: 112T951.

\end{document}